\documentclass[usenatbib,onecolumn]{mn2e}
\usepackage{amsmath, mathrsfs, amssymb}
\usepackage{graphicx}
\usepackage{hyperref}
\usepackage{rotating}
\usepackage{longtable,lscape}
\usepackage{morefloats}
\usepackage{color}
\usepackage[usenames,dvipsnames,svgnames,table]{xcolor}
\usepackage{caption}
\usepackage{verbatim}
\usepackage{color}
\usepackage{bm}
\usepackage[T1]{fontenc}
\usepackage{aecompl}

\title[Internal instabilities in magnetized jets]{Internal instabilities 
in magnetized jets}

\author[Das \& Begelman]{Upasana~Das$^{1}\thanks{E-mail: upasana.das@jila.colorado.edu}$ and
Mitchell~C.~Begelman$^{1,2}$ \\
$^{1}${JILA, University of Colorado and National Institute of Standards and Technology, 440 UCB, Boulder,
CO 80309-0440, USA.} \\
$^{2}${Department of Astrophysical and Planetary Sciences, University of Colorado, 391 UCB, Boulder, CO 80309-0391, USA.}}

\begin{document}
\label{firstpage}
\maketitle

%===========================================================================
%===========================================================================
% ABSTRACT
\begin{abstract}
We carry out an extensive linear stability analysis of magnetized 
cylindrical jets in a global framework. Foregoing the commonly invoked force-free limit, we 
focus on the small-scale, internal instabilities triggered in regions of the jet 
dominated by a toroidal magnetic field, with a weak vertical field and finite thermal pressure 
gradient. Such regions are likely to occur far from the jet source and boundaries, and 
are potential sites of magnetic energy dissipation that is essential to explain the 
particle acceleration and radiation observed from astrophysical jets. We validate the local 
stability analysis of Begelman by verifying that the eigenfunctions of the most 
unstable modes are radially localized. This finding allows us to propose a generic 
stability criterion in the presence of a weak vertical field.  A 
stronger vertical field with a radial gradient complicates the stability criterion, due to the competition 
between the destabilizing thermal pressure gradient and stabilizing 
magnetic pressure gradients. Nevertheless, we argue that the jet interiors generically should be subject 
to rapidly growing, small-scale instabilities, capable of producing 
current sheets that lead to dissipation.
We identify some new instabilities, not predicted by the local analysis, 
which are sensitive to the background radial profiles but have smaller growth rates than the 
local instabilities, and discuss the relevance of our work to the 
findings of recent numerical jet simulations.
\end{abstract}

 %and that these instabilities should tend to drive the jet towards a core-sheath structure.

%===========================================================================
%===========================================================================
% KEYWORDS
\begin{keywords}
instabilities - MHD - methods: numerical - stars: jets - galaxies: jets
\vspace{-6mm}
\end{keywords}

%===========================================================================
%===========================================================================
% INTRODUCTION
\section{Introduction}
\label{sec_intro}

Astrophysical jets are bright, collimated and fast moving outflows that emanate from a wide 
variety of systems including young stellar objects (YSOs), active galactic nuclei (AGN), microquasars 
%or stellar mass black holes in X-ray binaries 
 and gamma ray bursts (GRBs). 
Depending upon the source, these jets can be relativistic (e.g. AGN, microquasars, GRBs) or non-relativistic 
(e.g. YSOs).  
They traverse distances much larger than their initial radius, which is typically the 
size of their source, all the while maintaining a large-scale coherent structure. For instance, the jets 
from YSOs traverse
$\lesssim$ 1pc, which is $\sim 10^5-10^7$ times the size of the central star, and those from AGN 
traverse $\gtrsim 10^6$ pc, which is $\sim 10^9$ times the size of the central supermassive black hole
(see e.g. \citealt{2005AdSpR..35..908D,2010PhyU...53.1199B} for an observational overview).
%
%They traverse large distances while maintaining a coherent structure 
%($\lesssim$ 1pc for YSOs and $\gtrsim 10^6$ pc for AGNs), before depositing their energy into the ambient medium 
%without being disrupted appreciably. 
Magnetic fields 
are believed to play an important role in governing the structure, 
dynamics and origin of these jets (e.g. \citealt{1977MNRAS.179..433B}). 
%span different distance, velocity and luminosity scales. 
In spite of the differences in their environments and  scales, the jets display very similar overall properties, not 
all of which are well understood. One of the major open areas of inquiry in this field is 
the question of jet stability 
(see e.g. \citealt{2011IAUS..275...41H,2013EPJWC..6102001H} for reviews on this subject).
%The open questions that are still a matter of debate in this field revolve around the formation, collimation, 
%propagation and acceleration of jets.
%explaining all of which are still open questions in the field.
%indicating 
%a common physical mechanism driving them.

It has been long known from basic plasma physics theory and laboratory experiments
that cylindrical, magnetized plasmas are unstable to various kinds of instabilities 
(see e.g. \citealt{1978mit..book.....B,1982RvMP...54..801F}, especially 
in the context of controlled fusion in tokamaks). 
%Before we contemplate on how these can affect the real jets in nature, we briefly list 
%and define the commonly studied instabilities in this context. 
In the case of a jet, these instabilities can be either kinetically or magnetohydrodynamically driven. 
%In the former category we have the 
The {\it Kelvin-Helmholtz} instability 
epitomizes the former category, and arises mainly due to the velocity shear created by the bulk motion of the jet 
with respect to the ambient medium (see \citealt{2013EPJWC..6102001H}). The latter category is sub-classified as 
(see e.g. \citealt{2008LNP...754..131L}): (i) {\it current-driven} 
instabilities, which are driven by the current parallel ($j_{\parallel}$)
to the total magnetic field ${\bf B}$ and are most easily studied in the context of cold, pressureless jets; and 
(ii) {\it pressure-driven} instabilities, which are driven by 
the gradient of the background plasma pressure and 
the electric current perpendicular ($j_{\bot}$) to ${\bf B}$. The curvature of the magnetic field lines 
plays an important role in confining the plasma in this case, and instability occurs when the {\it destabilizing} 
plasma pressure gradient is strong enough to push the plasma out of this curvature. 
If both components of the current are present, along with a finite 
plasma pressure gradient, then the instabilities will be of a {\it mixed} 
nature \citep[see e.g.][]{2016MNRAS.462.2970S}. It can be roughly shown that pressure-driven 
effects dominate if $B_z \ll B_\phi$, which yields $j_{\bot} >j_{\parallel}$, 
and  current-driven effects dominate if $B_z \gtrsim B_\phi$, which yields $j_{\bot} <j_{\parallel}$, 
where $B_\phi$ and $B_z$ are the toroidal and vertical magnetic fields, respectively 
(see e.g. \citealt{2011A&A...525A.100B}).
 %Note that the distinction between current-driven and pressure-driven 
 %instabilities is often not made in the jet literature, and they are both referred to as 
 %``current-driven instabilities'', as long as the jet is sufficiently magnetized. The current work avoids this oversight.
%In this work, we will focus on the modes that are predominantly pressure-driven.
The two most commonly studied magnetohydrodynamic (MHD) instabilities in magnetized plasmas are the $m=0$ pinch mode 
and the $m=1$ kink mode, where $m$ is the azimuthal wavenumber. 
The former leads to a pinching distortion 
of the jet and triggers the so-called ``sausage instability,'' while the latter 
induces helical perturbations giving rise to the ``kink instability''  \cite[see e.g.][]{1979ApJ...234...47H}.
%In a real jet it is likely that the complex magnetic field structure, bulk motion and velocity gradients 
%make it even harder to distinguish between the various instabilities except at suitable limits \citep{2002ApJ...580..800B}.

Depending on the lengthscales at which they operate, 
the above instabilities can be further classified as {\it internal} or {\it external}. Internal 
instabilities are small-scale modes compared to the jet radius, which are confined deep within 
the jet interior as the perturbations have no displacement at the jet boundaries. By definition 
these induce local changes (in morphology and energetics) without destroying the jet as a whole. External instabilities, 
on the other hand, are large-scale modes of the order of the jet radius, such that the perturbations have a non-zero 
displacement at the jet boundaries. These can potentially disrupt the whole jet.
For instance, the external kink instability can bend and wiggle the entire jet off-axis, the 
external pinch instability can constrict the jet radius at various heights, and the 
Kelvin-Helmholtz instability due to the velocity shear at the jet surface can lead to mixing 
and interaction with the ambient medium.

The classical approach to understanding instabilities in jets is to perform a linear stability analysis 
of the ideal MHD equations. Many
studies have been carried out using this approach, while invoking different magnetic field profiles, boundary conditions 
(impenetrable, free, outflow, etc.), methods of solution and physical assumptions 
(inclusion/exclusion of rotation, velocity shear, plasma pressure, etc.).
 Most of the analyses, however, have been carried out 
under the force-free assumption, where the system is magnetically dominated and 
the effect of plasma pressure is neglected. 
These include both non-relativistic 
\citep{1996A&A...314..995A,2000A&A...355..818A} and 
relativistic  
\citep{1996MNRAS.281....1I,1999MNRAS.308.1006L,2001PhRvD..64l3003T,2009ApJ...697.1681N} analyses. 
The key developments in the linear stability analysis of cylindrical, relativistic, rotating and 
force-free jets can be summarized as follows. 
\citet{1996MNRAS.281....1I} showed 
that stability is achieved if $B_z$ is radially constant, while \citet{1999MNRAS.308.1006L} showed that 
a $B_z$ decreasing radially outward induces instability. \citet{2001PhRvD..64l3003T}, 
however, found a completely different criterion governed by the angular velocity 
of the magnetic field lines. \citet{2009ApJ...697.1681N} 
studied the effect of poloidal field curvature along with rigid rotation, with the aim 
of establishing a link between the various stability criteria. They found that their criterion is a 
generalization of the Kruskal-Shafranov stability criterion (which, however, 
does not account for rotation) for the $m=1$ kink mode 
\citep[most relevant for tokamaks, e.g.][]{1982RvMP...54..801F}, which 
states that instability ensues when $B_\phi/B_z < 2\pi R_l/L$, where 
$L$ is the length of the plasma column and $R_l$ its cylindrical radius.
Recent numerical investigations in this context 
have been carried out by 
\citet{2012ApJ...757...16M,2013MNRAS.434.3030B,2014ApJ...784..167M,2016MNRAS.462.3031B,2017MNRAS.468.4635S} 
and \citet{2018MNRAS.473.2813S}.
\citet[][hereafter B98]{1998ApJ...493..291B} was the first to consider the effect of the plasma 
pressure gradient balancing the magnetic gradients in 
a  regime dominated by toroidal magnetic field
%which is likely to occur farther down the jet 
(later works accounting for the plasma pressure include 
\citealt{2002ApJ...580..800B,2011A&A...525A.100B,2012MNRAS.422.1436O,2012MNRAS.427.2480N,2015MNRAS.450..982K}). 
He performed a local study of the linear properties of 
short-wavelength, internal instabilities, and proposed a corresponding stability criterion.
%which are often ignored in the literature compared to their current-driven counterparts occurring in force free jets. 
%that underlines the circumstances under which these instabilities are stabilized.
One of our aims in this work is to compare and validate the solutions from B98's purely local calculations 
with those from the linear eigenvalue analysis in our global framework. 

With the advancement of computational resources, numerical simulations of 
both non-relativistic and relativistic jets are now possible. 
These simulations are either local, where a small section of 
the 3D jet that is comoving with the bulk flow is modeled using periodic boundary conditions 
along the jet axis \citep{2009ApJ...700..684M,2012MNRAS.422.1436O,2012ApJ...757...16M,2015MNRAS.452.1089P}; 
semi-global, where a pre-existing jet is simulated in a non-periodic 
computational domain to better study the spatial development of various 
instabilities \citep{2014ApJ...784..167M,2016ApJ...824...48S};
or fully global, where the whole 3D jet is simulated starting from its injection/launch 
to large-scale propagation \citep{2009A&A...507.1203M,2009MNRAS.394L.126M,2010MNRAS.402....7M,2016MNRAS.456.1739B}. 
Numerical simulations are important in order to study the non-linear evolution and saturation of 
both the internal and external instabilities, as well as to visualize the 
morphological changes brought about by them in order to compare with observations. 
However, they still have limitations, as the results may depend significantly on the numerical solvers 
employed, resolution, launching mechanisms, etc. (see e.g. \citealt{2012MNRAS.422.1436O}).
Thus, linear analyses are still relevant and essential for interpreting the complex numerical results 
and for making useful predictions.

Stability is a fundamental issue for jet studies in two complementary ways. 
First, since we know observationally that astrophysical jets survive 
and propagate for large distances, we can infer that they are somehow stable to the external instabilities. 
Such stability may be achieved
due to (1) the jet speed becoming sufficiently supersonic or relativistic that  
external modes are suppressed or do not have enough time to grow; 
%or if the most disruptive modes gets stabilized, leaving the modes with small pitch angles that do little damage; 
(2) rapid sideways expansion of the jet that breaks causal connection across it; (3) the stabilizing influence of rotation and velocity shear; or 
(4) a high density contrast or other properties of the ambient medium that promote stability
%that breaks causal connection across it, thus disabling the instabilities from propagating 
(see e.g. \citealt{2011IAUS..275...41H} for a discussion and, also, 
\citealt{2009MNRAS.394L.126M,2015MNRAS.452.1089P,2016MNRAS.456.1739B}).
Second, some kind of 
internal instability is {\it needed} in order to explain several jet observations, 
although overall jet stability and coherence are still desirable.
In spite of various different mechanisms proposed (see references in \citealt{2016MNRAS.456.1739B}), 
there is 
significant debate over what is responsible for the multi-wavelength emission from 
relativistic jets in AGN, microquasars and GRBs. 
It has been shown that internal 
 kink-type instabilities are capable of triggering the dissipation of magnetic energy into 
 thermal energy and radiation 
 (see e.g. \citealt{2011ApJ...728...90M,2014MNRAS.438..278P} in the context of pulsar wind nebulae, and 
 \citealt{2003astro.ph.12347L,2016MNRAS.456.1739B} in the context of GRBs).
This most likely occurs due to the formation of thin current sheets at small scales that facilitate
 reconnection of oppositely directed magnetic field lines being pushed closer together 
 \citep[see e.g.][for a review]{2015SSRv..191..545K} --- as demonstrated by 
 MHD simulations \citep[e.g.][]{2016ApJ...824...48S} 
 as well as fully-kinetic particle-in-cell (PIC) simulations \citep[e.g.][]{2015MNRAS.450..183S}.
Also, 3D {\it resistive} MHD simulations have shown that magnetized plasma columns can become unstable 
to both pressure-driven and current driven instabilities, which lead to the fragmentation of  
current sheets into filaments, followed by magnetic reconnection at an enhanced rate \citep{2016MNRAS.462.2970S}.
The local dissipation zones created by internal instabilities can also serve as sites for particle acceleration, which 
is required to explain the observed
non-thermal synchrotron and inverse Compton emission from relativistic jets \citep{2005ApJ...625...72S}. 
This idea has gained considerable support from test particle acceleration studies that 
have been carried out 
in magnetic reconnection domains inside relativistic jets \citep[see e.g.][for a review]{2015ASSL..407..373D}. 
More, recently, efficient particle acceleration has also been demonstrated in both 
reconnection and turbulence studies using PIC simulations, and the results are in 
general consistent with jet observations
\citep[see e.g.][and references therein]{2016MNRAS.462.3325P,2016ApJ...818L...9G,2017PhRvL.118e5103Z,2018MNRAS.473.4840W}.
Furthermore, internal pinch and kink instabilities together can explain the rich, small-scale
radial structures observed in jets, such as equispaced emission knots, 
wiggles, etc. \citep{2005AdSpR..35..908D,2013EPJWC..6102001H}. 
A lot of work has been done in trying to understand what perpetuates large-scale jet integrity. 
However, a fundamental understanding of internal instabilities is equally important, in order to answer 
basic questions such as how jets radiate.

%
%For example, in order for the external modes to successfully disrupt a jet at large-scales, 
%a causal connection needs to be established across the jet such that 
%the instability can communicate between the physically separated parts. 
%This could be established by, for example, the fast magnetosonic waves in AGNs  (PK). 
%
%or the external modes might not have time to grow due to the jet 
%speed becoming supersonic or relativistic, or the most disruptive
% modes might be stabilized, leaving modes with small pitch angles that do little damage.  
%Not sure we should single out free expansion without referring to the others

%One possible way the causal connection could be broken is if the jet expands rapidly as it 
%propagates in an ambient medium with a steep enough pressure profile, thus making the 
%jet globally stable \citep{2015MNRAS.452.1089P}.
%Fully global simulations by TB also show similar large-scale jet stability, although they see the appearance of the 
%external kink for some cases, which leads to helical motions at large distances.
%

In this work, we perform a comprehensive study of the linear development of internal instabilities 
and the associated stability criteria,
by solving the global eigenvalue problem for a cylindrical magnetized plasma. 
Most linear studies of internal instabilities are carried out either in the force-free 
limit and/or in conjunction with studies of large-scale 
external instabilities that require the presence of an ambient medium. 
%However, the force-free limit is applicable only close to the central engine, well below the 
%fast magnetosonic point (i.e., where the relativistic flow velocity equals the fast magnetosonic velocity; see e.g. 
%\citealt{2009ApJ...697.1681N}), whereas boundary or surface effects are likely to contaminate the development of 
%the internal instabilities in the jet interior. 
%Also, in a real jet it is likely that the complex magnetic field structure, bulk motion and velocity gradients 
%make it even harder to distinguish between the various instabilities except at suitable limits \citep{2002ApJ...580..800B}.
%In our work, we drop the force-free assumption and, also, attempt to isolate the internal instabilities 
%In our work, we avoid both these drawbacks. 
%Our global framework comprises of a radially extended 
%region of a static jet bounded by rigid walls, which eliminates the Kelvin Helmholtz instabilities 
%as well as the external kink (the vertical direction is still assumed to be local). 
 %This mimics the lack of communication between the interior and its surrounding due to the supermagnetosonic jet-speed.  
In our work, we drop the force-free assumption, and 
focus exclusively on the internal instabilities that are confined deep within the jet interior. 
We primarily model regions having a dominant toroidal magnetic field, a sub-dominant 
vertical magnetic field and a non-zero thermal pressure gradient, which is more prone to 
trigger pressure-driven instabilities.
%(however, see below).
Most of the magnetic energy dissipation is supposed to take place in such a regime, 
which is likely to occur significantly away from the jet source.
%Such a regime is likely to occur significantly 
%away from the jet source, where most of the magnetic energy dissipation is supposed to take place. 
Our work thus expands on B98's local, analytical calculations by self-consistently 
including the effects of radial gradients, geometric and magnetic curvatures, and a more complex 
toroidal field (B98 considered a radial power-law). 
We find that the maximum growth rates of the unstable modes are in excellent accord 
with the local prediction. More importantly, 
the B98 instability criterion ---
which essentially states that when the vertical field is very weak, the thermal pressure 
gradient is sufficiently strong and $B_\phi$ falls slower than $1/r$, then the underlying region is 
unstable to both axisymmetric and higher-order non-axisymmetric modes --- 
holds true even in a global framework. 
This validation implies that dissipation in the jet interior is likely dominated by myriad local instabilities, 
i.e. radially localized modes, which tend to occur at large vertical wavenumbers 
and have the highest growth rates. 
Thus, we predict that {\it fast growing, 
internal instabilities will 
always be triggered inside a jet as long as the B98 instability criterion is met locally, irrespective of the 
exact nature of the background radial profiles}.
 Note that once triggered, the instabilities should have enough time to 
grow before the jet fluid moves to a significant distance from the source.
B98 showed that
these instabilities have growth times of the order of Alfv\'{e}n crossing times, 
which are short enough in a toroidal field dominated region to develop non-linearly. Since in this work 
we recover growth rates very similar to those predicted by B98, the same conclusion applies.
Not surprisingly, we also identify some intrinsically global instabilities, whose 
characteristics differ somewhat from the local predictions. These usually occur
at smaller vertical wavenumbers and have smaller growth rates than the local instabilities.
We clarify here that both the local and global instabilities in this work refer to small-scale, internal instabilities. 
In our context, local implies that the instabilities are insensitive to the background radial 
curvature and gradients, whereas the global instabilities are sensitive to these factors (i.e. global 
here is not to be confused with large-scale, 
external instabilities that we do not talk about in this work). 
We additionally test the effect of introducing a stronger vertical field 
and corresponding radial gradient. Such conditions 
tend to introduce aforementioned current-driven effects, and 
the stability criterion becomes more complex due to the 
interplay between the thermal and magnetic pressure gradients.
%, and cannot be solely 
%determined by equipartition, as is commonly believed.
%the instabilities are seen to stabilize when the vertical field strength becomes 
%comparable to or exceeds the toroidal field strength (for the pinch and kink modes).
We also compare our results with recently performed numerical jet simulations, in order 
to obtain more insight into the workings of real jets.

%and conclude that {\it stabilization of the internal instabilities is governed by the 
%complex interplay between the thermal and magnetic pressure gradients, rather than solely by the
%condition of equipartition between the thermal and magnetic pressures, as commonly believed}.

%We also explore the effect of a more complex toroidal field profile on 
%the pressure-driven instabilities (B98 assumed a constant vertical field and power-law toroidal field). 

%O'neill carried out local, periodic box simulations 

%Although many of these experiments were performed in the context of controlled fusion in 
%tokamaks, it
%which in principle can 
%destroy the coherent structure of jets in very short distance-scales. 
%This, however, is in contradiction to the largely stable jets observed in nature, as mentioned above.
%However, in reality, astrophysical jets are observed to survive for 
%very long timescales and distances. 

This paper is organized as follows. In \S \ref{sec_theory}, we lay out the 
basic properties of the jet model we study, along with the assumptions invoked 
and the underlying linearized MHD equations governing the problem. In \S \ref{sec_analytic}, we 
discuss previously carried out analytical stability analyses most relevant to our work. 
We recall the local dispersion relation and stability criterion proposed 
by B98, the local stability criterion of 
\citet{suydam1958stability} and the magnetic resonance condition, all of which shall be tested in 
our global framework. 
In \S \ref{sec_numerical}, we describe the numerical set-up and normalization scheme
used for our global eigenvalue analysis. In \S \ref{sec_global_solns}, 
we present the detailed global solutions 
for different background conditions appropriate for jet interiors, 
and compare them with the local predictions wherever applicable. 
In \S \ref{sec_disc}, we compare our results with those of existing 3D jet 
simulations, with the aim of better understanding what triggers as well as quenches
internal instabilities in jets.
We conclude
in \S \ref{sec_conc}, by summarizing our main results and their
implications for observed astrophysical jets.

\section{Theory}
\label{sec_theory}

The literature studying MHD instabilities in the context of astrophysical jets is  
vast and rather confusing. It is indeed a complex topic 
and only a glimpse of this is given in the Introduction. 
%In linear stability analyses alone, 
%different set-ups, boundary conditions and methods 
%have been invoked, along with various different physical assumptions, which often 
%obscure the bigger picture. 
Keeping this in mind, in this section we  outline the main rationale behind our jet 
model, justify our assumptions and lay out the basic equations underlying the problem.
%we clearly outline the details, objectives and the necessity of our jet model. 

%A detailed review is beyond the scope of this work.

\subsection{Jet model and assumptions}
\label{sec_jetmodel}
%In this section we first lay out the basic equations underlying the problem.

We carry out a linear stability analysis for the following jet model:

\begin{itemize}

\item We assume the jet to have a cylindrical geometry ($r, \phi, z$), threaded by 
an axisymmetric magnetic field ${\bf B} =(0, B_{0\phi}(r), B_{0z}(r))$, which 
is often referred to as 
a {\it screw-pinch} in the plasma physics literature. 
 Note that  we are particularly interested 
in studying the internal instabilities that may arise in a toroidal magnetic field dominated regime, 
which are predominantly pressure-driven in nature. Such a region may arise due to the following reasons.
For a non-relativistic jet, as it 
propagates away from the source along the $z$-axis, the poloidal field 
is expected to fall off faster ($\sim 1/r^2$) than the toroidal field ($\sim 1/r$) 
in the radial direction due to magnetic flux conservation during lateral expansion
\citep{1984RvMP...56..255B}. 
%In a relativistic (externally collimated) jet, however, the toroidal and poloidal 
%field tend to remain comparable in the jet frame (although the above scalings still hold in the observer frame). 
%In such a situation, a toroidally dominated region will arise if the outer flux 
%surfaces diverge away from the jet core, as demonstrated by \citet{1994ApJ...426..269B}, 
%which can further explain the puzzle behind converting Poynting flux into
%kinetic energy in a magnetically dominated jet.
In a relativistic jet, a toroidally dominated region can arise if the outer flux 
surfaces diverge away from the jet core, as demonstrated by \citet{1994ApJ...426..269B}, 
which can further explain the puzzle behind converting Poynting flux into
kinetic energy in a magnetically dominated jet.
We shall also briefly discuss the effect of a stronger poloidal field, likely to occur 
near the jet core, that makes current-driven effects important.

\item In a real jet, it is likely that the 
   complex magnetic field structure, bulk motion and velocity gradients make it quite hard 
   to distinguish between the various instabilities triggered except at suitable limits \citep{2002ApJ...580..800B}. 
In this work, however, our goal is to isolate the internal instabilities to the extent possible. 
 Our global framework thus comprises an annular
region of a static jet bounded by rigid walls (see \S \ref{sec_numsetup} for details), 
which eliminates contamination due to the Kelvin-Helmholtz instability
as well as the external kink acting at the jet boundaries (the vertical direction is still assumed to be local).
%This set-up mimics the lack of communication between the interior and its surrounding as the jet 
%moves through it supermagnetosonically.
 Note that our inner radial boundary does not extend all the way up to the $r=0$ singularity 
(i.e. we do not study the instabilities that may arise in the jet core itself). This 
restriction does not interfere with our objective of studying the internal instabilities or 
testing the local calculations. In fact, it allows us to 
solve the global eigenvalue problem more accurately, i.e. without making any assumptions about the asymptotic behavior
of the eigenfunctions at small radii (e.g. \citealt{2013MNRAS.434.3030B,2015MNRAS.450..982K}).

\item The more commonly invoked force-free equilibrium 
%applicable only close to the central engine, well below the fast magnetosonic point (i.e. where the relativistic
   %flow velocity equals the fast magnetosonic velocity; see e.g. \citealt{2009ApJ...697.1681N}).
is quite restrictive and, hence, we invoke the presence of a non-zero, 
thermal pressure gradient in the jet that balances the magnetic forces (following B98).
Also, as internal instabilities
are triggered in the jet, they are likely to dissipate magnetic energy into thermal energy, which 
further enhances the role of such a thermal pressure gradient.

\item We perform our analysis in the non-relativistic limit, and assume the jet to be non-rotating 
as well as static (i.e. the fluid velocity ${\bf v} = 0$). In other words, 
we work in the jet fluid-frame for simplicity. Note that we also ignore the effect of a comoving, 
poloidal velocity shear that may be present in the jet interior \citep[see e.g.][]{2012MNRAS.427.2480N}.

\end{itemize}

\subsection{Basic equations}
%\subsection{Linearized MHD equations of motion}
\label{eqs_sec}

We model our jet by invoking the ideal, non-relativistic 
MHD equations describing a magnetized, compressible fluid:
\begin{gather}
\frac{\partial \rho}{\partial t} + \nabla \cdot (\rho \mathbf{v}) = 0 ~, \label{mhd1} \\
\frac{\partial \mathbf{v}}{\partial t} + (\mathbf{v} \cdot \nabla ) \mathbf{v} + 
\frac{1}{\rho}\nabla \left(P + \frac{\mathbf{B}^2} {8\pi}\right) - \frac{1}{4\pi \rho}(\mathbf{B}\cdot \nabla)\mathbf{B} = 0 ~,
\label{mhd2}\\
\frac{\partial \mathbf{B}}{\partial t} - \nabla \times (\mathbf{v}\times  \mathbf{B}) = 0 ~, \label{mhd3} \\
\nabla \cdot {\mathbf B} = 0  ~~~{\rm or}~~~ {\mathbf B} = \nabla \times {\mathbf A}  ~.
\label{mhd4}
\end{gather}
where $\rho$ is the density, $P$ the plasma pressure and 
${\bf A}$ the magnetic vector potential. We substitute equation (\ref{mhd4}) 
in equation (\ref{mhd3}) to obtain
\begin{equation}
\frac{\partial \mathbf{A}}{\partial t} -  (\mathbf{v}\times  \mathbf{B}) = 0 ~,
\label{mhd5}
\end{equation}
which is a particularly useful form of the magnetic 
induction equation when solving for non-axisymmetric perturbations. 
Recall that 
${\bf B}$ is invariant under the gauge transformation ${\bf A} \rightarrow {\bf A} + \nabla \psi$, 
where $\psi$ is any scalar function.
%that needs to be suitably fixed.
%
%We neglect the 
%effect of gravity and assume the jet to be non-rotating as well as static 
%(i.e., ${\bf v} = 0$) for simplicity. 
%We consider 
%a cylindrically symmetric, axisymmetric background having 
%${\bf B} =(0, B_{0\phi}(r), B_{0z}(r))$ in the ($r, \phi, z$) co-ordinate system. 
We also define 
the  pitch ${\cal P}$ of the magnetic field, 
often invoked during the stability analysis of magnetized jet columns:
\begin{equation}
{\cal P} = r \frac{B_z}{B_\phi} ~.
\label{eq_pitch}
\end{equation}
The magnetic configurations studied in this work involve a wide range of pitch parameters.

%The above equations are also supplemented by an equation of state, $P = P(\rho)$.

Note that the vertical equilibrium is trivial as we do not consider any vertical gradients of 
the background quantities. The radial equilibrium is derived from equation (\ref{mhd2}), 
which on assuming an isothermal equation of state $P_0(r)= \rho_0 (r) c_s^2$ can be written as:
\begin{equation}
\partial_r \rho_0 = - \frac{B_{0\phi}}{4\pi c_s^2} \left[ \frac{B_{0\phi}}{r} + \partial_r B_{0\phi} \right] - 
\frac{B_{0z}}{4\pi c_s^2}\partial_r B_{0z}  ~,
\label{eq_radeqbm} 
\end{equation}
where $c_s$ is the sound speed and the subscript `0' denotes unperturbed 
background variables (see Appendix \ref{sec_linmhd}). Note that because of the isothermal approximation, the 
density and plasma pressure gradient are synonymous in this work.
The complete set of linearized MHD equations, obtained by perturbing equations 
(\ref{mhd1}), (\ref{mhd2}), (\ref{mhd4}) and 
(\ref{mhd5}), is given in Appendix \ref{sec_linmhd}.

%Unlike most works in the literature, we drop the 
%force-free assumption while modeling our jet such that the background density gradient 
%plays an important role in governing the stability of the system.

%The above form of the induction is especially useful when solving for 
%non-axisymmetric perturbations as it automatically imposes the $\nabla \cdot {\bf B_1} =0$ condition.

%We can now use the same linearized system of equations to derive a local dispersion relation, as well as
%solve the global eigenvalue problem.

%\section{Local Analysis}
%\label{sec_local}

%OR 
\section{Analytical stability analyses}
\label{sec_analytic}

In this work, 
we compare the outcome of our global 
analysis with the solutions from B98's local dispersion relation, which was derived for a magnetized, 
relativistic, non-rotating, compressible, cylindrically symmetric plasma (see eq. 3.32 of B98). 
%He assumed the system to be toroidally dominated, having a uniform $B_{0z} \ll B_{0\phi}$. 
%The non-relativistic limit of the dispersion relation is given in Appendix B of \cite{2018MNRAS.473.2791D}.  
%Unlike most jet studies in the literature, B98 considered the effect 
%of a non-zero thermal pressure gradient, which is the more general case, that becomes particularly 
%important as the jet expands laterally. 
%The non-relativistic limit of the dispersion relation is given in Appendix B of \cite{2018MNRAS.473.2791D}.  
In this section, we present only the relevant equations and final expression for this
dispersion relation, and 
refer the readers to B98 and 
\cite{2018MNRAS.473.2791D} for a detailed derivation. 
We also discuss the stability criteria proposed by B98, \citet{suydam1958stability} 
and the magnetic resonance condition.

\subsection{B98 local dispersion relation}
\label{sec_mcb_disp}

B98 considered a toroidally dominated configuration, having 
a weak, uniform vertical field, $B_{0z} \ll B_{0\phi}$ and power-law toroidal field, 
$B_{0\phi} \propto r^\alpha$. The
perturbed variables were assumed to have the form
\begin{equation}
\exp i(l r +  m \phi + k_z z -  \omega t) ~,
\label{eq_pert_form}
\end{equation}
where $\omega$ is the modal frequency, $t$ the time, $l$, $m$ and $k_z$
the radial, azimuthal and vertical wavenumbers respectively. The above form of the perturbations 
corresponds to the short wavelength approximation, i.e. $lr, k_zr \gg 1$, where
the amplitudes are treated as constant coefficients.
Note that $m$ can take only integer 
values such that $m=0,1,2$ etc., of which the $m=0$ (pinch) and $m=1$ (kink) modes are of 
particular interest to jet stability as discussed in the Introduction 
(the $m=2$ mode causes elliptical distortions). 
%The former leads to a pinching distortion 
%of the jet and triggers the so-called ``sausage instability'', while the latter 
%are helical perturbations giving rise to the ``kink instability'' in MHD plasmas \cite[see e.g.][]{1979ApJ...234...47H}.
%(the $m=2$ mode causes elliptical distortions) 
$m$ can be positive or negative, which indicates the sense of the spatial twist of the 
non-axisymmetric mode.

\begin{table*}
\large
\centering
%\vskip0.2cm
\renewcommand{\arraystretch}{1.8}
\caption{Summary of the parameters introduced 
in the local analysis of \S \ref{sec_mcb_disp} and \S \ref{sec_b98_stabcrit}.}
 \begin{tabular}{|c|c|}
 %\hline
 %\hskip1cm General & &  \hskip2cm General  \\ 
 \hline
 Parameters & Notes \\
 \hline
 $\{ \tilde{k}_z, \tilde{l} \} = r_0 \{k_z, l\}$ & dimensionless wavenumbers \\
  ${\tilde \omega} = \frac{r_0}{c_s} \omega$ & dimensionless modal frequency\\
  $\eta =  \tilde{k}_z \frac{B_{0z}}{B_{0\phi}}$ &  \\
$v_{A\phi} = \frac{B_{0\phi}}{\sqrt{4\pi \rho_0}}$ & toroidal Alfv\'{e}n velocity \\
$v_{Az} = \frac{B_{0z}}{\sqrt{4\pi \rho_0}}$ & vertical Alfv\'{e}n velocity \\
 %$\beta = \frac{8\pi P_0}{B_{0\phi}^2 + B_{0z}^2} = \frac{2 c_s^2}{v_{A\phi}^2 + v_{Az}^2}$ & plasma-beta \\ 
 $d = \frac{c_s^2}{v_{A\phi}^2}$ & \\
 $\beta = \frac{2 c_s^2}{v_{A\phi}^2 + v_{Az}^2} \approx 2d$ & plasma-beta for $v_{Az} \ll v_{A\phi}$\\ %for $B_{0\phi} \gg B_{0z}$ \\ 
$\alpha =\frac{\partial \ln B_{0\phi}}{\partial \ln r}$ &  \\
\hline
\end{tabular}
%\vspace{-4mm}
\label{tab_notation}
\end{table*}

The radial equilibrium given by 
equation (\ref{eq_radeqbm}) for the case considered by B98 simplifies to
\begin{equation}
\partial_r \rho_0 = - (1+\alpha)\frac{B_{0\phi}^2}{4\pi c_s^2 r}  ~.
\label{mcb_radeqbm} 
\end{equation}
Finally, the {\it non-relativistic} version of the local dispersion relation obtained by 
B98 can be written as (also, see Appendix B of \citealt{2018MNRAS.473.2791D}):
%\begin{align}
%\frac{1}{2} \biggl(1 + \frac{{\tilde l}^2}{{\tilde k_z}^2} \biggr) \biggl(1 + \frac{{\tilde v_{A\phi}^2} }{2} \biggr) \tilde{\omega}^4 
%&- \biggl[  \biggl(1 + \frac{l^2}{k_z^2} \biggr) \biggl(1 + \frac{m}{\eta} \biggr)^2 \biggl(1 + \frac{{\tilde v_{A\phi}^2} }{2} \biggr) {\tilde k}_z^2 {\tilde v_{Az}}^2  
% - \hat{B_{\phi}} (1 + {\tilde v_{A\phi}}^2  ) {\tilde v_{A\phi}}^2  + (1 - {\tilde v_{A\phi}}^2 ) {\tilde v_{A\phi}}^2  \biggr] \tilde{\omega}^2 
%\nonumber \\ &+  \biggl(1 + \frac{m}{\eta} \biggr)^2 {\tilde k_z}^2 {\tilde v_{Az}}^2\biggl[ 
%\frac{1}{2} \biggl(1 + \frac{{\tilde l}^2}{{\tilde k_z}^2} \biggr) \biggl(1 + \frac{m}{\eta} \biggr)^2 {\tilde k_z}^2 {\tilde v_{Az}}^2  
%- (1 + \hat{B_\phi}) {\tilde v_{A\phi}}^2 \biggr] = 0 ~,
%\label{mcb_dispeq}
%\end{align}
%
\begin{align}
\frac{1}{2} \biggl(1 + \frac{l^2}{k_z^2} \biggr) \biggl(1 + \frac{v_{A\phi}^2}{c_s^2} \biggr) \omega^4 
&- \biggl[  \biggl(1 + \frac{l^2}{k_z^2} \biggr) (m+ \eta)^2 \biggl(1 + \frac{v_{A\phi}^2}{2c_s^2} \biggr) \frac{v_{A\phi}^2}{r^2}
 - \alpha \biggl(1 +  \frac{v_{A\phi}^2}{c_s^2} \biggr) \frac{v_{A\phi}^2}{r^2}  
 + \biggl(1 -  \frac{v_{A\phi}^2}{c_s^2} \biggr) \frac{v_{A\phi}^2}{r^2}  \biggr] \omega^2 
\nonumber \\ &+  (m + \eta)^2 \frac{v_{A\phi}^2}{r^2} \biggl[ \frac{1}{2} \biggl(1 + \frac{l^2}{k_z^2} \biggr) (m + \eta)^2 \frac{v_{A\phi}^2}{r^2}
- (1 + \alpha) \frac{v_{A\phi}^2}{r^2} \biggr] = 0 ~,
\label{mcb_dispeq}
\end{align}
where all the new variables are defined in Table \ref{tab_notation}. 
This is a fourth-degree dispersion relation in $\omega$, obtained after neglecting the fast magnetosonic 
modes for simplicity. To derive this relation we have 
assumed, following B98, that the plasma-$\beta \approx 2c_s^2/v_{A\phi}^2$, since $v_{A\phi} \gg v_{Az}$. 
Hence, $v_{Az}$ only appears in the dispersion relation through $\eta$. 
 We non-dimensionalize this equation such that
all lengths are scaled by the local radial coordinate $r_0$, all wavenumbers by $1/r_0$, 
all velocities by $c_s$, all frequencies by $c_s/r_0$ and all densities by the local 
background density $\rho_0$ at $r_0$; which reduces equation (\ref{mcb_dispeq}) to 
the dimensionless form (see Table \ref{tab_notation} for the new notation):
%\begin{align}
%\frac{1}{2} \biggl(1 + \frac{l^2}{k_z^2} \biggr) (1 + v_{A\phi}^2) \omega^4 
%&- \biggl[  \biggl(1 + \frac{l^2}{k_z^2} \biggr) (m+ \eta)^2 \biggl(1 + \frac{v_{A\phi}^2}{2} \biggr) v_{A\phi}^2
% - \alpha (1 +  v_{A\phi}^2 )v_{A\phi}^2  
% + (1 - v_{A\phi}^2 ) v_{A\phi}^2 \biggr] \omega^2 
%\nonumber \\ &+  (m + \eta)^2 v_{A\phi}^2 \biggl[ \frac{1}{2} \biggl(1 + \frac{l^2}{k_z^2} \biggr) (m + \eta)^2 v_{A\phi}^2
% - (1 + \alpha) v_{A\phi}^2 \biggr] = 0 ~.
%\label{mcb_dispeq_dimles}
%\end{align}
%
%\begin{align}
%\frac{1}{2} \biggl(1 + \frac{l^2}{k_z^2} \biggr) \biggl(1 + \frac{1}{d} \biggr) \omega^4 
%&- \biggl[  \biggl(1 + \frac{l^2}{k_z^2} \biggr) (m+ \eta)^2 \biggl(1 + \frac{1}{2d} \biggr) \frac{1}{d}
% - \alpha \biggl(1 +  \frac{1}{d} \biggr) \frac{1}{d} 
% + \biggl(1 - \frac{1}{d}\biggr) \frac{1}{d} \biggr] \omega^2 
%\nonumber \\ &+  (m + \eta)^2 \frac{1}{d} \biggl[ \frac{1}{2} \biggl(1 + \frac{l^2}{k_z^2} \biggr) (m + \eta)^2 \frac{1}{d}
% - (1 + \alpha) \frac{1}{d} \biggr] = 0 ~.
%\label{mcb_dispeq_dimles}
%\end{align}
%
\begin{align}
\frac{1}{2} \biggl(1 + \frac{l^2}{k_z^2} \biggr) d(1 + d) {\tilde \omega}^4 
&- \biggl[  \biggl(1 + \frac{l^2}{k_z^2} \biggr) (m+ \eta)^2 (1 + 2d)
 - \alpha (1 +  d)  
 + (d - 1) \biggr] {\tilde \omega}^2 
\nonumber \\ &+  (m + \eta)^2 \biggl[ \frac{1}{2} \biggl(1 + \frac{l^2}{k_z^2} \biggr) (m + \eta)^2 
 - (1 + \alpha) \biggr] = 0 ~.
\label{mcb_dispeq_dimles}
\end{align}
%where $d= c_s^2/v_{A\phi}^2$.
%Note that the above dispersion relation is invariant under the transformations 

\subsection{B98 local stability criterion}
\label{sec_b98_stabcrit}

B98 found that relativistic effects have little effect on the 
growth rate of the unstable modes 
%(in fact, they do not appear in the instability criterion at all), 
and, hence, his findings are applicable in the non-relativistic formalism as well.
Seeking unstable modes such that $\omega^2<0$ and $\omega_I > 0$ (as per the convention in equation \ref{eq_pert_form}), 
B98 obtained a {\it necessary and sufficient} condition for instability to occur (see eq. 4.2 of B98)
%from his dispersion relation (eq. 3.32 of B98), 
such that
\begin{equation}
\frac{\partial \ln B_{0\phi}}{\partial \ln r} \equiv \alpha > \frac{1}{2} (m + \eta)^2 - 1 
\label{mcb_instab_crit1}
\end{equation}
or, equivalently, unstable modes exist for
\begin{equation}
0<(m+\eta)^2< 2(1+\alpha) ~.
\label{mcb_instab_crit2}
\end{equation}
The above inequality clearly indicates that $\alpha=-1$ is the case of marginal stability for all $m$ 
or, equivalently,  
{\it any local patch of the jet in which $B_{0\phi}$ decreases faster than $1/r$ 
should be stable with respect to both axisymmetric and higher-order non-axisymmetric modes.} 
%Note that \citet{1962JFM....14..463H} had obtained a sufficient stability criterion for an ideal Couette 
%flow, threaded by a constant $B_z$ and a radially varying $B_\phi$, and subjected to axisymmetric perturbations only. 
%Interestingly, this is identical to the B98 stability criterion (i.e. $\alpha \leq -1$) 
%in the limit of a static flow, which is applicable to non-axisymmetric perturbations as well.
%Thus, a weak but non-zero vertical field plays an important role in governing the stability of the jet.
%
From equation (\ref{mcb_instab_crit1}) we can also say that for a given $m$ and $\alpha>-1$, instability occurs 
in the range
\begin{equation}
-\sqrt{2(1+\alpha)} - m <\eta< \sqrt{2(1+\alpha)} - m  
\label{mcb_instab_crit3}
\end{equation}
or
\begin{equation}
(-\sqrt{2(1+\alpha)} - m)\frac{B_{0\phi}}{B_{0z}}  <\tilde{k}_{z}|_{\rm unst} < (\sqrt{2(1+\alpha)} - m) 
\frac{B_{0\phi}}{B_{0z}}  ~.
\label{eq_kzunst}
\end{equation}

Apart from the above instability criterion, B98 made certain estimates from
his analytical calculations, which we quote here in the non-relativistic limit. First, B98 
showed that the fastest growing modes (i.e. the modes having the most negative $\omega^2$) 
occur in the limit $l^2/k_z^2 \ll 1$. Thus, putting 
$l^2/k_z^2 \approx 0$, followed by minimizing $\omega^2$ with respect to $(m+\eta)^2$ 
in the dispersion relation given by equation (\ref{mcb_dispeq_dimles}), 
one obtains:
\begin{equation}
{\tilde \omega}^2_{\rm min} = - 8 d \biggl[ \biggl(1 - \frac{1+\alpha}{4d}\biggr) - 
\biggl(1 - \frac{1+\alpha}{2d} \biggr)^{1/2}  \biggr] ~,
\end{equation}
which is equivalent to eq. 4.8 of B98, when the relativity parameter $e$ therein is assumed to be $e \ll 1$. 
Thus, the maximum growth rate of the system is given by:
\begin{equation}
%{\tilde \omega}_{I {\rm max}} 
{\tilde \omega}_{g}|_{\rm max}= \sqrt{8 d \biggl[ \biggl(1 - \frac{1+\alpha}{4d}\biggr) - 
\biggl(1 - \frac{1+\alpha}{2d} \biggr)^{1/2}  \biggr]} ~,
\label{eq_wmax}
\end{equation}
which occurs at
\begin{equation}
(m + \eta)^2_{\rm min} = 1 + \alpha + \frac{1}{2}(1 + 2d) {\tilde \omega}^2_{\rm min} ~,
\end{equation}
which in turn is equivalent to eq. 4.9 of B98 for $e \ll 1$. Alternatively, the vertical wavenumbers corresponding 
to the most unstable modes are given by
\begin{equation}
\tilde{k}_{z}|_{\rm max} = \biggl(\pm \sqrt{1 + \alpha + \frac{1}{2}(1 + 2d) {\tilde \omega}^2_{\rm min}} -m \biggr) 
\frac{B_{0\phi}}{B_{0z}} ~.
\label{eq_kzmax}
\end{equation}
We note from equation (\ref{eq_wmax}) that the maximum growth rate is independent of $m$, which is a 
characteristic of pressure-driven instabilities.
This is in contrast with current-driven instabilities, whose maximum growth rate is inversely 
proportional to $m$ \citep[e.g.][]{2000A&A...355..818A}.

\subsection{Magnetic resonance condition}
\label{sec_resonance}

For non-axisymmetric perturbations in a cylindrical plasma column 
having the form given by equation (\ref{eq_pert_form}), the magnetic resonance condition 
is defined as ${\bf k \cdot B}=0$, where ${\bf k}$ is the wave-vector with components $(l,m,k_z)$. 
Consequently, one can define a  {\it resonant surface} $r= r_{\rm res}$, where this condition 
is satisfied such that
\begin{equation}
k_z B_z (r_{\rm res}) + \frac{m}{r_{\rm res}} B_\phi = 0 ~.
\label{eq_reson}
\end{equation}
The restoring force due to the magnetic tension is supposed to be minimum at these 
surfaces, which are hence more likely to trigger instabilities (see e.g. \citealt{2008LNP...754..131L}). 
By studying the global eigenfunctions of the most unstable modes, 
%in our global analysis (see \S \ref{sec_global_solns}), 
we shall test whether the above condition influences the instabilities in this work (see \S \ref{sec_global_solns}).

\subsection{Suydam's local stability criterion}
\label{sec_suydam}

%displacement in the vicinity of the magnetic
%resonance of an (m, k) mode, and looks under which conditions this displacement
%makes δW negative; the condition turns out to be Suydam criterion for
%a well-chosen displacement.

\citet{suydam1958stability} proposed a necessary condition for stability for the 
non-axisymmetric ($m \neq 0$) modes of a screw-pinch such that
\begin{equation}
\frac{r}{4} \biggl(\frac{1}{{\cal P}}\frac{d {\cal P}}{dr} \biggr)^2 + \frac{8\pi}{B_z^2} \frac{dP}{dr} \geq 0 ~,
\label{eq_suydam}
\end{equation}
where ${\cal P}$ is the magnetic pitch given by equation (\ref{eq_pitch}).
%, and the converse of this theorem gives the condition for instability.
Note that this criterion was derived from a variational energy principle, which states that an equilibrium is 
unstable if the radial displacement of a mode makes the change in potential energy negative. Furthermore, 
this is a {\it local stability criterion} as the displacement of the concerned ($m,k_z$) modes 
is assumed to be localized only about the magnetic resonance surface (see equation \ref{eq_reson}), for reasons cited 
in \S \ref{sec_resonance}.
The thermal pressure gradient (second term in the inequality \ref{eq_suydam}) has a negative sign and
is known to drive the system to instability. Thus, Suydam's criterion basically states that the  
{\it magnetic shear} (first term in the inequality \ref{eq_suydam}) must be strong enough to counteract this effect 
to maintain stability.
We shall revisit the applicability of this criterion when discussing our global solutions in \S \ref{sec_global_solns}.
We note that although Sudyam's criterion is local, it conveys the crucial point 
that a competition between the thermal and magnetic pressure gradients plays an important role in determining the 
stability of the system.

%\section{Global eigenvalue analysis}
%\label{sec_global}

\section{Global Eigenvalue Analysis}
\label{sec_numerical}

\subsection{Numerical Set-up}
\label{sec_numsetup}

In order to construct global solutions from the linearized axisymmetric
system of equations (\ref{eq_cont})-(\ref{eq_az}),
we consider the jet to  be extended in
radius $r \in [R_{\rm in}, R_{\rm out}]$, to account for the curvature terms. 
As mentioned before, since we are only interested in the internal instabilities 
characterizing the jet, we assume this domain to be located far from the influence of 
the ambient medium as well as from the central axis (i.e. $R_{\rm in} \neq 0$). 
We solve equations (\ref{eq_cont})-(\ref{eq_az}) as a linear eigenvalue problem using
the pseudo-spectral code Dedalus\footnote{Dedalus is available at
\url{http://dedalus-project.org}.} (Burns et al., in preparation).
The solutions are decomposed on a basis of Chebyshev polynomials along the radial grid
and on a Fourier basis in the vertical direction such that the perturbed variables have a 
form 
\begin{equation}
f(r)\exp i(k_z z + m\phi -\omega t) ~,
\end{equation}
and the linear eigenvalue problem becomes
\begin{equation}
 {\cal M}(r) {\boldsymbol \xi}(r)  = \omega {\cal I} {\boldsymbol \xi}(r) ~,
 \label{eq_eigen}
\end{equation}
where $\omega = \omega_R + i \omega_I$ is the complex eigenvalue; ${\cal I}$ the identity
operator; ${\cal M}(r)$ the MHD linear differential operator and ${\boldsymbol \xi}(r)$ the eigenfunction
constituted of the perturbed variables in the system.
% $k_z$ the vertical wavenumber and $m$ the azimuthal wavenumber. 
The growth rate of an unstable mode is given by $\omega_I>0$, 
whereas $\omega_R$ is an indicator of the overstability of the mode.

%The real part $\omega_R$ is a measure of the overstability of the mode, such that $\omega_R=0$
%implies a non-propagating mode and $\omega_R \neq 0$ implies a traveling mode.

%The vertical thickness of the disc is assumed to be very small
%and, hence, the dynamics along the vertical direction can be considered to be local.
% Don't you need to say that we are first Fourier transforming into plane waves in z?

%The solution is decomposed on a basis of Chebyshev polynomials along a radial grid, while
%the (local) vertical direction is characterized by a constant vertical wavenumber $k_z$.

%Note that Dedalus does not allow over-specification of the boundary conditions and, 
%hence, one needs to supply exactly as many boundary conditions as there are first-order, 
%independent Chebyshev derivatives (which in our case is $\partial_r$ only). 
%For the axisymmetric problem defined above, this number is four.

We impose rigid/impenetrable boundary conditions at our inner and outer jet boundaries, such that 
(also, see \citealt{2018MNRAS.473.2791D}): 
%This facilitates the study of internal instabilities in isolation and removes contamination 
%from instabilities of the Kelvin-Helmholtz kind, 
%which arise due to the velocity shear between the jet and ambient medium (for e.g.,....). Thus, 
%we impose (also, see \citealt{2018MNRAS.473.2791D}):
\begin{equation}
v_{1r}  = 0  ~~~ {\rm at} ~~~ r = R_{\rm in}, R_{\rm out}  \nonumber
\end{equation}
\begin{equation}
\partial_r B_{1z} = 0  ~~~ {\rm at} ~~~ r = R_{\rm in}, R_{\rm out} ~,
\label{eq_bcded}
\end{equation}
where the subscript `1' indicates perturbed variables (see Appendix \ref{sec_linmhd}). 
%where the second condition is motivated by \citet{2004ApJ...602..892K}. 
These 4 boundary conditions suffice for the case with the Weyl gauge, however, for the Coulomb 
gauge we additionally impose
\begin{equation}
A_{1r}  = 0  ~~~ {\rm at} ~~~ r = R_{\rm in}, R_{\rm out}
\end{equation}
to close the system of equations. We also impose boundary conditions 
on the background pressure/density, which we shall describe in \S \ref{sec_global_solns}.
We choose our fiducial resolution to be either $N_r=200$ 
or $256$, where $N_r$ is the number of
radial grid points. The fiducial 
radial extent of the jet is set to be $r \in [1, 5]R_{\rm in}$, unless otherwise mentioned.

\subsection{Normalization scheme}

\begin{table*}
\large
\centering
%\vskip0.2cm
\renewcommand{\arraystretch}{1.8}
\caption{Summary of the parameters introduced 
in the global analysis of \S \ref{sec_numerical} and \S \ref{sec_global_solns}.}
 \begin{tabular}{|c|c|}
 %\hline
 %\hskip1cm General & &  \hskip2cm General  \\ 
 \hline
 Parameters & Definitions \\
 \hline
 $r \equiv \frac{r}{R_{\rm in}} $ & dimensionless radius \\
 $k_z \equiv k_z R_{\rm in}$ & dimensionless vertical wavenumber \\
$\rho \equiv \frac{\rho}{\rho_{\rm in}}$  & dimensionless density \\ 
$\omega_{I,R} \equiv \omega_{I,R} \frac{R_{\rm in}}{c_s}$ & dimensionless eigenvalues (I $\equiv$ imaginary; R $\equiv$ real)\\
$v_{1r,1\phi,1z} \equiv \frac{v_{1r,1\phi,1z}}{c_s}$ & dimensionless perturbed velocities\\
$v_{A\phi,Az} \equiv \frac{B_\phi,z}{\sqrt{4\pi \rho_{\rm in}}} \frac{1}{c_s}$ & dimensionless background Alfv\'{e}n velocities\\
$v_{A\phi 0, Az0}$ & background Alfv\'{e}n velocities at $r=R_{\rm in}$\\
$\alpha =\frac{\partial \ln v_{A\phi}}{\partial \ln r}$ & dimensionless parameter in $v_{A\phi}$-PL models \\
$a_1$ & dimensionless parameter in $v_{A\phi}$-Gen models \\
$\epsilon$ & dimensionless parameter in $v_{Az}$-Var models;
$\frac{dP}{dr} = \frac{\epsilon}{1-\epsilon} \frac{d}{dr} \biggl(\frac{B_z^2}{8\pi}\biggr)$ \\
%$B_1 = v_{A\phi 0}(1+a_1)$ & $v_{A\phi}$-Gen model \\ 
\hline
\end{tabular}
%\vspace{-4mm}
\label{tab_notation_global}
\end{table*}

We nondimensionalize our  linearized
equations (\ref{eq_cont})-(\ref{eq_az}) for the global analysis by using 
quantities at the inner radial jet boundary $R_{\rm in}$ (unless otherwise 
mentioned).  We scale all lengthscales by $R_{\rm in}$, 
all wavenumbers by $1/R_{\rm in}$, all velocities by the 
sound speed $c_s$, all frequencies by $c_s/R_{\rm in}$ 
and all densities by the inner radial density $\rho_{\rm in}$. 
In order to solve the linearized equations, we replace $\partial_z \rightarrow ik_z$ and 
$\partial_t \rightarrow -i\omega$, 
and define dimensionless {\it perturbed} Alfv\'{e}n velocities as
%
%We also assume an adiabatic background such that
%equation (\ref{eq_pert_eos}) holds true with a constant sound speed $c_s$.
%We consider the exact
%background equilibrium given by equation (\ref{eq_radeq_nograd}), which when nondimensionalized
%according to this scheme becomes
%
%\begin{equation}
%\Omega = r^{-3/2}(1 + r v_{A\phi}^2)^{1/2} ~,
%\label{dimles_radeq}
%\end{equation}
%where $v_{A\phi}$ now represents the dimensionless background toroidal Alfv\'{e}n velocity.
%Thus, the final set of dimensionless, axisymmetric, linearized equations that we solve are
%
\begin{equation}
v_{A1 r,A1\phi,A1z} \equiv \frac{B_{1r,1\phi,1z}}{\sqrt{4 \pi \rho_{\rm in}}} \biggl(\frac{1}{c_s} \biggr) ~,
\end{equation}
dimensionless {\it perturbed} vector potential components as
\begin{equation}
A_{1 r,\phi,z} \equiv \frac{A_{1r,\phi,z}}{\sqrt{4 \pi \rho_{\rm in}}} \biggl(\frac{1}{c_s R_{\rm in}} \biggr)
\end{equation}
and dimensionless {\it background} Alfv\'{e}n velocities as
\begin{equation}
v_{A \phi, Az} \equiv \frac{B_{\phi,z}}{\sqrt{4 \pi \rho_{\rm in}}} \biggl(\frac{1}{c_s} \biggr) ~.
\label{eq_bg_alfven}
\end{equation}
Note that from here onwards we will describe the magnetic field in terms of the definition 
given by equation (\ref{eq_bg_alfven}) (also, see Table \ref{tab_notation_global} 
for a summary of the definitions to be used for the global analysis).
Finally, depending on the gauge condition imposed, 
we solve for the set of 
eigenfunctions ${\boldsymbol \xi}(r)|_{Weyl}=
\{\rho_1, v_{1r}, v_{1\phi}, v_{1z}, v_{A1\phi}, v_{A1z}, A_{1r}, A_{1\phi}, A_{1z}\}$ or 
${\boldsymbol \xi}(r)|_{Coulomb} =\{\rho_1, v_{1r}, v_{1\phi}, v_{1z}, v_{A1\phi}, v_{A1z}, 
A_{1r}, A_{1\phi}, A_{1z}, \psi\}$.

%Note that all the variables appearing henceforth in this work (i.e., in both the global eigenvalue
%solutions and the PLUTO simulations) are nondimensionalized according to scheme explained in this section.

%\subsection{Cases}

%points --  dedalus eq scaling - no neeed; eigentools bad evalues - mention in results.

\section{Global Solutions}
\label{sec_global_solns}

In this section, we present the results of our global stability analysis. We divide the results into 
three categories depending upon the transverse magnetic field profiles adopted for the jet interior. 
There is a lot of uncertainty regarding the nature of magnetic fields inside jets, which has led
different groups to adopt a wide range of field prescriptions (see references in the Introduction). However, 
our analysis reveals certain fundamental properties of the internal instabilities, which
stand out irrespective of the complex nature of the magnetic field imposed.

\begin{table*}
\large
\centering
%\vskip0.2cm
\renewcommand{\arraystretch}{1.5}
\caption{Summary of runs for case $v_{A\phi}{\rm-PL-}v_{Az}{\rm-Cons}$ presented in \S \ref{sec_Bphi_PLcons_Bz_cons}. 
%$v_{A\phi} = v_{A\phi0} r^\alpha$, 
%$v_{Az}=v_{Az0}$, $\rho = 1 + \frac{(1+\alpha)v_{A\phi0}^2}{2\alpha}(1 - r^{2\alpha})$.
}
 \begin{tabular}{|c|c|c|c|c|c|c|}
 %\hline
 %\hskip1cm General & &  \hskip2cm General  \\ \hline
 \hline
 $m$ & $\alpha$ & $v_{A\phi0}$ & $v_{Az0}/v_{A\phi0}$ & $N_r$ & B.C.s imposed on $\rho(r)$ & Other notes \\
  \hline 
  %\hline
 $0,1,2$  & $-0.3$ & $0.926$  & $0.005$  & $256$  & 
 $\rho (1) = 1$; $\rho (r \rightarrow \infty)=0$ & Fiducial $\alpha<0$ model; 
 $v_{A\phi0}= \sqrt{\frac{-2\alpha}{(1+\alpha)}}$  \\ 
 %$v_{A\phi0}= \sqrt{\frac{-2\alpha}{(1+\alpha)}}$; 
 \hline
 $0$ & $-0.1$ & $0.6$  & $0.008$  & $300$  & $\rho (1) = 1$  &  $\rho>0$ $ \forall$ $r \in [1,5]$\\
 $0$ & $-0.3$ & $0.6$  & $0.008$  & $200$  & $\rho (1) = 1$   & $\rho>0$ $ \forall$ $r \in [1,5]$\\
 $0$ & $-0.7$ & $0.6$  & $0.008$  & $200$  & $\rho (1) = 1$ &  $\rho>0$ $ \forall$ $r \in [1,5]$  \\
 $0$ & $-0.9$ & $0.6$  & $0.008$  & $200$  & $\rho (1) = 1$ & $\rho>0$ $ \forall$ $r \in [1,5]$  \\
 $0$ & $-0.5$ & $0.2$  & $0.025$  & $300$  &  $\rho (1) = 1$  & $\rho>0$ $ \forall$ $r \in [1,5]$  \\
 $0$ & $-0.5$ & $0.6$  & $0.008$  & $300$  & $\rho (1) = 1$ &  $\rho>0$ $ \forall$ $r \in [1,5]$  \\
 $0$ & $-0.5$ & $1.0$  & $0.005$  & $200$  &  $\rho (1) = 1$ & $\rho>0$ $ \forall$ $r \in [1,5]$  \\
 $0$ & $-0.5$ & $1.25$  & $0.004$ & $200$  &  $\rho (1) = 1$ & $\rho>0$ $ \forall$ $r \in [1,5]$  \\
 $0$ & $-0.5$ & $1.5$  & $0.003$  & $300$  & $\rho (1) = 1$  & $\rho>0$ $ \forall$ $r \in [1,5]$  \\
 \hline
 $1$  & $-0.1$ & $0.6$  & $0.008$  &  $200$  & $\rho (1) = 1$ & $\rho>0$ $ \forall$ $r \in [1,5]$  \\
 $1$  & $-0.25$ & $0.6$  & $0.008$  & $200$  &  $\rho (1) = 1$  & $\rho>0$ $ \forall$ $r \in [1,5]$   \\
 $1$  & $-0.3$ & $0.6$  & $0.008$   & $200$  & $\rho (1) = 1$ & $\rho>0$ $ \forall$ $r \in [1,5]$  \\
 $1$  & $-0.4$ & $0.6$  & $0.008$   & $200$  &  $\rho (1) = 1$ & $\rho>0$ $ \forall$ $r \in [1,5]$  \\
 $1$  & $-0.5$ & $0.6$  & $0.008$    & $200$  &  $\rho (1) = 1$ & $\rho>0$ $ \forall$ $r \in [1,5]$  \\
 $1$  & $-0.6$ & $0.6$  & $0.008$    & $200$  &  $\rho (1) = 1$ & $\rho>0$ $ \forall$ $r \in [1,5]$  \\
 $1$  & $-0.7$ & $0.6$  & $0.008$   & $200$  &  $\rho (1) = 1$ & $\rho>0$ $ \forall$ $r \in [1,5]$  \\
 $1$  & $-0.9$ & $0.6$  & $0.008$   & $200$  &  $\rho (1) = 1$ & $\rho>0$ $ \forall$ $r \in [1,5]$  \\
 $1$  & $-0.5$ & $\sqrt{2}$  & $0.005$   & $256$  & $\rho (1) = 1$ $\rho (r \rightarrow \infty)=0$ & 
 Pressure equipartition; $v_{A\phi0}= \sqrt{\frac{-2\alpha}{(1+\alpha)}}$\\
 $1$  & $-0.3$ & $0.926$ & $0.001$   & $200$ & $\rho (1) = 1$; $\rho (r \rightarrow \infty)=0$ & \\
 $1$  & $-0.3$ & $0.926$ & $0.01$   & $200$ & $\rho (1) = 1$; $\rho (r \rightarrow \infty)=0$ & \\
 $1$  & $-0.3$ & $0.926$ & $0.1$   & $200$ & $\rho (1) = 1$; $\rho (r \rightarrow \infty)=0$ & \\
 $1$  & $-0.3$ & $0.926$ & $0.5$   & $200$ & $\rho (1) = 1$; $\rho (r \rightarrow \infty)=0$ & \\
 $1$  & $-0.3$ & $0.926$ & $0.75$   & $200$ & $\rho (1) = 1$; $\rho (r \rightarrow \infty)=0$ & \\
 $1$  & $-0.3$ & $0.926$ & $1.0$   & $200$ & $\rho (1) = 1$; $\rho (r \rightarrow \infty)=0$ & \\
 $1$  & $-0.3$ & $0.926$ & $0.005$   & $256$ &  $\rho (r \rightarrow \infty)=10$ & constant density background\\
 $1$  & $-0.3$ & $0.926$ & $0.005$   & $256$ &  $\rho (r \rightarrow \infty)=100$ & constant density background\\
% $1$  & $0.4$ & $0.447$ & $0.022$   & $256$ & $\rho (1) = 1$ & $\alpha>0$ model; $\rho>0$ $ \forall$ $r \in [1,5]$  \\
\hline
\end{tabular}
%\vspace{-4mm}
\label{tab_results_PL1}
\end{table*}

\subsection{Results for a power-law toroidal field and constant vertical field: $v_{A\phi}{\rm-PL-}v_{Az}{\rm-Cons}$}
\label{sec_Bphi_PLcons_Bz_cons}

We begin by choosing a power-law for the toroidal field, 
$v_{A\phi}= v_{A\phi0}r^\alpha$, and a constant vertical field, $v_{Az}=v_{Az0}$, 
where $v_{A\phi0,Az0}$ are the respective quantities at the inner jet radius ($r=1$). 
This case provides a direct comparison with the local analysis of B98 and 
allows us to test the validity of the stability criterion proposed therein. 
%We carry out a detailed stability analysis 
%by varying all the free parameters of the problem, namely, $m,\alpha,v_{A\phi0,Az0}$.
The magnetic pitch for this case follows from equation (\ref{eq_pitch}) such that
\begin{equation}
{\cal P} = r^{1-\alpha} \frac{v_{Az0}}{v_{A\phi0}} ~.
\end{equation}
Thus, the pitch is an increasing function of radius if $\alpha<1$, a decreasing 
function if $\alpha>1$ and a constant if $\alpha=1$. All the runs carried out 
for this case are summarized in Table \ref{tab_results_PL1}.

The background density for this case is obtained by integrating equation (\ref{eq_radeqbm}). Keeping 
our normalization scheme in mind (see Table \ref{tab_notation_global}), this yields
\begin{equation}
\rho(r) = - \frac{(1+\alpha)v_{A\phi0}^2}{2\alpha}r^{2\alpha}  + \rho_{\rm cons} ~,
\label{eq_rhodimful_PL}
\end{equation}
where $\rho_{\rm cons}$ is the integration constant.
Now, in order to determine $\rho_{\rm cons}$, we impose the {\it inner} boundary condition 
\begin{equation}
\rho=1 ~~{\rm at}~~ r=1 ~,
\label{cond_rhoin1}
\end{equation}
which yields
\begin{equation}
\rho(r) = 1 + \frac{(1+\alpha)v_{A\phi0}^2}{2\alpha}(1-r^{2\alpha}) ~.
\end{equation}
The plasma-beta can be written as
\begin{equation}
\beta = \frac{2\rho(r)}{v_{A\phi0}^2 r^{2\alpha} + v_{Az0}^2} ~.
\label{eq_beta_PL}
\end{equation}
However, one needs to keep in mind that if $v_{A\phi0}$ becomes too large, and/or $\alpha$ assumes 
a large positive value, then the density may become negative beyond a certain radius. 
This puts a restriction on the computational domain for certain parameter values. 
If $\alpha<0$ on the other hand, 
then an additional {\it outer} boundary condition can be imposed such that 
\begin{equation}
\rho \rightarrow 0 ~~{\rm as}~~ r \rightarrow \infty ~.
\label{cond_rhoinf0}
\end{equation}
The above condition sets $\rho_{\rm cons}=0$ and 
self-consistently assures $\rho>0$ throughout the domain,  yielding
\begin{equation}
 \rho(r) = r^{2\alpha}  ~~~{\rm and}~~~ v_{A\phi0} = \sqrt{\frac{-2\alpha}{(1+\alpha)}} ~.
\label{eq_rhobg_PL}
\end{equation}
Furthermore, on assuming $v_{Az0} \ll v_{A\phi}$, equation (\ref{eq_beta_PL}) reduces to
\begin{equation}
\beta \approx -\frac{1+\alpha}{\alpha} = {\rm ~constant} ~.
\end{equation}
%This is the most self-consistent way to assure $\rho>0$ throughout the domain. 

Our fiducial model corresponds to $\alpha=-0.3$, $v_{A\phi0}=0.926$, 
$v_{Az0}=0.005$ and $\beta \approx 2.33$.
For this model, we impose {\it both} conditions 
(\ref{cond_rhoin1}) and (\ref{cond_rhoinf0}) on the background density. However, in this case, 
$v_{A\phi0}$ depends on $\alpha$ and is no longer a free parameter (see equation \ref{eq_rhobg_PL}). Thus, for some cases 
(e.g. when varying $\alpha$ for a given $v_{A\phi0}$, 
when $\alpha>0$, etc.), we impose {\it only} the condition (\ref{cond_rhoin1}). 
We use $r \in [1,5]$ for all the cases listed in Table \ref{tab_results_PL1},  
which also assures $\rho>0$ throughout the domain. 
We shall now discuss the properties of the axisymmetric $m=0$, 
and the non-axisymmetric $m=1$ and $2$ modes, corresponding to the cases listed in 
Table \ref{tab_results_PL1}, separately. 

%Figure \ref{fig_PLfidu_profiles} shows the 
%different background profiles as a function of $r$ for our fiducial power-law case 
%having and $\rho$ given by equation (\ref{eq_rhobg_PL}). 

%Note that in this section we will mostly consider $\alpha<0$ (i.e., with increasing pitch).
%which is a more realistic scenario.

%This is not necessarily true at small r; my fiducial models for 
%the Crab with Li have alpha ~ +1 ar small r, turning over at large r.

%\begin{figure*}
%\centering
%\includegraphics[width=0.7\columnwidth]{PL_fidu_profiles.eps}
 %      \caption{}
  %       \label{fig_PLfidu_profiles}
%\end{figure*}

\subsubsection{Axisymmetric $m=0$ modes}
\label{sec_PL_m0}

%\vspace*{\floatsep}

  %\begin{tabular}{@{}cccc@{}}
    %\includegraphics[width=\textwidth]{growth_draft_all_onlyaxiBp0.3_ref.eps} %&
 %      \end{tabular}
 %\includegraphics[width=0.9\textwidth]{growth_draft_all_onlyaxiBp0.30_w4disp_l10.png}

\begin{figure*}
%\hspace{-0.7cm}
%\captionsetup{width=1.1\textwidth}
\centering
  %\begin{tabular}{@{}cccc@{}}
    %\includegraphics[width=\textwidth]{growth_draft_all_onlyaxiBp0.3_ref.eps} %&
 %      \end{tabular}
 %\includegraphics[width=0.9\textwidth]{growth_draft_all_onlyaxiBp0.30_w4disp_l10.png}
\includegraphics[width=\columnwidth]{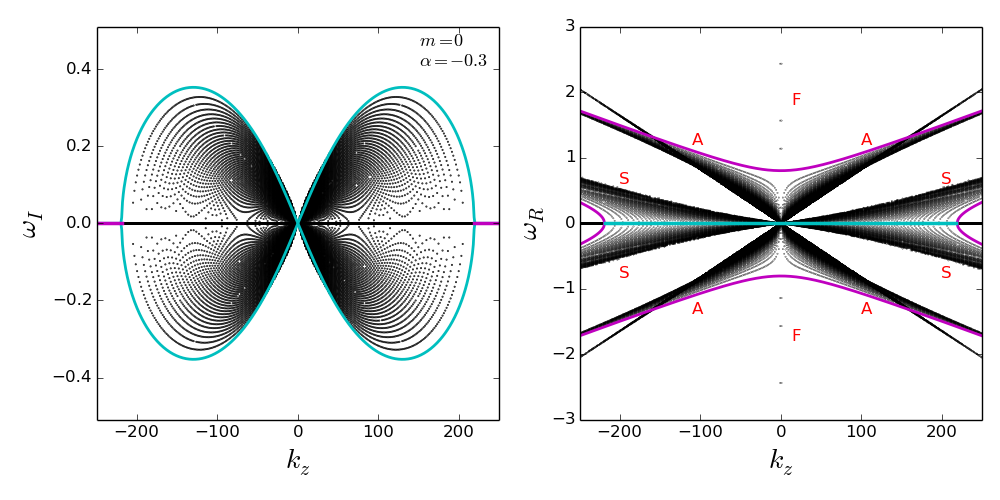}
       \caption{Global eigenvalue solutions for the fiducial $m=0$ case of $v_{A\phi}{\rm-PL-}v_{Az}{\rm-Cons}$, 
with $\alpha=-0.3$, $v_{A\phi0} = 0.926$ and $v_{Az}= v_{Az0} =0.005$.
Left panel: The imaginary part $\omega_I$ of the modal frequency or,
the growth rate, as a function of $k_z$.  Right panel:  
The real part $\omega_R$ of the modal frequency as a function of $k_z$. 
The global eigenvalue solutions are shown in black.
The cyan (unstable modes with $\omega_I > 0$ and stable modes with $\omega_I<0$ in 
left panel, and the corresponding $\omega_R$ in right panel) 
and magenta (stable modes with $\omega_I = 0$ in left panel and the 
corresponding $\omega_R$ in right panel) lines represent the solutions of the local dispersion relation, 
equation (\ref{mcb_dispeq_dimles}) with $l^2/k_z^2=0$.
The letters F, S and A denote the fast, slow and
Alfv\'{e}n modes, respectively. The global problem is solved on a radial
grid $r \in [1,5]$ with resolution $N_r$ = 256.}
         \label{fig_PLm0_wir}
\end{figure*}      
%\vspace*{\floatsep}

\begin{figure*}
\centering
\includegraphics[width=\columnwidth]{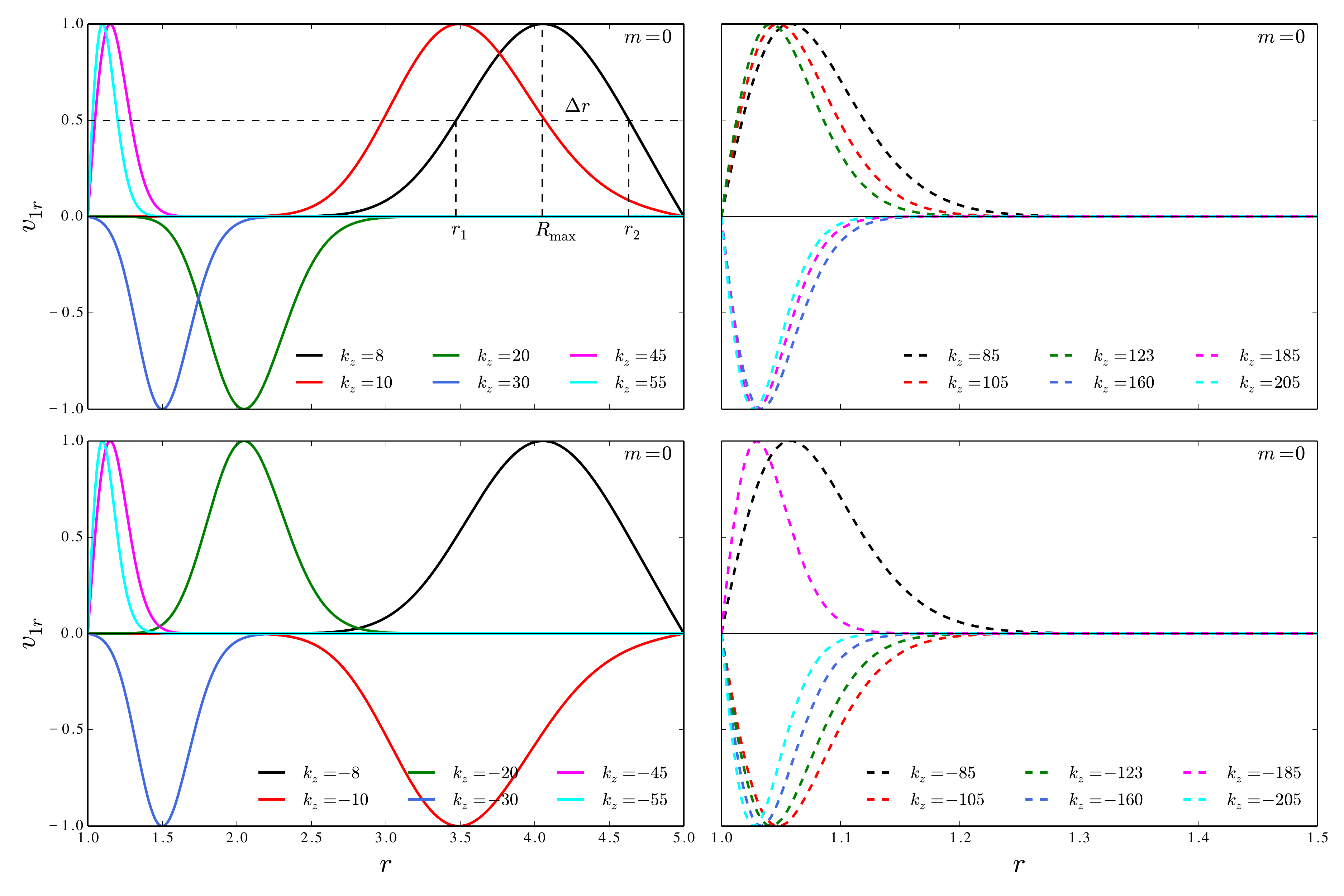}
       \caption{Normalized radial eigenfunction $v_{1r}$ for the most unstable modes 
       as a function of radius $r$, for the case discussed in Figure \ref{fig_PLm0_wir}. 
Top panels: $k_z>0$ modes.  Bottom panels: $k_z<0$ counterparts of the respective top panels 
(see colored legends for $k_z$ values; 
note that $|k_z|$ increases from right to left in all panels).
$\Delta r = r_2-r_1$ is the full-width at half maximum and $R_{\rm max}$ 
the radial peak location of the eigenfunctions. The 
radial axes in the right panels have been zoomed close to the inner boundary for clarity.}
         \label{fig_PLm0_efuncs}
\end{figure*}

The main findings of this subsection, listed below, are summarized by Figures 
\ref{fig_PLm0_wir}, \ref{fig_PLm0_efuncs} and \ref{fig_PLm0_most}:

\begin{enumerate}

\item The left panel of Figure \ref{fig_PLm0_wir} shows the imaginary part $\omega_I$ 
of the eigenvalues, or growth rate, as a 
function of the vertical wavenumber $k_z$ for the fiducial $\alpha=-0.3$ case (see Table \ref{tab_results_PL1}). 
The right panel shows the real part $\omega_R$ of the corresponding eigenvalues. 
We also demarcate the fast magnetosonic, Alfv\'{e}n and slow
magnetosonic modes in the right panel, indicated by the letters F, A and S, respectively.
Note that the total number of modes in the global problem is proportional 
to the radial resolution. The cyan ($\omega_I \neq 0$) and magenta ($\omega_I=0$) 
lines in this figure represent the corresponding local solutions
predicted by the B98 dispersion relation, which are given by equation (\ref{mcb_dispeq_dimles}) 
with $l^2/k_z^2=0$.

% The cyan line in the right panel indicates the 
%corresponding $\omega_R$ of the most unstable modes, while
%, showing that these are purely growing modes with zero phase velocity. 
%the magenta lines indicate the $\omega_R$ of the stable modes.
%, which are purely oscillatory. The unstable modes appear to be destabilized slow modes. 
%Note that the fast modes (stable) do not have a local counterpart as they are 
%excluded from the local analysis (which yields a fourth-degree dispersion relation instead of sixth; see equation 
%\ref{mcb_dispeq_dimles}). 

From the left panel of Figure \ref{fig_PLm0_wir}, we see 
that the $m=0$ modes are symmetric about the $k_z$-axis  (also, see equation \ref{mcb_dispeq_dimles}), 
with $\omega_I=0$ at $k_z=0$.
We  also find that the $\omega_I>0$ cyan lines are 
in very good agreement with the growth rates of the 
{\it most} unstable modes of the global eigenvalue solution.

\item Some of the measurable quantities predicted by the local dispersion relation for this case are: 
the maximum growth rate $\omega_g|_{\rm max} \approx 0.35$ (see equation \ref{eq_wmax}), the corresponding 
vertical wavenumbers $k_z |_{\rm max} \approx \pm 129$ 
(see equation \ref{eq_kzmax}) and the range of instability $-219 \lesssim k_z|_{\rm unst}\lesssim 219$ 
(see relation \ref{eq_kzunst}). All the corresponding global quantities have comparable but 
slightly lower values as seen from Figure \ref{fig_PLm0_wir}: 
$\omega_g|_{\rm max} \approx 0.33$, $k_z |_{\rm max} \approx \pm 124$ and $-206\lesssim k_z|_{\rm unst}\lesssim 206$.
%There are, however, a few subtle differences between the local and global solutions. For instance,
The growth rates of the most unstable modes in the range 
$0 \leq k_z \lesssim 60$ are also slightly larger than the local prediction (see point (v) below).

\item On the right panel of Figure \ref{fig_PLm0_wir}, note that the 
the cyan line indicates $\omega_R$ corresponding to the most unstable modes of the left panel 
($\omega_I>0$), while
%, showing that these are purely growing modes with zero phase velocity. 
the magenta lines indicate the $\omega_R$ corresponding to the stable modes ($\omega_I=0$ and $\omega_I<0$). From the 
global solutions we find that the most unstable modes are purely growing with zero phase velocity $\omega_R=0$, 
while the stable modes are purely oscillatory, corroborating the local predictions. 
In fact, the unstable modes appear to be destabilized slow modes. 
Note that the (stable) fast modes do not have a local counterpart as they are 
excluded from the local analysis (which yields a fourth-degree dispersion relation instead of sixth; see equation 
\ref{mcb_dispeq_dimles}). 
A subtle difference between the local and global solutions is as follows: 
in the global solutions, as $k_z \rightarrow 0$, $\omega_R \rightarrow 0$ for {\it all} the Alfv\'{e}n modes, unlike 
the local solutions, which have $\omega_R \neq 0$ for $k_z=0$. Also, for $|k_z|>160$ and $|\omega_R|>1.25$ 
(where the black lines intersect and then diverge away from the magenta lines), additional modes (possibly Alfv\'{e}nic) 
appear in the global solutions that do not have a local counterpart.

%a bit unclear in what sense these are additional modes.  Is this where the black and magenta lines intersect?

\item We next discuss Figure \ref{fig_PLm0_efuncs}, which
shows the eigenfunction $v_{1r}$ as a function of radius
for the {\it most} unstable mode at different $k_z$ corresponding to Figure \ref{fig_PLm0_wir}. 
Note that the eigenfunctions are normalized with respect to their respective maximum amplitudes. 
Let us define two characteristics of the radial eigenfunctions, which will help us better 
understand the nature of the instabilities in this work. These 
are $\Delta r = r_2-r_1$, which is the full-width at half maximum of the eigenfunction, 
and $R_{\rm max}$, 
which gives the peak position of the eigenfunction as measured from the inner boundary $r=1$ 
(demonstrated with $k_z=8$ in the top left panel of Figure \ref{fig_PLm0_efuncs}).
Based on these, we can broadly classify any instability in the current 
work either as: (1) {\it Local}
or (2) {\it Global}. A local instability has very narrow eigenfunctions (i.e. $\Delta r \lesssim 0.25$), 
which are highly concentrated (or localized) towards the inner boundary (i.e. $R_{\rm max} \rightarrow 1$). 
These properties 
indicate that the instability does not care about the radial variation of the background 
quantities like magnetic field and density and, hence, can be fully described by a local 
calculation (i.e. radially localized). A global instability, on the other hand, is affected by the 
background radial profiles and, hence, can only be captured in a global framework
(i.e. radially extended). 
These can be of three kinds, having eigenfunctions that are: either wide and peak significantly away from the inner 
boundary (Global I, i.e. $\Delta r > 0.25$ and $R_{\rm max} \nrightarrow 1$); or
wide and peak towards the inner 
boundary (Global II, i.e. $\Delta r > 0.25$ and $R_{\rm max} \rightarrow 1$); or
narrow and peak significantly away from the inner 
boundary (Global III, i.e. $\Delta r \lesssim 0.25$ and $R_{\rm max} \nrightarrow 1$). Table \ref{tab_instabilities} 
summarizes the characteristics of the various instabilities.
We emphasize that although we distinguish between local and global instabilities here, they 
are both still {\it internal}, small-scale instabilities, which do not 
affect the large-scale properties of the jet.

\begin{table*}
\large
\centering
%\vskip0.2cm
\renewcommand{\arraystretch}{1.5}
\caption{Classification of the {\it internal} instabilities based on the properties of their radial eigenfunctions.} 
%$v_{A\phi} = v_{A\phi0} r^\alpha$, 
%$v_{Az}=v_{Az0}$, $\rho = 1 + \frac{(1+\alpha)v_{A\phi0}^2}{2\alpha}(1 - r^{2\alpha})$.
 \begin{tabular}{|c|c|c|}
 %\hline
 %\hskip1cm General & &  \hskip2cm General  \\ \hline
 \hline
 Full-width at half maximum & Radial peak location & Type of instability \\
 \hline
 $\Delta r \lesssim 0.25$ & $R_{\rm max} \rightarrow 1$ & Local \\
 $\Delta r > 0.25$ & $R_{\rm max} \nrightarrow 1$ & Global I \\
 $\Delta r > 0.25$ & $R_{\rm max} \rightarrow 1$ & Global II \\
 $\Delta r \lesssim 0.25$ & $R_{\rm max} \nrightarrow 1$ & Global III \\
 \hline
\end{tabular}
%\vspace{-4mm}
\label{tab_instabilities}
\end{table*}

\begin{figure*}
%\hspace{-0.7cm}
%\captionsetup{width=1.1\textwidth}
\centering
  %\begin{tabular}{@{}cccc@{}}
    %\includegraphics[width=\textwidth]{growth_draft_all_onlyaxiBp0.3_ref.eps} %&
 %      \end{tabular}
 %\includegraphics[width=0.9\textwidth]{growth_draft_all_onlyaxiBp0.30_w4disp_l10.png}
\includegraphics[width=\columnwidth]{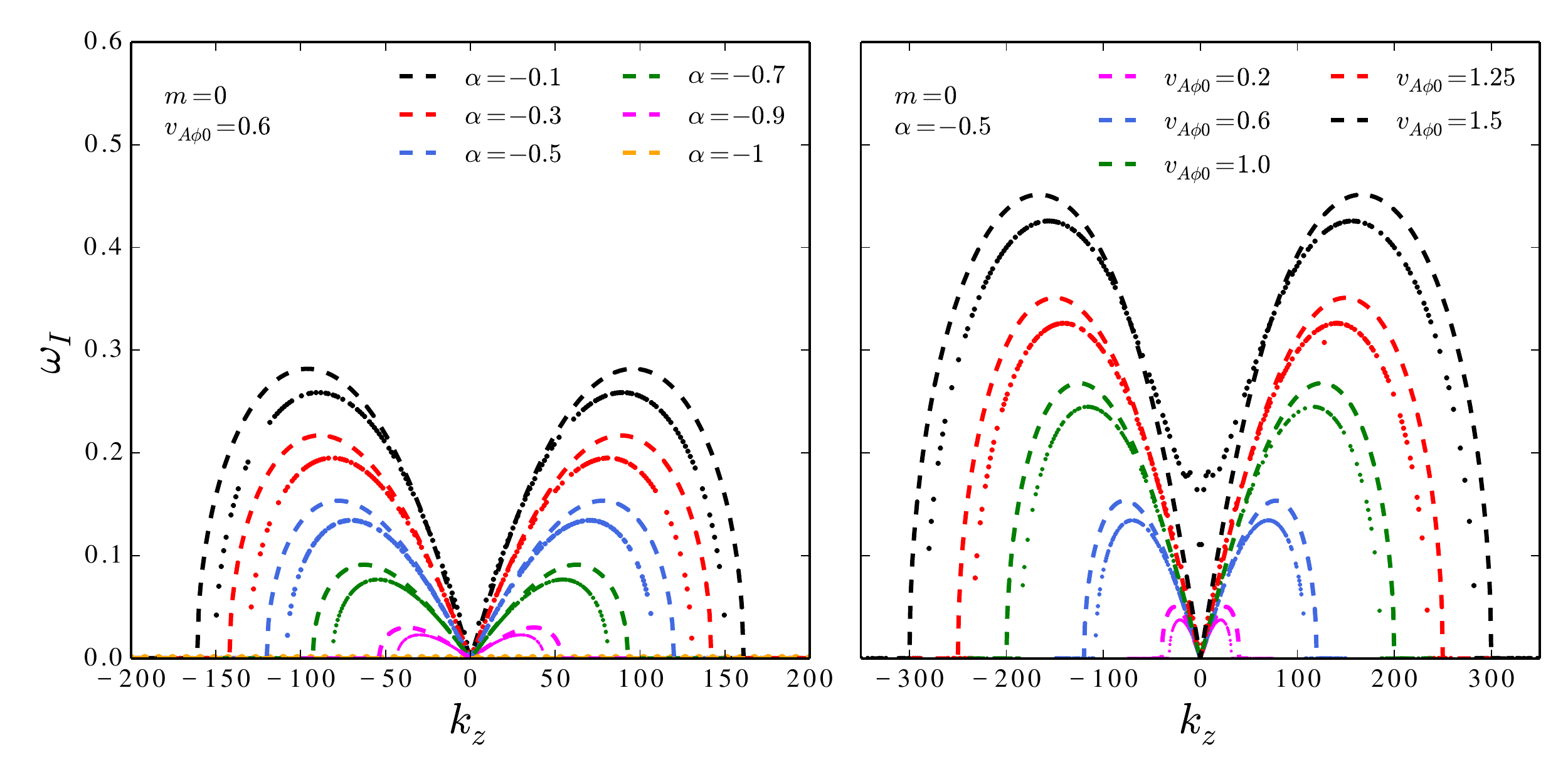}
       \caption{Growth rate $\omega_I$ as a function of $k_z$ for the $m=0$ case of $v_{A\phi}{\rm-PL-}v_{Az}{\rm-Cons}$
with $v_{Az0}= 0.005$. Left panel: Cases when $v_{A\phi0}=0.6$ is fixed and $\alpha$ is varied 
(see colored legends; note that $\alpha$ decreases from top to bottom). Right panel: Cases when $\alpha=-0.5$ 
is fixed and $v_{A\phi0}$ is varied (see colored legends; note that $v_{A\phi0}$ decreases from top to bottom).
The dotted and dashed lines represent the global eigenvalue and local (equation \ref{mcb_dispeq_dimles}) solutions  respectively.}
         \label{fig_PLm0_most}
\end{figure*}

\item Keeping the above characteristics in mind, we can now look at the different eigenfunctions in 
Figure \ref{fig_PLm0_efuncs}. The top left panel shows the eigenfunctions for $8 \leq k_z \leq 55$, the top 
right for $85 \leq k_z \leq 205$, the bottom left for $-55 \leq k_z \leq -8$ 
and the bottom right for $-205 \leq k_z \leq -85$ (the bottom panels 
thus representing the $k_z<0$ counterparts of the respective top panels). 
Note that the radial axes of the right panels of Figure \ref{fig_PLm0_efuncs} have been 
zoomed close to the inner boundary $r \in [1,1.5]$ for clarity, 
whereas the left panels show the full radial domain $r \in [1,5]$.
We see that the eigenfunctions in the bottom panels are identical to their top-counterparts 
(except for a 180$^\circ$ phase shift displayed by some), which again reflects the 
symmetry of the $m=0$ modes about the $k_z$-axis. 
The most unstable modes are found to be
nodeless in general, across the entire range of unstable $k_z$.
%Keeping in mind that the radial axes of the right panels of Figure \ref{fig_PLm0_efuncs} have been 
%zoomed close to the inner boundary $r \in [1,1.5]$ for clarity, 
%whereas the left panels show the full radial domain $r \in [1,5]$, 
%we make the following observations. 
%
The modes shown in the left panels of 
Figure \ref{fig_PLm0_efuncs} have smaller $|k_z|$ 
than those in the right panel. They represent Global I and II instabilities.
The eigenfunctions span a large radial extent, and as $|k_z| \rightarrow 0$, 
the modes seem to peak towards the outer boundary (Global I), 
gradually shifting towards the inner boundary as $|k_z|$ increases (Global II).  
This explains
why the global solution in Figure \ref{fig_PLm0_wir} deviates from the local prediction 
as $k_z \rightarrow 0$.
On the other hand, the larger $|k_z|$ modes in the right panels 
are Local instabilities, having narrow eigenfunctions that are
highly localized close to the inner boundary. {\it In general, 
the modes become more localized towards $r=1$ as $|k_z|$ increases}.

\item In Figure \ref{fig_PLm0_most}, we study some interesting trends displayed by the 
$m=0$ modes. The left panel 
shows the evolution of 
the most unstable $m=0$ modes for a fixed $v_{A\phi0}=0.6$ 
(and $\beta_0 \approx 2/v_{A\phi0}^2 = 5.55$), when $\alpha$ is varied. 
The dotted and dashed lines represent the global and local solutions 
respectively, which are in fairly good agreement.
We find that as $\alpha$ decreases, the growth rates of the 
modes (in both local and global solutions) keep decreasing till they are completely stabilized at $\alpha=-1$, thus 
validating the B98 local stability criterion (see \S \ref{sec_b98_stabcrit}). 
%a trend exhibited by both the local and global solutions. 
%The reason the local B98 stability criterion holds true even for the 
%global solutions is that the $m=0$ modes are essentially Local instabilities 
%across most of the unstable $k_z$ range.
%{\it This verifies the B98 stability criterion for the $m=0$ modes in the global scenario}.
%
The right panel of Figure \ref{fig_PLm0_most}, 
on the other hand, shows the corresponding evolution for a fixed $\alpha=-0.5$, when 
$v_{A\phi0}$ is varied. We find that as $v_{A\phi0}$ increases, the growth rates 
of the most unstable modes increase. We also see significant deviation from the local solution 
for $v_{A\phi0}=1.5$ ($\beta_0 =0.89$) at low $|k_z|$ values.
%, which is similar to the deviation observed in the left panel of Figure \ref{fig_PLm0_wir}. 
This is possibly a consequence of the stronger toroidal field, 
%in both cases, 
which makes the radial curvature effects more important than the corresponding weaker field cases
and, hence, necessitates a global analysis.

\end{enumerate}

%Thus, each value of kz corresponds to multiple modes and the mode with the 
%maximum growth rate is the most unstable mode at that kz. 

\subsubsection{Non-axisymmetric $m=1$ modes}
\label{sec_PL_m1}

The main findings of this subsection, described below, are summarized by Figures 
\ref{fig_PLm1_wir}-\ref{fig_PLm1_Bzeff}:

\begin{enumerate}

\begin{figure*}
%\hspace{-0.7cm}
%\captionsetup{width=1.1\textwidth}
\centering
  %\begin{tabular}{@{}cccc@{}}
    %\includegraphics[width=\textwidth]{growth_draft_all_onlyaxiBp0.3_ref.eps} %&
 %      \end{tabular}
 %\includegraphics[width=0.9\textwidth]{growth_draft_all_onlyaxiBp0.30_w4disp_l10.png}
\includegraphics[width=\columnwidth]{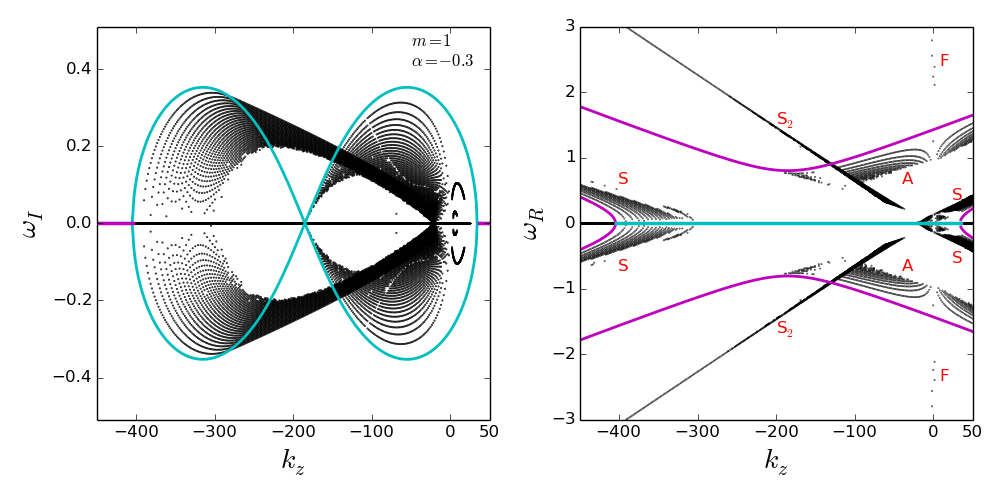}
       \caption{Global eigenvalue solutions for the fiducial $m=1$ case of $v_{A\phi}{\rm-PL-}v_{Az}{\rm-Cons}$, 
       with everything else the same as Figure \ref{fig_PLm0_wir}. In the right panel, S$_2$ denotes a new branch of slow modes 
       that appears in the global eigenvalue solutions.}
       \label{fig_PLm1_wir}
%\vspace*{\floatsep}
\end{figure*}

\item Figure \ref{fig_PLm1_wir} is similar to Figure \ref{fig_PLm0_wir}, except that it represents the
fiducial $\alpha=-0.3$ case for the $m=1$ modes. All the symbols and colors have the same meaning 
in both figures. We note that the $m=1$ modes are no longer symmetric about the $k_z$-axis, as shown by both the 
local and global solutions. Also, non-axisymmetry seems to induce a somewhat greater 
deviation from the local prediction as compared to the axisymmetric case.

First, we discuss the growth rates of the modes with $k_z>0$. 
From the left panel of Figure \ref{fig_PLm1_wir} we see that 
these modes have a much smaller 
growth rate than the local prediction. The local solution predicts a non-zero growth rate 
$\omega_I \approx 0.28$ at $k_z=0$ and instability in the range $0 \leq k_z|_{\rm unst} \lesssim 34$. 
The growth rate in the global solution, on the other hand, has
$\omega_g|_{\rm max} \approx 0.1$ at $k_z |_{\rm max} \approx 9$ and instability for
$1 \lesssim k_z|_{\rm unst} \lesssim 20$.

\item  Next, we discuss the growth rates of the modes with $k_z<0$. 
Note that, as already mentioned at the beginning of \S \ref{sec_mcb_disp}, the sign of $m$ 
determines the direction of the helical twist. However, if $m$ is assumed to be 
positive, as we do in this work, then 
a negative $k_z$ in conjunction with a positive 
$m$ can be interpreted as the helical perturbation, or kink, twisting the 
magnetic field in a sense opposite to itself. On the other hand, 
a positive $k_z$ with a positive $m$ twists the 
magnetic field in the same direction as the kink. This is because the local dispersion 
relation (equation \ref{mcb_dispeq}) is invariant under the transformation $(-m,k_z) \rightarrow (m,-k_z)$ or 
$(-m,\eta) \rightarrow (m,-\eta)$ .

The local solution predicts two peaks with 
{\it identical} maximum growth rates $\omega_g|_{\rm max} \approx 0.35$ 
(see cyan lines in the left panel of Figure \ref{fig_PLm1_wir}). The first instability 
peak occurs at 
$k_z |_{\rm max} \approx -55$ and the second at $k_z |_{\rm max} \approx -315$, 
with the instability in the range $-404 \lesssim k_z|_{\rm unst} \leq 0$. 
The double-peaked structure is 
due to the {\it local} magnetic resonance occurring at $r =r_{\rm res}=1$ for
$k_z|_{\rm res} = -m v_{A\phi0}/v_{Az0} \approx -185$ (see equation \ref{eq_reson}), 
which segregates the two $k_z<0$ instability peaks. 
One can thus interpret the instabilities to be originating 
from $r_{\rm res}=1$ (see \S \ref{sec_resonance}) in the local scenario. However, an important point to 
note is that magnetic resonance is an intrinsically global condition. Thus, in 
order to truly understand if it plays any role, we need to check if the 
unstable eigenfunctions have significant amplitude at the corresponding resonant radius. This can be only verified 
in a global framework, as demonstrated in point (vi) below.
%The resonance segregates the two $k_z<0$ instability peaks of the local solution, and

%such that the first stabilizes at $k_z|_{\rm res}$ and the second originates from $k_z|_{\rm res}$.

The global solution also exhibits two peaks in the growth rate, which are, however, 
{\it not} identical unlike the local solution (left panel of Figure \ref{fig_PLm1_wir}). 
The peak values and $k_z$ locations differ 
slightly from the local values, matching more closely for the second peak: 
$\omega_g|_{\rm max} \approx 0.32$ at $k_z |_{\rm max} \approx -63$ and $\omega_g|_{\rm max} \approx 0.34$ at 
$k_z |_{\rm max} \approx -304$ and instability in the range 
$-392 \lesssim k_z|_{\rm unst} \leq 0$. 
%Interestingly, $k_z \approx -185$ in the global solution also corresponds to the local magnetic resonance at $r=1$. 
%However, the behavior at resonance is very different in the two cases. 
The behavior at $k_z \approx -185$, where local magnetic resonance is satisfied, is also very different.
In the global solution, although the first instability peak 
stabilizes at $k_z \approx -185$, the
second peak does not originate from $k_z \approx -185$, but from $k_z=0$. This leads to an {\it interaction 
between the unstable modes} in the global scenario, which appears as a band 
between the two instability peaks (these are slow-mode-slow-mode interactions).
%Thus, the global solution has $\omega_I \neq 0$ at $k_z|_{\rm res}$ and $\omega_I=0$ at $k_z=0$, unlike the local solution.

%about $k_z|_{\rm res}$.

%which is clearly absent in the local solution.

 %\includegraphics[width=0.9\textwidth]{growth_draft_all_onlyaxiBp0.30_w4disp_l10.png}
\begin{figure*}
\centering
\includegraphics[width=\columnwidth]{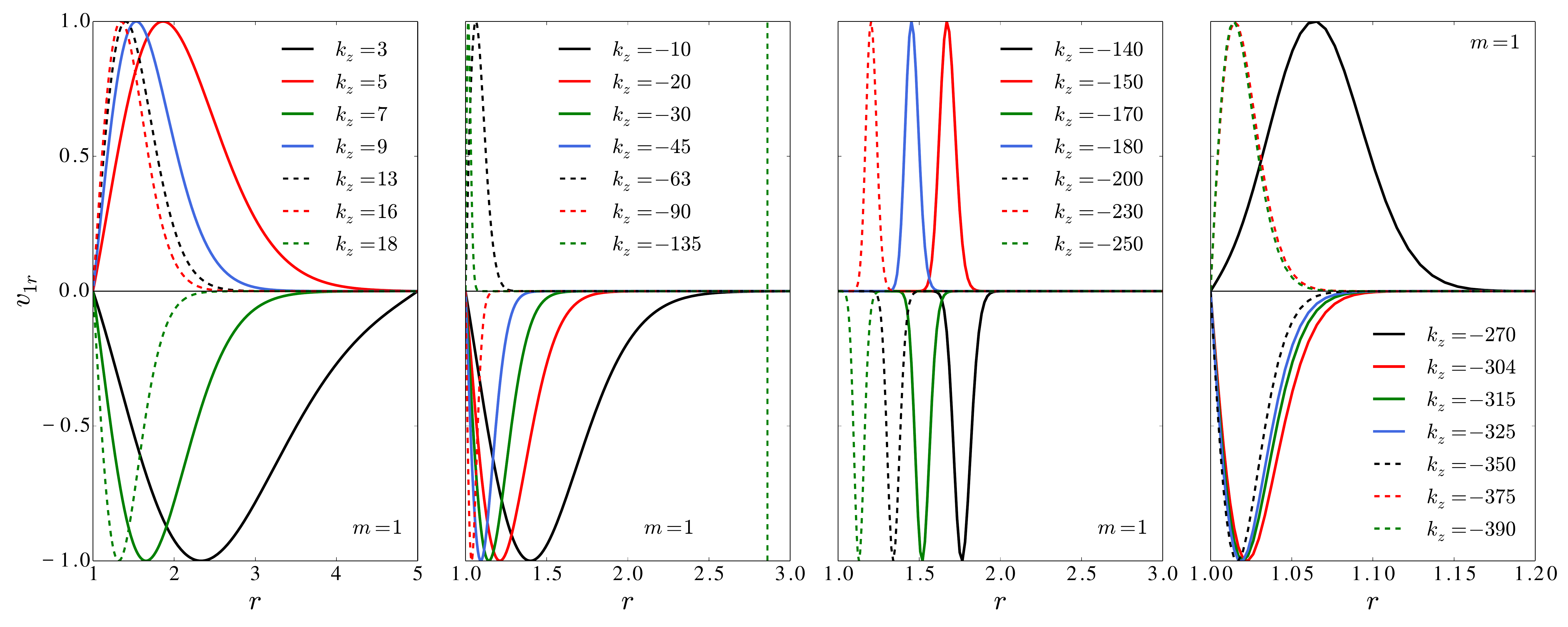}
       \caption{Normalized radial eigenfunction $v_{1r}$ for the most unstable modes as a function of radius $r$, for the case 
discussed in Figure \ref{fig_PLm1_wir} (see colored legends for $k_z$ values; 
note that $|k_z|$ increases from right to left in all panels). The green dashed vertical line, in the second panel 
from left, indicates the resonant radius $r_{\rm res}=2.86$ for $k_z=-135$ (green dashed curve), 
given by equation (\ref{eq_reson}). Note that the radial range shown 
differs from panel to panel for clarity, although the global problem is solved on a grid $r \in [1,5]$.}
         \label{fig_PLm1_efuncs}
\end{figure*}

\item On the right panel of Figure \ref{fig_PLm1_wir}, 
we find that the most unstable modes have zero phase velocity ($\omega_R=0$) 
like the $m=0$ case, 
and are likely to be destabilized slow modes. 
For $k_z>0$, the local solutions seem to be tracing the slow and Alfven modes of 
the global solutions fairly well. 
%Unlike the $m=0$ case, however, the Alfv\'{e}n modes in both the local and global solutions have $\omega_R \neq 0$ for $k_z=0$. 
For $k_z<0$, we see the appearance of 
possibly another branch of slow modes (marked as S$_2$) that does not have a local counterpart, which also 
seems to be interacting with the Alfv\'{e}n modes.

  \begin{figure*}
%\hspace{-0.7cm}
%\captionsetup{width=1.1\textwidth}
\centering
  %\begin{tabular}{@{}cccc@{}}
    %\includegraphics[width=\textwidth]{growth_draft_all_onlyaxiBp0.3_ref.eps} %&
 %      \end{tabular}
 %\includegraphics[width=0.9\textwidth]{growth_draft_all_onlyaxiBp0.30_w4disp_l10.png}
\includegraphics[width=\columnwidth]{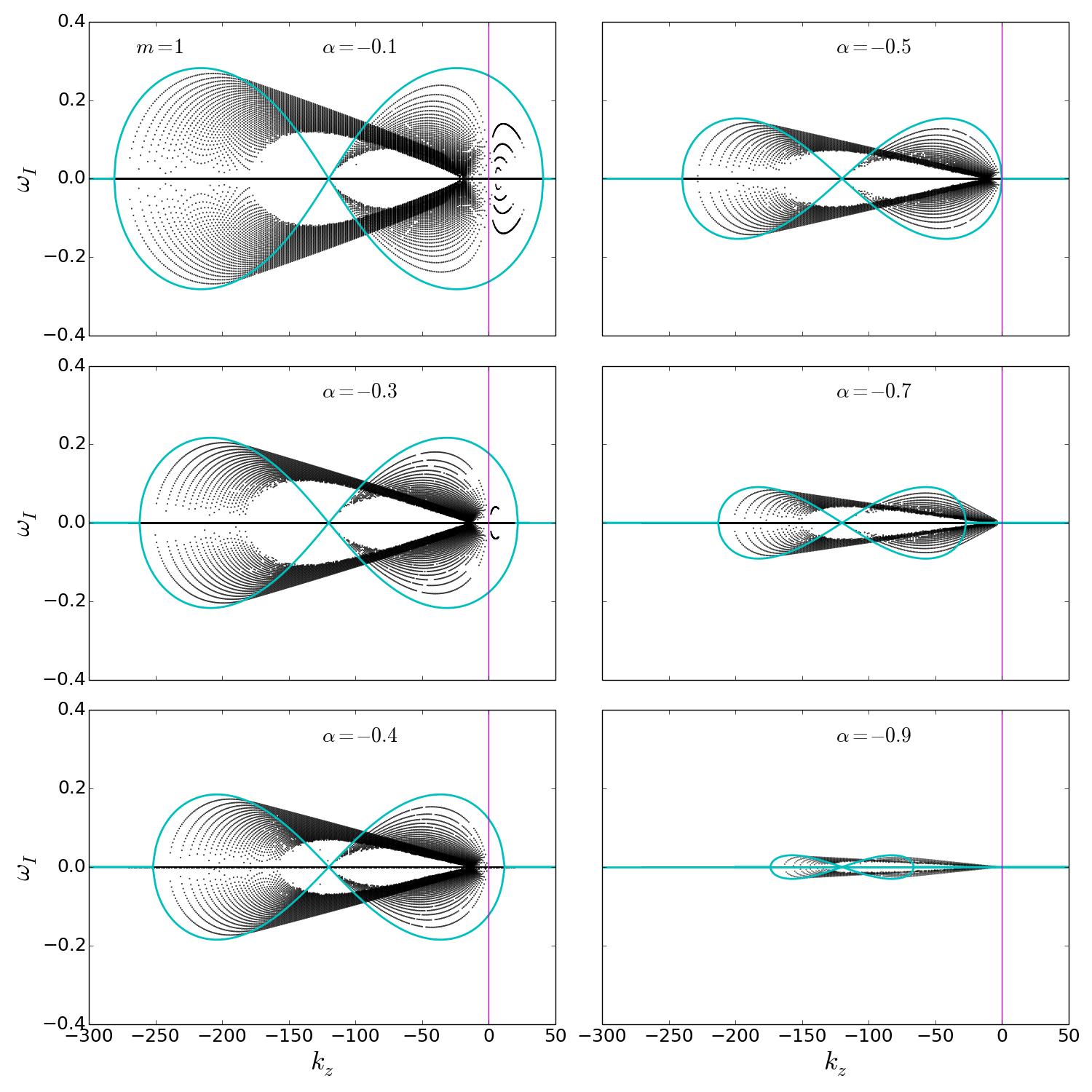}
       \caption{Growth rate $\omega_I$ as a function of $k_z$ for the $m=1$ case of $v_{A\phi}{\rm-PL-}v_{Az}{\rm-Cons}$, 
       when $v_{A\phi0}=0.6$ and $v_{Az0}=0.005$ are fixed, and $\alpha$ is varied. The magenta vertical line indicates
    the $k_z=0$ line in each panel. The global eigenvalue solutions are shown in black. 
    The cyan lines are the corresponding local solutions from the dispersion relation 
    given by equation (\ref{mcb_dispeq_dimles}) with $l^2/k_z^2=0$.}
         \label{fig_PLm1_alphaeff}
\end{figure*}

\item For a detailed understanding of the nature of the $m=1$ instabilities, we need to look at the 
corresponding eigenfunctions. Figure \ref{fig_PLm1_efuncs} 
shows the eigenfunction $v_{1r}$ as a function of radius
for the range of unstable $k_z$ corresponding to Figure \ref{fig_PLm1_wir}. 
%Note that even if a given $k_z$ corresponds to multiple values of $\omega_I$, 
The eigenfunction shown here is that of 
the {\it most} unstable mode at a given $k_z$.
Starting from left, the first panel shows the eigenfunctions for $3 \leq k_z \leq 18$, the
second for $-135 \leq k_z \leq -10$, the third for $-250 \leq k_z \leq -140$ 
and the last for $-390 \leq k_z \leq -270$, plotted on radial axes, $r \in [1,5]$, 
$r \in [1,3]$, $r \in [1,3]$ and $r \in [1,1.2]$, respectively, for clarity. 
We now classify the instabilities according to Table \ref{tab_instabilities}.

The first panel of Figure \ref{fig_PLm1_efuncs} shows that the eigenfunctions for $k_z>0$ 
represent Global I ($k_z<5$) and Global II ($5<k_z<18$) modes. 
This explains the
deviation of the global solution from the local prediction in this $k_z$-range, as
seen in the left panel of Figure \ref{fig_PLm1_wir}. The decrease of $v_{A\phi}$ with $r$
 suppresses the growth rate of the global $k_z>0$ modes compared to the local value. 

The modes with $k_z<0$  exhibit 
mixed behaviors. The second panel shows that the modes with $k_z \geq -45$ are Global II
instabilities, which progress to becoming Local instabilities for $-135 \leq k_z \leq -63$. 
This is reflected in 
the left panel of Figure \ref{fig_PLm1_wir}, where we see that the global eigenvalues approach 
the local prediction (cyan line) as $k_z \rightarrow -135$.

The eigenfunctions in the third panel 
represent Global III instabilities and 
are unique
in the sense that they are shifted away from both the radial boundaries. 
These eigenfunctions represent the interaction between the two sets of
$k_z<0$ instabilities mentioned in point (ii) above (also see left panel of Figure \ref{fig_PLm1_wir}).
As $k_z \rightarrow -250$, the eigenfunctions 
tend to get localized close to the inner boundary again.

Finally, the eigenfunctions in the last panel 
represent Local instabilities, 
and hence the global eigenvalue solution for this range of $k_z$ matches well with the local prediction, as 
seen in the left panel of Figure \ref{fig_PLm1_wir}. 
Overall, we find that the eigenfunctions become more localized 
towards $r=1$ as $|k_z|$ increases, similar to the $m=0$ case.

\item  Figure \ref{fig_PLm1_alphaeff} shows the effect of varying $\alpha$ on 
the growth rates of the $m=1$ modes, for a fixed $v_{A\phi0}=0.6$ and $v_{Az0}=0.005$ (compare with the left panel 
of Figure \ref{fig_PLm0_most} for the equivalent $m=0$ case). As $\alpha$ decreases, the growth rates 
of all the modes decrease for both the local and global solutions. The $k_z>0$ 
unstable modes stabilize first. In the global eigenvalue solution these 
modes stabilize at $\alpha=-0.4$, slightly higher than 
the local prediction of stabilization at $\alpha=-0.5$ (see equation \ref{eq_kzunst}).
%, which yields $-240 <k_z |_{\rm unst} < 0$ for $\alpha=-0.5$). 
This is probably due to the fact that the $k_z>0$ modes are Global 
instabilities, as discussed above and, hence, they stabilize earlier due to the 
rapid fall in $v_{A\phi}$ with radius as $\alpha$ decreases. 
Nevertheless, {\it we may still consider $\alpha=-0.5$ as an absolute stability criterion for the $k_z>0$ modes, 
when $m=1$}.
As $\alpha$ decreases beyond $-0.5$, 
instability exists only for $k_z<0$.
We see that the onset of instabilities as per the local solution keeps shifting to lower $k_z$. 
The instabilities in the global solution also seem to pull away from $k_z=0$ slightly as 
$\alpha \rightarrow -1$, however at a much slower rate than the local solutions. 
The small $|k_z|$ modes thus correspond to Global instabilities 
(I or II; see point (iv) above), which deviate from the local prediction. 
Finally, at $\alpha=-1$ there is complete stabilization of all the $m=1$ modes, again validating B98's local 
stability criterion.
%The physical reason behind the onset of stability at  (when $v_{Az}$ is constant) 
%, which is due to the fact that the (destabilizing) pressure gradient vanishes at this value} (see equation \ref{mcb_radeqbm}).

%, which establishes this as a {\it global stability criterion}.
 
%\begin{figure*}
%\centering
%\includegraphics[width=0.7\columnwidth]{PL_m1_Bp0p92_rhoc_effect.png}
 %      \caption{}
  %       \label{fig_PLm1_rhoceff}
%\end{figure*}

\item We now discuss the resonant character of the $m=1$ modes 
from a global perspective. For the power-law profile in this section, 
the $m=1$ 
modes satisfying the magnetic resonance condition (equation \ref{eq_reson}) 
are given by $k_z = -v_{A\phi0}r^\alpha/v_{Az0}$. 
Thus, for our fiducial domain $1 \leq r \leq 5$, the most 
unstable modes with $-185 \lesssim k_z \lesssim -114$ 
are supposed to be resonant and the ones outside this range are non-resonant. 
We now look at the eigenfunctions of these modes.

Note that the {\it most} unstable modes in the range $-185 \lesssim k_z \lesssim -140$, 
shown in the third panel of Figure \ref{fig_PLm1_efuncs}, 
are the interacting slow modes mentioned in point (ii) above, and are unlikely to be resonant. 
The most unstable modes in the range $-135 \lesssim k_z \lesssim -114$, 
shown in the second panel, could in principle be resonant. In order to verify this, we need to compare the 
the peak locations of these eigenfunctions with the corresponding resonant radius $r_{\rm res}$ 
from equation (\ref{eq_reson}).

%Regarding the peak locations of the resonant eigenfunctions, 
Consider the example of $k_z=-135$ in the 
second panel of Figure \ref{fig_PLm1_efuncs}. It is highly localized close to the inner boundary, 
peaking at $r \sim 1.02$, i.e. well inside
the resonant surface for this mode at 
$r_{\rm res}= (-k_z v_{Az0}/v_{A\phi0})^{1/\alpha} \sim 2.86$ 
(green dashed vertical line in second panel of Figure \ref{fig_PLm1_efuncs}). Thus, the 
amplitude of the eigenfunction is effectively zero at $r_{\rm res}$, which implies that 
magnetic resonance is not actually triggering these modes. 
Note that the role of magnetic resonance might be sensitive to the background 
as well as boundary conditions.
For example, \cite{2013MNRAS.434.3030B} indeed found unstable modes that peak at the 
corresponding resonant radius in their stability analyses carried out for 
relativistic, force-free, pressureless jets with surface velocity shear. Alternatively, \cite{2015MNRAS.450..982K} 
showed that the presence of a finite pressure dilutes the importance of the magnetic resonance condition 
in triggering instabilities.
%as opposed to the case of purely current-driven instabilities. 

%As $k_z$ decreases, the 
%corresponding resonant surface shifts to smaller radii, till  $k_z=-185$, where the two coincide at $r=1$. 

\item With reference to the above discussion on magnetic resonance, we 
can further conclude that Suydam's criterion  
is not applicable to check for instability in this case, as the unstable modes are 
{\it not} localized about $r_{\rm res}$ (see \S \ref{sec_suydam}). 
We do, however, find an interesting result for $\alpha=-1$.
%Interestingly, however, we find that 
%However, Suydam's criterion should be still satisfied if the system is stable. For the power-law model used in this 
%section, 
The left hand side of the Suydam's stability criterion (inequality \ref{eq_suydam}) reduces, in our units and notation, to
\begin{equation}
\frac{1}{4r}(1-\alpha)^2 - \frac{2}{r}(1+\alpha)\frac{v_{A\phi}^2}{v_{Az}^2} ~,
\end{equation} 
which is $\geq 0$ for $\alpha \leq -1$. This happens to be 
consistent with the stability condition for the $m=1$ modes established above (i.e. B98 stability criterion).

\begin{figure*}
%\hspace{-0.7cm}
%\captionsetup{width=1.1\textwidth}
\centering
  %\begin{tabular}{@{}cccc@{}}
    %\includegraphics[width=\textwidth]{growth_draft_all_onlyaxiBp0.3_ref.eps} %&
 %      \end{tabular}
 %\includegraphics[width=0.9\textwidth]{growth_draft_all_onlyaxiBp0.30_w4disp_l10.png}
\includegraphics[width=0.7\columnwidth]{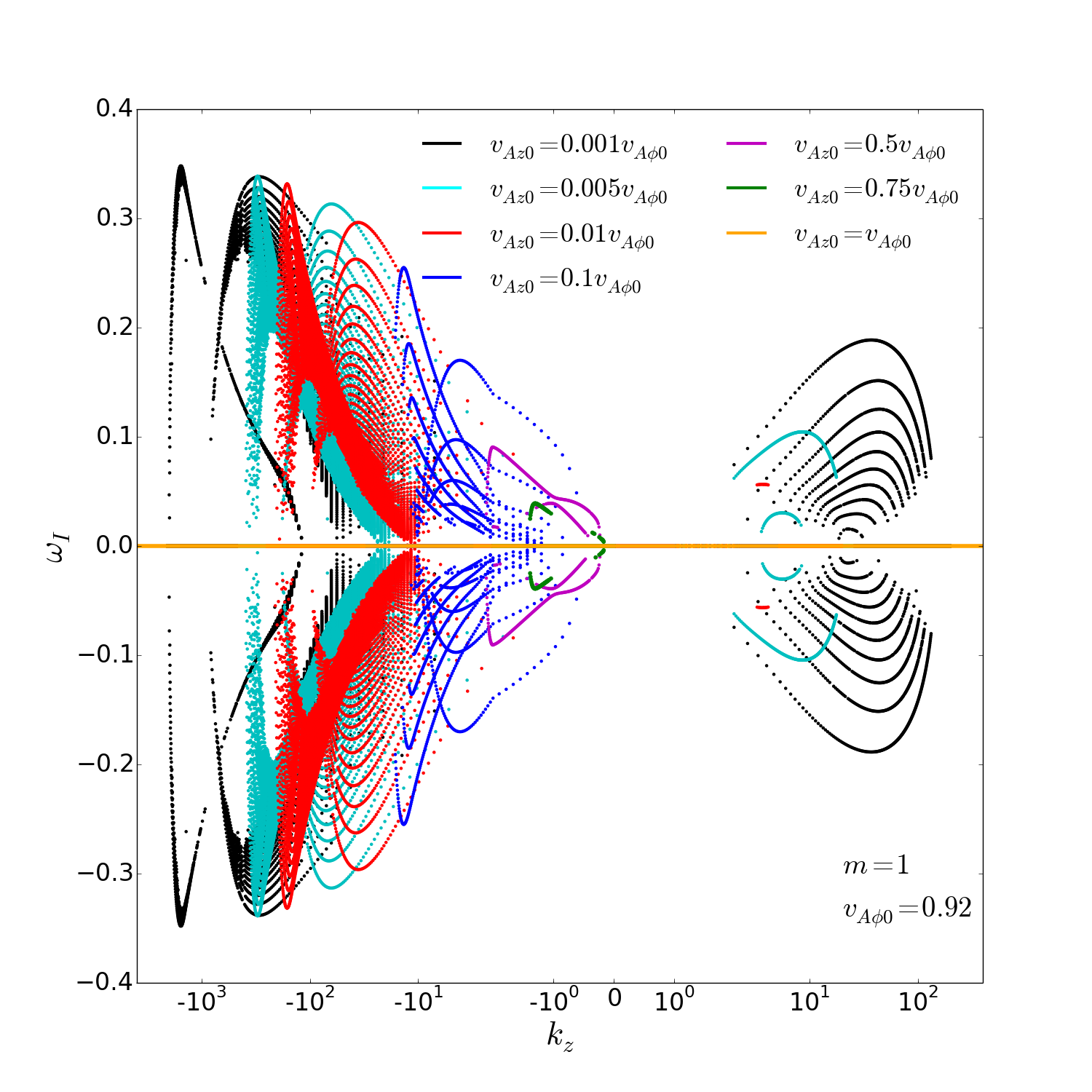}
       \caption{Growth rate $\omega_I$ as a function of $k_z$ for the $m=1$ case of $v_{A\phi}{\rm-PL-}v_{Az}{\rm-Cons}$, 
       when $v_{A\phi0}=0.926$ and $\alpha=-0.3$ are fixed, and $v_{Az0}$ is varied (see colored legends; note that 
       maximum growth rate decreases as $v_{Az0}$ increases). Only global eigenvalue solutions are shown in this figure.
       The $k_z$-axis is in log scale.}
         \label{fig_PLm1_Bzeff}
\end{figure*}

\item The $m=1$ instabilities are likely to be responsible for the dissipation of magnetic energy 
into thermal energy in the jet due to the formation of current sheets (as mentioned in the Introduction). 
This  may lead to an enhanced background pressure, which in turn 
can affect the instabilities. In order to 
study this, we add a constant background pressure (which in our case corresponds to a 
constant density) to our fiducial model. We test two cases: 
$\rho_{\rm cons}=\rho(r \rightarrow \infty)=10$ and $\rho_{\rm cons}=\rho(r \rightarrow \infty)=100$ 
(see equation \ref{eq_rhodimful_PL} and Table \ref{tab_results_PL1}).
Note that the above densities are  normalized using the 
same inner density $\rho_{\rm in}$ as our fiducial model, i.e. when 
$\rho_{\rm cons}=\rho(r \rightarrow \infty) = 0$, in order to maintain the same 
magnetic field profile for all three cases.
We find that an enhancement 
in the background pressure leads to a suppression of the maximum growth 
rates, which can be roughly approximated as $\propto \omega_{\rm fidu}\rho_{\rm cons}^{-0.5}$, 
where $\omega_{\rm fidu}$ is the maximum growth rate of the  fiducial case. 
However, the stability criteria remains unchanged, and  we observe the same range of unstable 
$k_z$, as well as the same $k_{z\rm max}$, for all the cases. We also verified that 
even in the presence of an enhanced background pressure, the system gets stabilized as $\alpha \rightarrow -1$.

%Figure \ref{fig_PLm1_rhoceff} shows the growth rates for the three different constant density 
%values as mentioned above (only global eigenvalue solutions are shown). 

\item So far, we have restricted our discussion to the limit $v_{Az} \ll v_{A\phi}$ 
(see Introduction and \S \ref{sec_jetmodel} for the motivation behind this) and, hence, 
to predominantly pressure-driven instabilities.
%This is because one of our primary goals in this work is to compare with and verify 
%the local calculations of B98, which has been carried out in this limit. 
However, if the vertical field becomes strong enough, it can have a stabilizing effect 
by counteracting the hoop stress due to the toroidal field. Also, as mentioned towards 
the beginning of the Introduction, a stronger vertical field is likely to introduce current-driven 
effects, which may induce a change in the behavior of the instabilities. 

Figure \ref{fig_PLm1_Bzeff} shows the growth rates of the $m=1$ modes as 
$v_{Az0}$ is increased with respect to $v_{A\phi0}$ 
for our fiducial $\alpha=-0.3$ case (only global eigenvalue solutions are shown). 
For $0.001v_{A\phi0} \leq v_{Az0} \leq 0.01v_{A\phi0}$, the maximum growth rate 
of the $m=1$ modes remain fairly similar and this is the limit up to which the local calculations are valid. 
As $v_{Az0}$ increases further, we see that the maximum growth rate starts decreasing, although it is still 
appreciable for $v_{Az0} = 0.1v_{A\phi0}$. There is a sharp 
drop in the maximum growth rate for $v_{Az0} \geq 0.5v_{A\phi0}$ and  complete  
stabilization for $v_{Az0}=v_{A\phi0}$, where the vertical field dominates over the 
toroidal field everywhere in the domain (since $v_{Az0}$ is a constant and $v_{A\phi0}$ decreases with radius). 
 Nevertheless, it has been pointed out that 
even a strong vertical field may lead to instability if $m$ is sufficiently high 
(see \citealt{2011PhRvE..84e6310B} for a discussion).
Note that when
$v_{Az0} = 0.1v_{A\phi0}$, the $k_z>0$ modes stabilize for $\alpha=-0.3$ in 
 contrast to the case shown in Figure \ref{fig_PLm1_alphaeff}  ($v_{Az0} = 0.005 v_{A\phi 0}$), where 
 they stabilize for
$\alpha=-0.4$. Thus, with an increase in the vertical field,
the $k_z>0$ modes are marginally stabilized at higher $\alpha$. Similarly, there is 
complete stabilization for {\it all} $k_z$-modes at a higher value of $\alpha$, 
%when the vertical field is strong enough, 
and $\alpha \leq -1$ reduces to a sufficient (and no longer necessary) stability criterion.
%Thus, the $\alpha=-1$ stability criterion, although never 
%violated, appears to be more closely applicable when $v_{Az} \ll v_{A\phi}$.

\end{enumerate}

\subsubsection{Non-axisymmetric $m=2$ modes}

The main findings of this subsection, discussed below, are summarized by Figures 
\ref{fig_PLm2_wir} and \ref{fig_PLm2_efuncs}:

\begin{enumerate}

\item Figure \ref{fig_PLm2_wir} is similar to Figures \ref{fig_PLm0_wir} and \ref{fig_PLm1_wir}, 
with the same color-scheme and symbolization, except that it represents the $m=2$ modes. 
Like the $m=1$ modes, the $m=2$ modes are not symmetric about the $k_z$-axis. 
There is also significant deviation from the local prediction.

First, we discuss the growth rates of the $m=2$ modes in the left panel of Figure \ref{fig_PLm2_wir}. 
We see that there is no instability for $k_z>0$, as shown by both the 
local and global solutions. This is true for a wide range of $\alpha$, namely $\alpha \leq 1$, 
as seen from equation (\ref{eq_kzunst}).
The local solution predicts two peaks with 
the same maximum growth rate $\omega_g|_{\rm max} \approx 0.35$. From the cyan lines 
we see that the first instability 
peak occurs at 
$k_z |_{\rm max} \approx -240$ and the second at $k_z |_{\rm max} \approx -500$, 
with the instability occupying a total range $-590 \lesssim k_z|_{\rm unst} \leq -151$. 
The local magnetic resonance corresponds to $k_z|_{\rm res} = -m v_{A\phi0}/v_{Az0} \approx -370$ 
at $r=1$, which again separates the two peaks in the local solution.

\begin{figure*}
%\hspace{-0.7cm}
%\captionsetup{width=1.1\textwidth}
\centering
  %\begin{tabular}{@{}cccc@{}}
    %\includegraphics[width=\textwidth]{growth_draft_all_onlyaxiBp0.3_ref.eps} %&
 %      \end{tabular}
 %\includegraphics[width=0.9\textwidth]{growth_draft_all_onlyaxiBp0.30_w4disp_l10.png}
\includegraphics[width=\columnwidth]{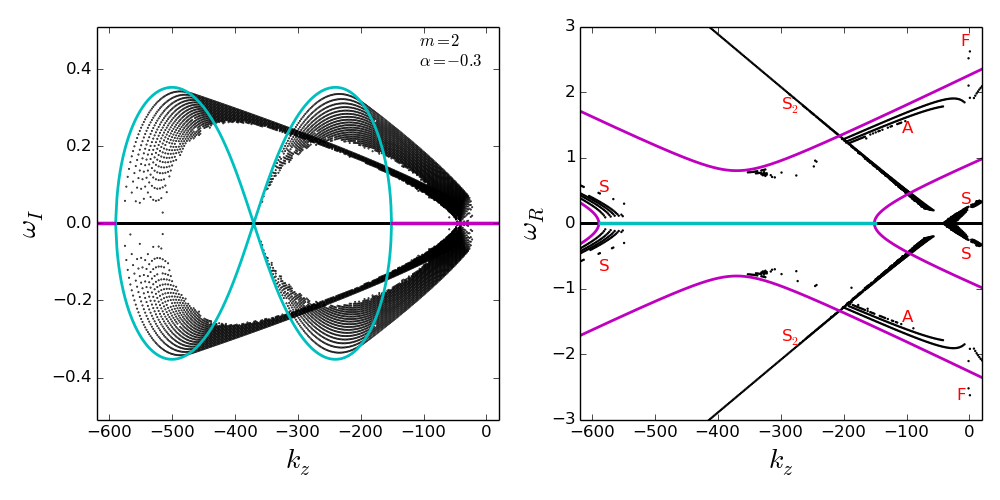}
       \caption{Global eigenvalue solutions for the fiducial $m=2$ case of $v_{A\phi}{\rm-PL-}v_{Az}{\rm-Cons}$, 
       with everything else same as Figures \ref{fig_PLm0_wir} and \ref{fig_PLm1_wir}.}
         \label{fig_PLm2_wir}
%\vspace*{\floatsep}
\end{figure*}

The global solution also exhibits two peaks in the growth rate, which match 
fairly well with the local prediction (the first peak has a slightly smaller $\omega_I \approx0.34$ 
than the second peak). 
%Similar to the $m=1$ case,
%$k_z \approx -370$ in the global solution also corresponds to the local magnetic resonance at $r = 1$.
However, the instability onsets at a much lower $|k_z|$ in the global solution, 
yielding a total range $-580 \lesssim k_z|_{\rm unst} \leq -20$. Also, both the 
instability peaks originate from $k_z \approx -20$, contrary to the local solution. 
%This in turn leads to interacting (slow) modes. 

Examining the right panel of Figure \ref{fig_PLm2_wir}, 
the global solutions have similar characteristics to the $m=1$ case (see right panel of Figure \ref{fig_PLm1_wir}). 
The most unstable modes have zero phase velocity like both the $m=0$ and $m=1$ cases, 
and are likely to be destabilized slow modes. There is significant deviation from the 
local solutions, with a new branch of slow modes (marked S$_2$) appearing and interacting with the 
Alfv\'{e}n modes, just like in the $m=1$ case.

\begin{figure*}
\centering
\includegraphics[width=\columnwidth]{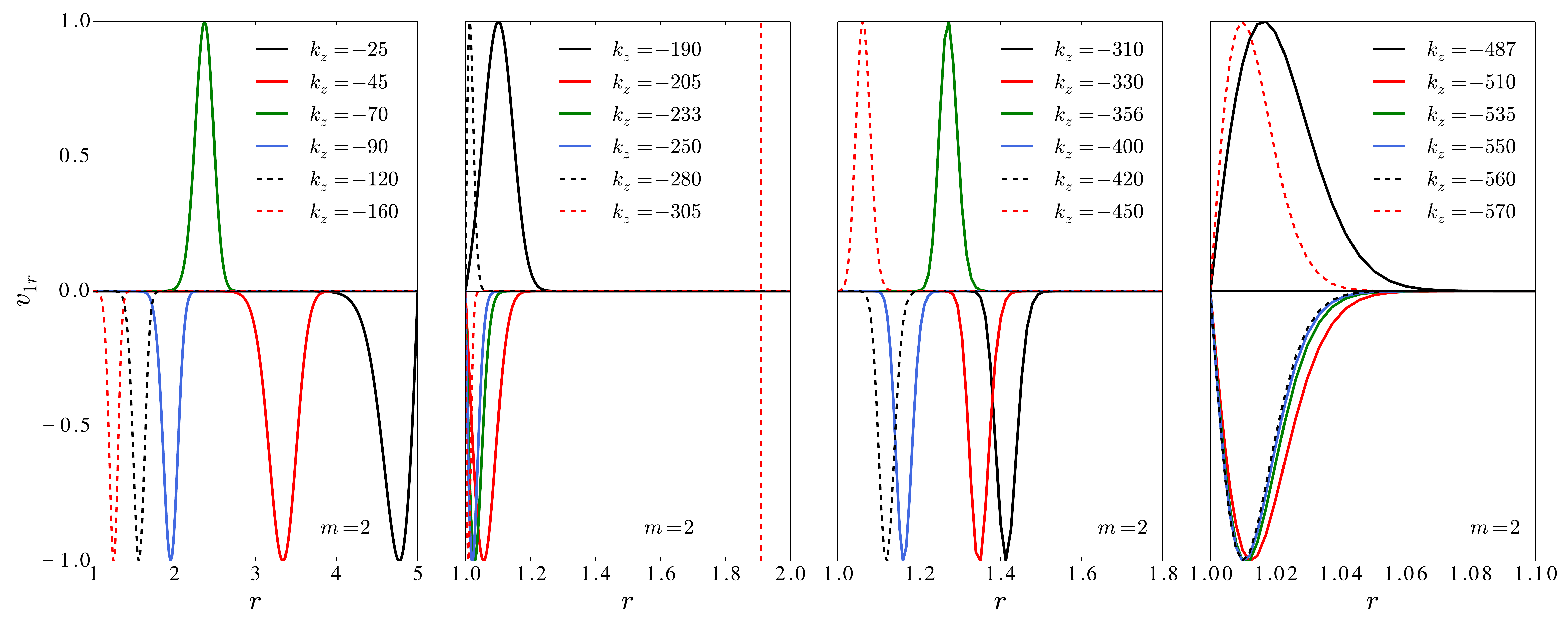}
       \caption{Normalized radial eigenfunction $v_{1r}$ for the most unstable modes as a function of radius $r$, for the case 
discussed in Figure \ref{fig_PLm2_wir} (see colored legends for $k_z$ values; 
note that $|k_z|$ increases from right to left in all panels). The red dashed vertical line, in the second panel 
from left, indicates the resonant radius $r_{\rm res}=1.91$ for $k_z=-305$ (red dashed curve), 
given by equation (\ref{eq_reson}).
Note that the radial axes has been plotted for different extents in the 
four panels for clarity, although the global problem is solved on a grid $r \in [1,5]$.}
         \label{fig_PLm2_efuncs}
\end{figure*}

\item Figure \ref{fig_PLm2_efuncs} shows 
the eigenfunctions $v_{1r}$ for a range of unstable $k_z$. 
%corresponding to the $m=2$ modes in Figure \ref{fig_PLm2_wir}. 
Note that we 
only plot the eigenfunction corresponding to the most unstable mode at a given 
$k_z$ to avoid confusion. 
Starting from left, the first panel shows the eigenfunctions for $-160 \leq k_z \leq -25$, the
second for $-305 \leq k_z \leq -190$, the third for $-450 \leq k_z \leq -310$ 
and the last for $-570 \leq k_z \leq -487$, with different radial axes for clarity, 
namely, $r \in [1,5]$, $r \in [1,2]$, $r \in [1,1.8]$ and $r \in [1,1.1]$ respectively. 
We now classify the instabilities according to Table \ref{tab_instabilities}.
%We can again classify the modes as {\it global}, {\it local} or {\it interacting}, as described in the context of 
%the $m=0$ and $m=1$ modes.

From the first panel of Figure \ref{fig_PLm2_efuncs}, we see that the 
eigenfunction for $k_z=-25$ peaks towards the outer boundary. We have verified that all modes 
in the range $-25<k_z<0$ exhibit a similar behavior and, hence, represent Global I instabilities.
These then transform to Global III instabilities in the range $-160< k_z <-25$, which is an 
indication of slow-mode-slow-mode interaction.
%, which seems to appear only for the $m=2$ case and is absent in the $m=1$ case. 
%Note that all the eigenfunctions have $\Delta r \ll 0.5$.
The eigenfunctions in the second panel clearly represent Local instabilities. The 
eigenfunctions in the third panel represent Global III instabilities, 
signifying the slow-mode-slow-mode interaction in the range $-450<k_z<-310$. 
The eigenfunctions in the last panel again represent Local instabilities.
 Note that there are {\it two} sets of interacting instabilities for $m=2$ 
(first and third panel of Figure \ref{fig_PLm2_efuncs}), 
unlike the $m=1$ case, which has only {\it one} (third panel of Figure \ref{fig_PLm1_efuncs}).
%, which is similar to that seen in the $m=1$ case (see third panel of Figure \ref{fig_PLm1_efuncs}). 

%Note that the global eigenvalue solution characterizing the most unstable Global instabilities 
%deviates significantly from the local prediction (see left panel of Figure \ref{fig_PLm2_wir}).

\item %A note here about the resonant character of the $m=2$ modes in this section. 
The $m=2$ 
modes satisfying the magnetic resonance condition (equation \ref{eq_reson}) 
are given by $k_z = -2v_{A\phi0}r^\alpha/v_{Az0}$. 
Thus, for our fiducial domain $1 \leq r \leq 5$, the modes with $-370 \lesssim k_z \lesssim -228$ 
are supposed to be resonant. 
%All the conclusions regarding the nature of the resonant modes discussed for the $m=1$ case still hold true.
%
The most unstable 
modes in the range $-370 \lesssim k_z \lesssim -310$ 
are interacting in nature and, hence, non-resonant (third panel of Figure \ref{fig_PLm2_efuncs}).
The modes in the range $-305 \lesssim k_z \lesssim -228$, in the second panel, 
are supposed to be resonant, but they are localized far away from the corresponding resonant 
radii.
%, which keeps shifting towards the inner boundary 
%as $|k_z|$ increases (just as in the $m=1$ case). For example, the eigenfunction for $k_z=-305$ peaks at $r \sim 1.01$, which 
%is well inside  $r_{\rm res}= (-k_z v_{Az0}/2v_{A\phi0})^{1/\alpha} \sim 1.91$ (cyan 
%dashed vertical line 
%in the second panel of Figure \ref{fig_PLm2_efuncs}). 
Thus, again magnetic resonance does not 
play any role in triggering the $m=2$ instabilities.
The inapplicability of 
Suydam's criterion remains the same as for the $m=1$ case, since the inequality (\ref{eq_suydam}) 
holds true for all $m \neq 0$.

%The $m=2$ resonant modes also represent Local instabilities, 
%whose eigenfunctions are indistinguishable from the Local instabilities 
%represented by the non-resonant modes (compare the 
%$-305 \lesssim k_z \lesssim -228$ modes in the second panel with those in the last panel 
%of Figure \ref{fig_PLm2_efuncs}). The eigenfunctions of the resonant modes 
%peak inside the corresponding resonant radius, which keeps shiftinging towards the inner boundary 
%as $k_z$ decreases. For example, the eigenfunction for $k_z=-280$ 
%(second panel of Figure \ref{fig_PLm2_efuncs}) peaks at $r=1.01$, which 
%is inside  $r= (-k_z v_{Az0}/2v_{A\phi0})^{1/\alpha} =2.54$. 

%However, just like the $m=1$ case, these 
%modes are not truly resonant as their eigenfunctions 
%do not peak at the corresponding radial positions. 
%For example, consider the eigenfunction for $k_z=-280$ in the 
%second panel of Figure \ref{fig_PLm2_efuncs}, which lies in 
%the aforementioned range of resonant modes. It clearly peaks at $r=1.01$, whereas according to the 
%resonant condition it is supposed to peak at $r= (-k_z v_{Az0}/2v_{A\phi0})^{1/\alpha} =2.54$. 
%As another example, consider the eigenfunction for $k_z=-356$ in the 
%third panel of Figure \ref{fig_PLm2_efuncs}. It actually peaks at $r=1.27$ instead of at the predicted value of $r=1.14$.

\end{enumerate}

The global solutions yield very similar {\it maximum} growth rates 
for the $m=0,1,2$ cases studied in this section (see left panels of Figures \ref{fig_PLm0_wir}, 
\ref{fig_PLm1_wir} and \ref{fig_PLm2_wir}). This is in accordance with the local prediction 
for pressure-driven instabilities (see equation \ref{eq_wmax}).
Although not shown here, we have verified that the $m=2$ modes also get completely stabilized 
as $\alpha \rightarrow -1$. Physically this condition means that the destabilizing thermal pressure gradient 
vanishes when $\alpha=-1$ (see equation \ref{mcb_radeqbm}). 
Interestingly, the background equilibrium becomes 
force-free but not pressureless when $\alpha=-1$. 
Thus, $\alpha \leq -1$ can be interpreted as a {\it global stability criterion} 
for {\it all} axisymmetric and non-axisymmetric modes, for a nearly constant vertical field such that 
$v_{Az} \ll v_{A\phi}$. 
%in the presence of a thermal pressure gradient (when $v_{Az} \ll v_{A\phi}$).
The reason the local B98 stability criterion holds true even for the 
global solutions is because the fastest growing modes are purely Local instabilities 
(occurring at large $|k_z|$) irrespective of $m$.

%which are governed only by the local magnetic field strength, and hence are likely 
%to be present 
%are insensitive to the details of the background magnetic field, and are hence

%[gen. comments on m=0,1,2]

\subsection{Results for a power-law toroidal field and varying vertical field: $v_{A\phi}{\rm-PL-}v_{Az}{\rm-Var}$}
\label{sec_Bphi_PLcons_Bz_var}

\begin{table*}
\large
\centering
%\vskip0.2cm
\renewcommand{\arraystretch}{1.5}
\caption{Summary of runs for case $v_{A\phi}{\rm-PL-}v_{Az}{\rm-Var}$ with $\alpha=-0.3$ 
presented in \S \ref{sec_Bphi_PLcons_Bz_var}.
%$v_{A\phi} = v_{A\phi0}r^\alpha$, $v_{Az} = v_{Az0}r^\alpha$, 
%$\frac{dP}{dr} = \frac{\epsilon}{1-\epsilon} \frac{d}{dr} \biggl(\frac{B_z^2}{8\pi}\biggr)$, 
%$\rho = r^{2\alpha}$, $\alpha=-0.3$.
}
 \begin{tabular}{|c|c|c|c|c|c|c|}
 %\hline
 %\hskip1cm General & &  \hskip2cm General  \\ \hline
 \hline
 $m$ & $\epsilon$ & $v_{A\phi0} = \sqrt{\frac{-2\alpha}{\epsilon(1+\alpha)}}$ & 
 $v_{Az0} = \sqrt{\frac{2(1-\epsilon)}{\epsilon}}$ & $v_{Az0}/v_{A\phi0}$ & $N_r$ & B.C.s imposed on $\rho(r)$ \\
  \hline 
$0$ & $0.999$  & $0.926$ & $0.045$ & $0.048$ & 200 &  $\rho (1) = 1$; $\rho (r \rightarrow \infty)=0$ \\
$0$ & $0.99$  & $0.930$ & $0.142$ & $0.153$ & 200 &  $\rho (1) = 1$; $\rho (r \rightarrow \infty)=0$ \\
$0$ & $0.9$  & $0.976$ & $0.471$  & $0.483$ & 200 &  $\rho (1) = 1$; $\rho (r \rightarrow \infty)=0$ \\
$0$ & $0.8$  & $1.035$  & $0.707$ & $0.683$ & 200 &  $\rho (1) = 1$; $\rho (r \rightarrow \infty)=0$ \\
\hline
 $1$ & $0.999$  & $0.926$ & $0.045$ & $0.048$ & 256 &  $\rho (1) = 1$; $\rho (r \rightarrow \infty)=0$ \\
$1$ & $0.99$  & $0.930$ & $0.142$ & $0.153$ & 256 &  $\rho (1) = 1$; $\rho (r \rightarrow \infty)=0$ \\
$1$ & $0.9$  & $0.976$ & $0.471$ & $0.483$ & 256 &  $\rho (1) = 1$; $\rho (r \rightarrow \infty)=0$ \\
$1$ & $0.8$  & $1.035$  & $0.707$ & $0.683$ & 256 &  $\rho (1) = 1$; $\rho (r \rightarrow \infty)=0$ \\
$1$ & $0.7$  & $1.106$ & $0.926$ & $0.837$ & 200 &  $\rho (1) = 1$; $\rho (r \rightarrow \infty)=0$ \\
%$1$ & $0.6$  & $1.195$ & $1.155$ & $0.966$ & 300 &  $\rho (1) = 1$; $\rho (r \rightarrow \infty)=0$ \\
$1$ & $0.4$  & $1.464$ & $1.732$ & $1.18$ & 300 &  $\rho (1) = 1$; $\rho (r \rightarrow \infty)=0$ \\
\hline
\end{tabular}
%\vspace{-4mm}
\label{tab_resultsPL_Bzvar}
\end{table*}

In this section, we discuss the effect of 
a {\it radially varying} vertical field
on the instabilities (for a power-law toroidal field). 
In order to do so, we make the following assumption about the relation between 
the thermal pressure gradient and the vertical field gradient:
\begin{equation}
\frac{d}{dr} \biggl(\frac{B_z^2}{8\pi}\biggr) = \biggl(\frac{1-\epsilon}{\epsilon} \biggr) \frac{dP}{dr} ~,
\label{eq_dpdreps}
\end{equation}
where $0<\epsilon<1$. Thus, a larger value of $\epsilon$ indicates a stronger thermal pressure gradient 
and vice versa. Note that equation (\ref{eq_dpdreps}) is an idealization and may 
not hold true inside a jet column. Nevertheless, this formalism 
helps us understand the basic contribution of a vertical field gradient towards the stability of the system. 

Using the above equation and assuming $v_{A\phi} = v_{A\phi0} r^\alpha$, with $\alpha<0$, we can 
integrate equation (\ref{eq_radeqbm}) to obtain the background density 
(or equivalently, thermal pressure) as a function of radius. We 
impose the boundary conditions (\ref{cond_rhoin1}) and (\ref{cond_rhoinf0}) to obtain:
\begin{equation}
\rho(r) = r^{2\alpha}  ~~{\rm and}~~ v_{Az}(r) = v_{Az0}r^{\alpha}  ~,
\label{eq_rho_pldbzdr}
 \end{equation}
such that 
\begin{equation}
v_{Az0} = \sqrt{\frac{2(1-\epsilon)}{\epsilon}}  ~~{\rm and}~~ 
v_{A\phi 0} = \sqrt{\frac{-2\alpha}{(1+\alpha)\epsilon}} ~.
\label{eq_vaz0vaphi0}
 \end{equation}
 Note that we also assumed $v_{Az} \rightarrow 0$ as $r \rightarrow \infty$.
 The plasma-beta for this case is a constant with respect to radius (for a given $\alpha$ and $\epsilon$) such that
 \begin{equation}
 \beta = \frac{2}{(v_{A\phi0}^2 + v_{Az0} ^2)} = \frac{\epsilon(1+\alpha)}{1 - \epsilon(1+\alpha)}~.
 \label{eq_beta_pl_dbzdr}
 \end{equation}
The magnetic pitch for this case can be written using equation (\ref{eq_pitch}) as
\begin{equation}
{\cal P} = r\frac{v_{Az0}}{v_{A\phi0}} = r \sqrt{\frac{(1-\epsilon)(1+\alpha)}{-\alpha}} ~,
\end{equation}
which always increases with radius.
We shall now study the effect of varying $\epsilon$ on the stability of the $m=0$ and $m=1$ modes for 
$\alpha=-0.3$. The cases are listed in Table \ref{tab_resultsPL_Bzvar}. 
We use $r \in [1,5]$ for all cases.

\begin{figure*}
%\hspace{-0.7cm}
%\captionsetup{width=1.1\textwidth}
\centering
  %\begin{tabular}{@{}cccc@{}}
    %\includegraphics[width=\textwidth]{growth_draft_all_onlyaxiBp0.3_ref.eps} %&
 %      \end{tabular}
 %\includegraphics[width=0.9\textwidth]{growth_draft_all_onlyaxiBp0.30_w4disp_l10.png}
\includegraphics[width=\columnwidth]{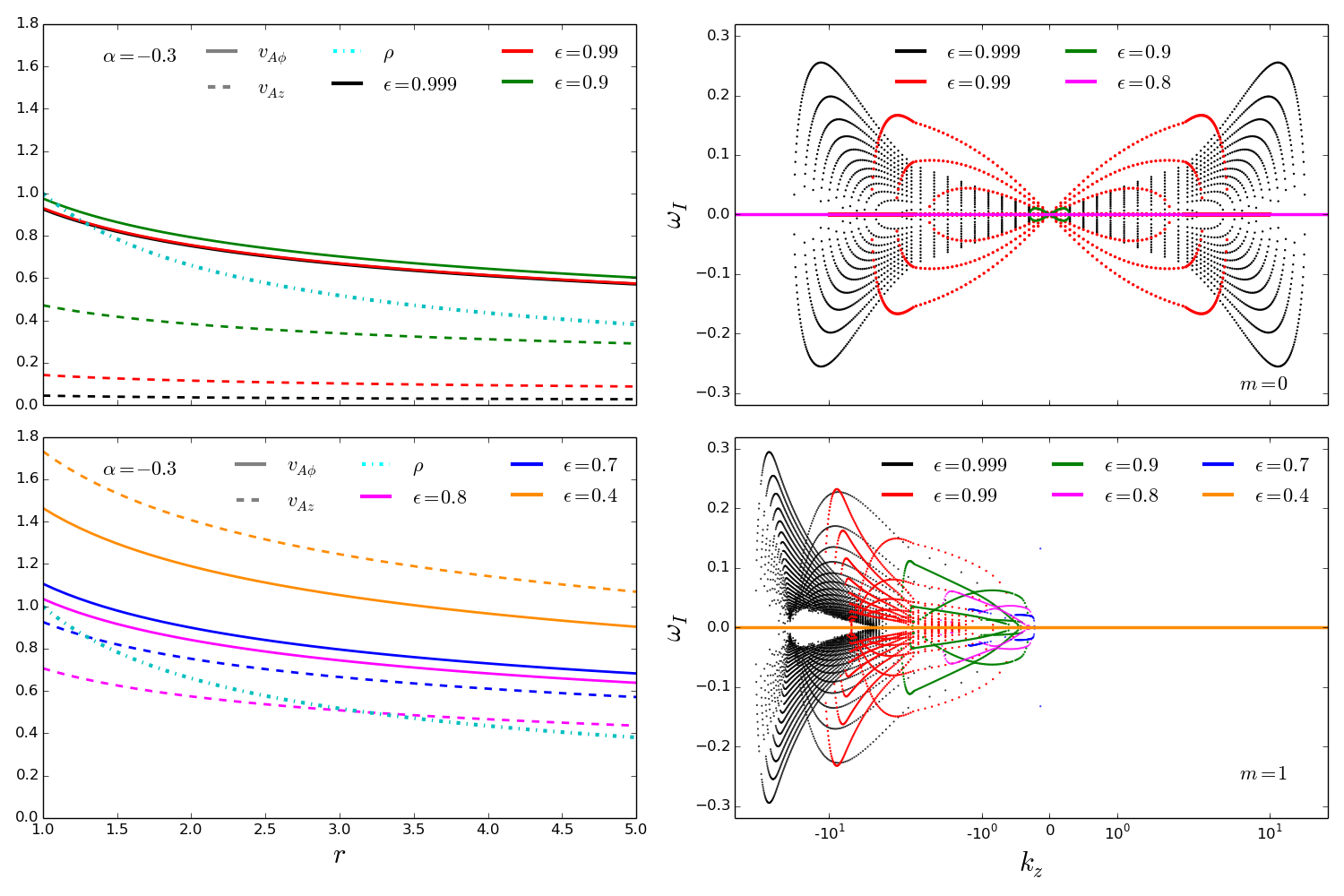}
       \caption{Global eigenvalue solutions for the case $v_{A\phi}{\rm-PL-}v_{Az}{\rm-Var}$ with $\alpha=-0.3$. 
       Left panels: Background radial profiles of $v_{A\phi}$ (solid lines), $v_{Az}$ 
       (dashed lines) and $\rho$ (cyan dot-dashed line) for different $\epsilon$ given by equation (\ref{eq_dpdreps}) 
    (see colored legends; note that $\epsilon$ increases from top to bottom for the dashed curves).
   Right panels: Growth rate $\omega_I$ as a function of $k_z$ for the $m=0$ (top) 
       and $m=1$ (bottom) cases, corresponding to the background profiles in the Left panels. Note that the maximum growth rate 
       as well as the maximum unstable $|k_z|$ decreases as $\epsilon$ decreases.}
         \label{fig_PL_varBz}
\end{figure*}

\subsubsection{Effect on $m=0$  and $m=1$ modes}

The left panels of Figure \ref{fig_PL_varBz} shows the profiles for 
the toroidal (solid) and vertical (dashed) fields as $\epsilon$ is varied, 
keeping $\alpha=-0.3$ fixed (top left: $\epsilon=0.999,0.99,0.9$; bottom left: $\epsilon=0.8,0.7,0.4$). 
The background density (and pressure) 
is independent of $\epsilon$ and is shown by the dot-dashed cyan line in the left panels 
(see equation \ref{eq_rho_pldbzdr}).
We see from the figure that for $\epsilon=0.999$, 
$v_{Az}$ is very subdominant compared to $v_{A\phi}$. 
Note that although $v_{A\phi}$ for $\epsilon=0.999$ and $\epsilon=0.99$ coincides (black 
and red solid lines in top left panel of Figure \ref{fig_PL_varBz}), 
$v_{Az}$ is different (black and red dashed lines in the same figure).
As $\epsilon$ decreases, 
$v_{Az}$ approaches $v_{A\phi}$ quite steeply, and finally dominates over $v_{A\phi}$ when $\epsilon=0.4$. 
Note that
since $v_{Az}$ follows the same power law as $v_{A\phi}$, the two profiles are identical when 
$v_{A\phi0}=v_{Az0}$. This happens when  
$\epsilon = (1+2\alpha)/(1+\alpha)$, giving $\epsilon=0.57$ for $\alpha=-0.3$ and 
$\beta=0.667$. Thus, for $\epsilon<0.57$, the vertical field is stronger than the toroidal field.

The top right panel of Figure \ref{fig_PL_varBz} shows the growth rates 
as a function of $k_z$ (from the 
global eigenvalue analysis) for the $m=0$ modes,  
when $\epsilon=0.999,0.99,0.9,0.8$. The bottom right panel shows the same for the 
$m=1$ modes,  when $\epsilon=0.999,0.99,0.9,0.8,0.7,0.4$. 
%Only global eigenvalue solutions are shown here as the effect of a radially varying field cannot by definition 
%be captured in a local calculation. 
We see that for both the axisymmetric 
and non-axisymmetric cases, as the vertical field  gradient and strength become stronger 
compared to the thermal pressure gradient (which is independent of $\epsilon$),
the growth rates decrease drastically.
%thus counteracting the destabilizing effect of the thermal pressure gradient. 
The $m=0$ modes appear to stabilize completely at a critical value, 
$\epsilon = \epsilon_{\rm cr} \sim 0.8$, contrary to the 
$m=1$ modes, which seem to persist under stronger vertical fields, and 
stabilize completely when $\epsilon_{\rm cr} \sim 0.4$.
The $k_z>0$ modes for the $m=1$ case, however,
%appear to be very sensitive to the presence of a vertical field gradient and become 
stabilize for $\epsilon=0.999$. This is because 
the vertical field strength for this case is
almost a factor of 10 higher at the inner boundary, i.e. $v_{Az0}=0.045$,
compared to the fiducial 
$v_{A\phi}{\rm-PL-}v_{Az}{\rm-Cons}$ case shown in Figure \ref{fig_PLm1_wir} with $v_{Az0}=0.005$ 
(although the toroidal field profile remains similar for both the cases).
This behavior is very similar to the stabilizing effect of a strong (constant) vertical 
field shown in Figure \ref{fig_PLm1_Bzeff}, and is likely due to  
current-driven effects.  

Note that as the vertical field gradient, and hence strength, gets stronger, 
the instabilities shift to smaller $|k_z|$ and become 
increasingly Global in nature (and thus can no longer be captured in a local analysis). This 
makes determining a global stability criterion for this case more difficult, as the B98 stability criterion is 
not applicable in this limit (unlike in \S \ref{sec_Bphi_PLcons_Bz_cons}). Absolute stability 
is attained for a critical 
combination of the magnetic pressure gradients and the thermal pressure gradient in the jet, 
as indicated by $\epsilon_{\rm cr}$ (for a given $m$ and $\alpha$).
In the case of $m \neq 0$ modes, this is similar to the statement of Suydam's stability criterion, which however 
is not directly applicable in this case because of its local nature (see \S \ref{sec_suydam}).
Alternatively, it has been suggested by \cite{2016MNRAS.456.1739B} that the internal kink instability 
(equivalent to the $m=1$ modes in this work) 
stabilizes when there is equipartition between the thermal and magnetic pressures.
However,  that seems to be neither a necessary nor a sufficient condition from our analysis.
For instance, equipartition is attained when (in our units) 
\begin{equation}
\rho  = \frac{1}{2}( v_{A\phi}^2 + v_{Az}^2  ) 
\label{eq_rhoequip}
\end{equation}
or, equivalently, $\beta=1$ in equation (\ref{eq_beta_pl_dbzdr}),
%which, using equations (\ref{eq_rho_pldbzdr}) and (\ref{eq_vaz0vaphi0}), 
which yields $\epsilon =0.71$ for $\alpha=-0.3$. However, from Figure \ref{fig_PL_varBz} we see that 
the $m=0$ modes attain stability for $\epsilon > 0.71$, when the thermal pressure dominates 
over the total magnetic pressure; while the $m=1$ modes attain stability for $\epsilon <0.71$, 
when the total magnetic pressure dominates over the thermal pressure. Although $\epsilon$ 
is just a parametrization, our results indicate that pressure balance alone 
may not guarantee stability, and one needs to consider the interplay between the 
magnetic and thermal pressure {\it gradients} as well 
(also, see \S \ref{sec_disc} below).

%The $m=1$ modes are  completely stabilized when 
%$v_{Az} \geq v_{A\phi}$, i.e., for $\epsilon \leq 0.57$. 

%Thus, the presence of a significant thermal pressure gradient is destabilizing (i.e., larger $\epsilon$).

\begin{figure*}
%\hspace{-0.7cm}
%\captionsetup{width=1.1\textwidth}
\centering
  %\begin{tabular}{@{}cccc@{}}
    %\includegraphics[width=\textwidth]{growth_draft_all_onlyaxiBp0.3_ref.eps} %&
 %      \end{tabular}
 %\includegraphics[width=0.9\textwidth]{growth_draft_all_onlyaxiBp0.30_w4disp_l10.png}
\includegraphics[width=0.7\columnwidth]{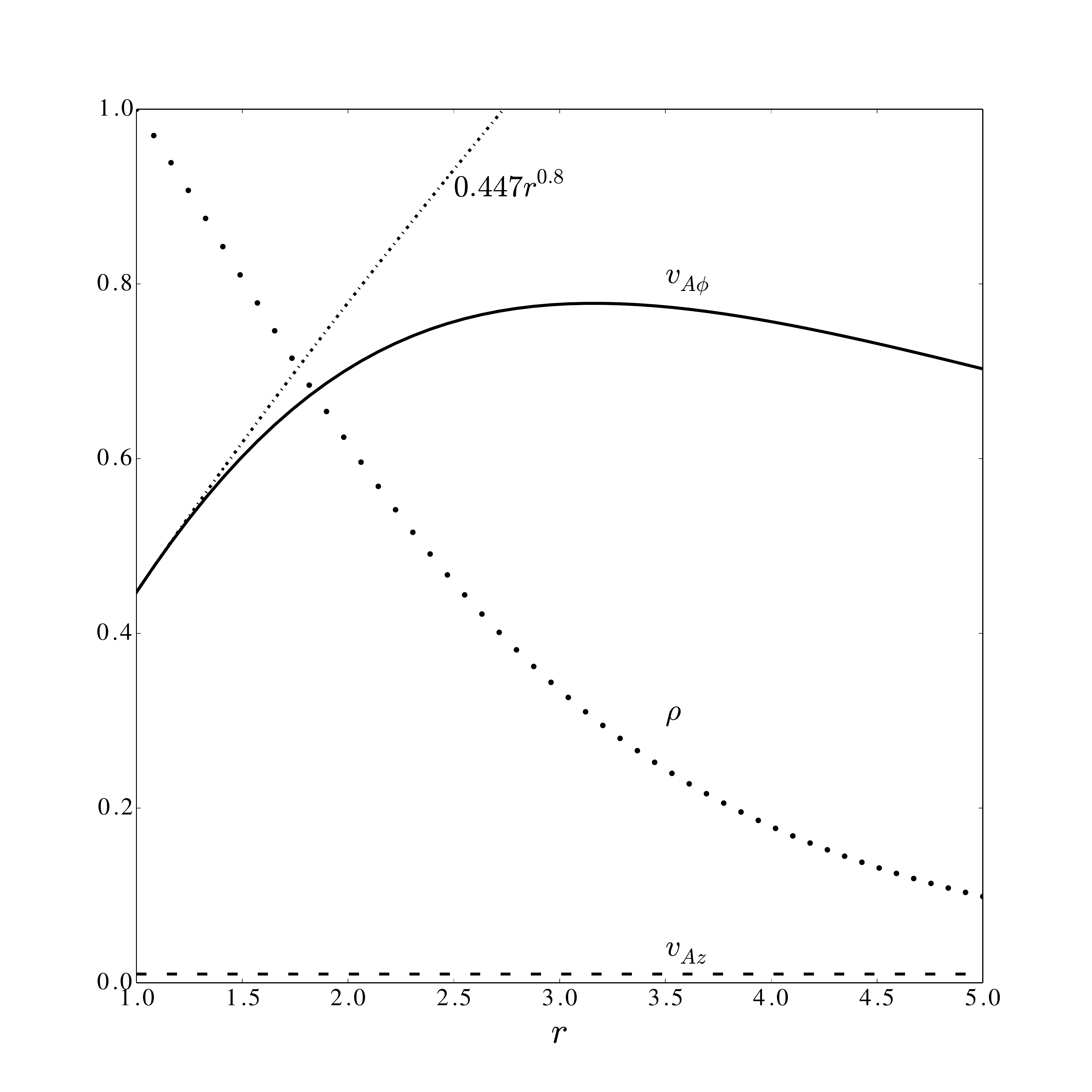}
       \caption{Background radial profiles for the fiducial $v_{A\phi}{\rm-Gen-}v_{Az}{\rm-Cons}$ case, 
       with $v_{A\phi}$ (solid line) given by equation 
       (\ref{eq_bphigen}), $\rho$ (dotted line) given by equation (\ref{eq_rho_nfw_fidu}), $v_{A\phi0}=0.447$, 
       $a_1=0.1$ and $v_{Az} = {\rm constant} =0.01$ (dashed line). The dot-dashed line shows a power-law profile for 
       $v_{A\phi} = 0.447r^{0.8}$.}
         \label{fig_NFW_fiduprof}
\end{figure*}

\subsection{Results for a generic toroidal field and constant vertical field: $v_{A\phi}{\rm-Gen-}v_{Az}{\rm-Cons}$}
\label{sec_Bphigen_Bzcons}

In this section, we consider a more generic profile for the background toroidal field such that 
\begin{equation}
v_{A\phi} = v_{A\phi 0}\frac{(1+a_1)r}{1+ a_1r^2} ~,
\label{eq_bphigen}
\end{equation}
where $a_1$ is a constant parameter governing the radial location of the peak of $v_{A\phi}$. 
The above equation mimics the magnetic field profiles obtained by 
\cite{1992ApJ...397..187B} for their Crab Nebula models (see also \citealt{2000A&A...355..818A,2012MNRAS.422.1436O}).

\begin{table*}
\large
\centering
%\vskip0.2cm
\renewcommand{\arraystretch}{1.5}
\caption{Summary of runs for case $v_{A\phi}{\rm-Gen-}v_{Az}{\rm-Cons}$ presented in \S \ref{sec_Bphigen_Bzcons}. 
%$v_{A\phi} = \frac{v_{A\phi 0}(1+a_1)r}{1+ a_1r^2}$, $v_{Az}=v_{Az0}$, 
%$\rho = 1 + \frac{v_{A\phi0}^2(1+a_1)^2}{2 a_1} \biggl( \frac{1}{(1 + a_1 r^2)^2}  - \frac{1}{(1+a_1)^2} \biggr)$.
}
 \begin{tabular}{|c|c|c|c|c|c|c|c|}
 %\hline
 %\hskip1cm General & &  \hskip2cm General  \\ \hline
 \hline
 $m$ &  $a_1$ & $v_{A\phi0}$ & $v_{A\phi0}(1+a_1)$ & $v_{Az0}$ & $N_r$ & B.C.s imposed on $\rho(r)$ & Other notes \\
  \hline 
$0,1,2$ &  $0.1$ & $0.447$ & $0.492$ & $0.01$ & $200$ &  $\rho (1) = 1$; $\rho (r \rightarrow \infty)=0$ & Fiducial model; 
$v_{A\phi0}= \sqrt{2a_1}$ \\
\hline
%$1$ & $0.537$ & $0.2$ & $0.447$ &$0.01$ & $200$ &  $\rho (1) = 1$ & $b_0= \sqrt{0.2}(1+a_0)$; $\rho \geq 0$ $ \forall$ $r$  \\
$1$ &  $0.2$ & $0.447$ & $0.537$ & $0.01$ & $200$ &  $\rho (1) = 1$ & $\rho \geq 0$ $ \forall$ $r$  \\
$1$ & $0.3$ & $0.447$ & $0.581$ & $0.01$ & $200$ &  $\rho (1) = 1$ &  $\rho \geq 0$ $ \forall$ $r$ \\
$1$ &  $0.4$ & $0.447$ & $0.626$ & $0.01$ & $200$ &  $\rho (1) = 1$ &  $\rho \geq 0$ $ \forall$ $r$ \\
$1$ &  $0.5$ & $0.447$ & $0.671$ & $0.01$ & $200$ &  $\rho (1) = 1$ &  $\rho \geq 0$ $ \forall$ $r$  \\
%$1$ & $1.0$ & $0.447$ & $0.894$ &  $0.01$ & $200$ &  $\rho (1) = 1$ & $\rho \geq 0$ $ \forall$ $r$  \\
\hline
\end{tabular}
%\vspace{-4mm}
\label{tab_results_NFW_cons}
\end{table*}

\begin{figure*}
%\hspace{-0.7cm}
%\captionsetup{width=1.1\textwidth}
\centering
  %\begin{tabular}{@{}cccc@{}}
    %\includegraphics[width=\textwidth]{growth_draft_all_onlyaxiBp0.3_ref.eps} %&
 %      \end{tabular}
 %\includegraphics[width=0.9\textwidth]{growth_draft_all_onlyaxiBp0.30_w4disp_l10.png}
\includegraphics[width=\columnwidth]{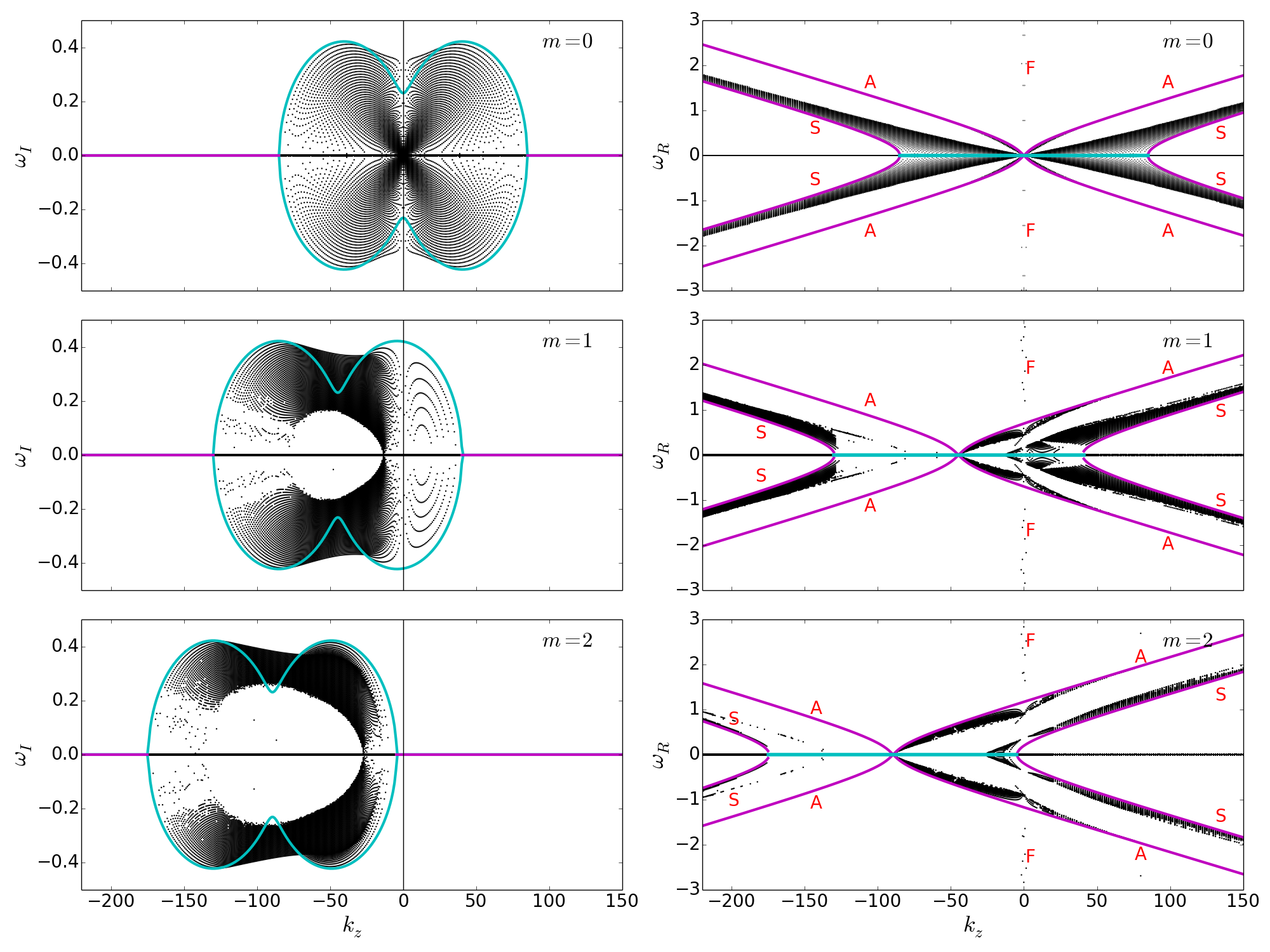}
       \caption{Global eigenvalue solutions for the fiducial $v_{A\phi}{\rm-Gen-}v_{Az}{\rm-Cons}$ case, 
       with $v_{A\phi0}=0.447$, $a_1=0.1$ and $v_{Az} =0.01$. The top, middle and bottom panels 
       represent the $m=0,1$ and $2$ cases respectively. Left panels: Growth rate $\omega_I$ as a function of $k_z$. 
        Right panels: The real part of the eigenvalue $\omega_R$ as a function of $k_z$. The
cyan (unstable modes with $\omega_I>0$ and stable modes with $\omega_I < 0$ in the left panels, 
and the corresponding $\omega_R$ in the right panels) 
and magenta (stable modes with $\omega_I=0$ in the left panels and the corresponding $\omega_R$ 
in the right panels) lines represent 
the solutions of the local dispersion relation, equation (11), for $v_{A\phi} = 0.447r^{0.8}$, $v_{Az}=0.01$ and $l^2/k_z^2=0$.        
        The global problem is solved on a radial grid $r \in [1, 5]$ with resolution $N_r = 200$.
All symbols have the same meaning as in Figure \ref{fig_PLm0_wir}.}
         \label{fig_NFW_m012}
\end{figure*}

\begin{figure*}
%\hspace{-0.7cm}
%\captionsetup{width=1.1\textwidth}
\centering
\includegraphics[width=\columnwidth]{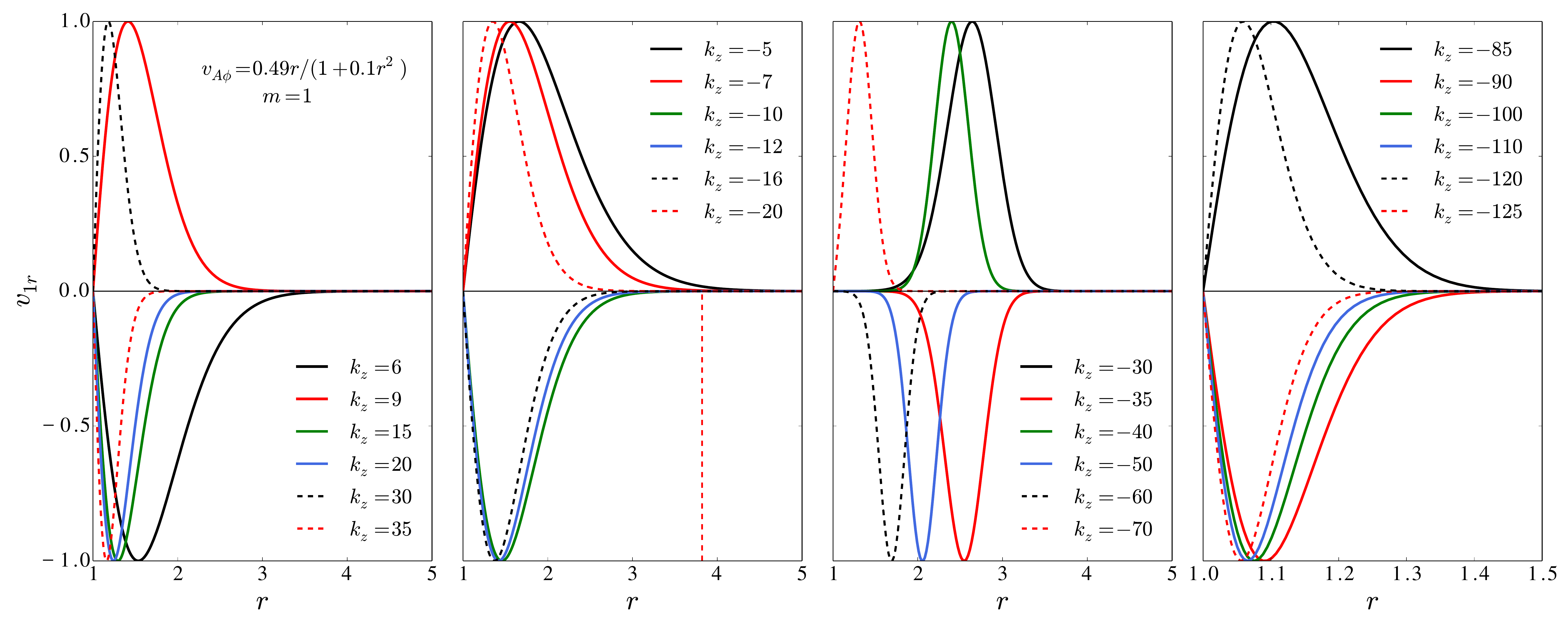}
       \caption{Normalized radial eigenfunction $v_{1r}$ for the most unstable modes 
       as a function of radius $r$, for the $m=1$ case shown in 
Figure \ref{fig_NFW_m012} (see colored legends for $k_z$ values; 
note that $|k_z|$ increases from right to left in all panels). The red dashed vertical line, in the second panel from left, 
indicates the resonant radius $r_{\rm res} = 3.82$ for $k_z = -20$ (red dashed curve), 
given by equation (\ref{eq_reson}). Note that the radial axis in the last panel has been zoomed close 
to the inner boundary for clarity.}
         \label{fig_NFW_efuncs}   
         \end{figure*}

\begin{figure*}
%\hspace{-0.7cm}
%\captionsetup{width=1.1\textwidth}
\centering
  %\begin{tabular}{@{}cccc@{}}
    %\includegraphics[width=\textwidth]{growth_draft_all_onlyaxiBp0.3_ref.eps} %&
 %      \end{tabular}
 %\includegraphics[width=0.9\textwidth]{growth_draft_all_onlyaxiBp0.30_w4disp_l10.png}
\includegraphics[width=\columnwidth]{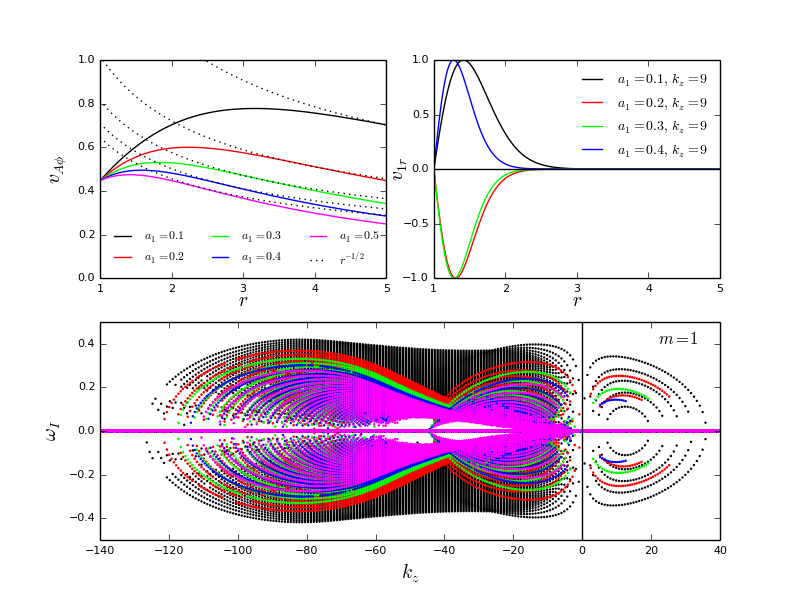}
       \caption{Global eigenvalue solutions for the $m=1$ case of $v_{A\phi}{\rm-Gen-}v_{Az}{\rm-Cons}$, when 
       $v_{A\phi0}= 0.447$ and $v_{Az}=0.01$ are fixed, and $a_1$ is varied. 
       Top left panel: Background radial profiles of $v_{A\phi}$ given by equation 
       (\ref{eq_bphigen}), with different $a_1$ (see colored solid curves; note that $a_1$ increases from top to bottom). 
       The dotted lines represent the $r^{-1/2}$ profiles tangential to each $v_{A\phi}$.
       Bottom panel: Growth rate as a function $k_z$ for different $a_1$ (the color scheme is same
       as in top left panel; as $a_1$ increases, the maximum growth rate decreases).
       Top right panel: Normalized radial eigenfunction $v_{1r}$ as a function of $r$, 
       corresponding to the most unstable mode at $k_z=9$ 
       for different $a_1$ (note that $a_1$ increases from right to left).}
         \label{fig_NFW_diffa1}
\end{figure*}

We consider a constant vertical field $v_{Az}=v_{Az0}$, and obtain the density by integrating
equation (\ref{eq_radeqbm}) subject to the boundary condition (\ref{cond_rhoin1}), 
\begin{equation}
\rho(r) = 1 + \frac{v_{A\phi0}^2(1+a_1)^2}{2 a_1} \biggl( \frac{1}{(1 + a_1 r^2)^2}  - \frac{1}{(1+a_1)^2} \biggr) ~.
\end{equation}
For our fiducial model, we also impose the boundary condition (\ref{cond_rhoinf0}) to obtain
\begin{equation}
\rho(r) =  \frac{(1 +a_1)^2}{(1 + a_1 r^2)^2} ~~{\rm and}~~ v_{A\phi0}= \sqrt{2a_1} ~.
\label{eq_rho_nfw_fidu}
\end{equation}
The plasma-beta for the fiducial model, on assuming $v_{Az0} \ll v_{A\phi}$, then becomes
\begin{equation}
\beta \approx \frac{1}{a_1 r^2} ~.
\end{equation}
The pitch for this magnetic configuration follows from equation (\ref{eq_pitch}):
\begin{equation}
{\cal P} = \frac{v_{Az0}(1+a_1r^2)}{v_{A\phi0}(1+a_1)} ~,
\end{equation}
which increases with radius.
Figure \ref{fig_NFW_fiduprof} shows the different background profiles as a function of $r$ for our 
fiducial case having $a_1=0.1$, $v_{A\phi0} = 0.447$ and $v_{Az0}=0.01 \ll v_{A\phi0}$. 
%We study different cases by varying $a_1$ but keeping $v_{A\phi0}$ same as the fiducial case. 
We use $r \in [1,5]$ for all the cases 
studied in this section, which are listed in Table \ref{tab_results_NFW_cons}.
Figure \ref{fig_NFW_m012} shows $\omega_I$ (left panels) and $\omega_R$ (right panels) as a function of 
$k_z$ for the $m=0$ (top), $m=1$ (middle) and $m=2$ (bottom) modes described by the fiducial model 
shown in Figure \ref{fig_NFW_fiduprof}. The global eigenvalue solutions are in black, while the cyan 
($\omega_I \neq 0$) and magenta ($\omega_I=0$) 
lines indicate local solutions derived using the B98 dispersion relation with $l^2/k_z^2=0$ (equation \ref{mcb_dispeq_dimles}) 
for a power-law toroidal field $v_{A\phi} = 0.447r^{0.8}$. 
%(i.e., $\alpha=0.8$ and $v_{A\phi0}=0.447$, same as the generic profile). 
Note that there is no fixed $\alpha$ corresponding to the generic profile, however, a slope of $\alpha=0.8$ is
tangent to the generic profile at the inner radius (see Figure \ref{fig_NFW_fiduprof}). 
%This well approximates the 
%toroidal field close to the inner boundary (see Figure \ref{fig_NFW_fiduprof}), which gives us the 
This gives us the opportunity to directly compare the solutions of the generic and
power-law toroidal fields. Note that all colors and symbols have the same meaning as in Figure \ref{fig_PLm0_wir}.
The main findings of this section, discussed below, are summarized by Figures 
\ref{fig_NFW_m012}, \ref{fig_NFW_efuncs} and \ref{fig_NFW_diffa1}:

\begin{enumerate}

\item The growth rates of the $m=0$ modes are symmetric about 
the $k_z$-axis, as expected. The global solution matches well with the 
local solution (representing the most unstable modes) 
for $|k_z| \gtrsim 30$. 
The local power law solution predicts a maximum growth rate $\omega_g|_{\rm max} \approx 0.42$ 
(see equation \ref{eq_wmax}), the corresponding 
vertical wavenumbers $k_z |_{\rm max} \approx \pm 40$ 
(see equation \ref{eq_kzmax}) and the range of instability $-85 \lesssim k_z|_{\rm unst}\lesssim 85$ 
(see relation \ref{eq_kzunst}). All the corresponding global quantities are in very good agreement but with 
slightly lower values. 
However, the growth rates are somewhat higher than the local prediction 
for $|k_z| < 30$. Note that unlike the $m=0$ case for $\alpha=-0.3$ shown in Figure \ref{fig_PLm0_wir}, 
the most unstable modes do not have $\omega_I \rightarrow 0$ as $k_z \rightarrow 0$. This can be explained by the 
fact that the present solution is roughly approximated by a $\alpha>0$ power-law toroidal field, 
which tends to give higher growth rates because $v_{A\phi}$ is increasing with radius. Thus, the 
solutions behave differently from our fiducial $\alpha<0$ case (see \S \ref{sec_PL_m0}).

\item The envelope of the global solutions for both the $m=1$ and $m=2$ cases, i.e. the most unstable modes, 
seem to match reasonably well with the local prediction, with some differences.
%The global solutions for the $m=1$ case differ significantly from the 
%corresponding local prediction. 
%The local solutions for $m=1$ have a double peaked structure such that
%$\omega_g|_{\rm max} \approx 0.42$ at $k_z |_{\rm max} \approx -4$ and $-85$, 
%and $-130 \lesssim k_z|_{\rm unst}\lesssim 40$. 
The local solutions for both $m=1$ and $2$ have a double peaked structure.
This is again due to the occurrence of local magnetic resonance 
at $r=1$ for $m=1$ with $k_z|_{\rm res} = - v_{A\phi0}/v_{Az0} \approx -45$; 
and for $m=2$ with $k_z|_{\rm res} = - 2v_{A\phi0}/v_{Az0} \approx -89$. 
However, 
$\omega_I \neq 0$ at $k_z |_{\rm res}$, unlike the behavior of 
the local resonance discussed in \S \ref{sec_PL_m1}.
The two peaks in the global solution are barely distinguishable due to slow-mode-slow-mode interaction, 
leading to a band of higher growth rate about $k_z \approx -45$ for $m=1$ and about 
$k_z \approx -89$ for $m=2$. 
For $m=1$, the local solution slightly overestimates the 
growth rates for the $k_z>0$ modes, as well as that of the first peak.
For $m=2$, the $k_z>0$ modes are completely stabilized, 
however the $k_z<0$ modes start slightly further away from the axis than the local prediction. 
In both cases, the local solution predicts the 
total range of instability extremely well.

%We discuss the eigenfunctions for this case in point (vi) below.

%The global solutions for the $m=2$ case seem to match more closely with the local prediction than the 
%$m=1$ case. The local solutions again have a double peaked structure with
%$\omega_g|_{\rm max} \approx 0.42$ at $k_z |_{\rm max} \approx -49$ and $-130$, 
%and $-174 \lesssim k_z|_{\rm unst}\lesssim -5$. The local magnetic resonance occurs at 
%$k_z|_{\rm res} = - 2v_{A\phi0}/v_{Az0} \approx -89$ when $r=1$.

%The maximum growth rates and the instability range of the global solution is in accord with the local prediction. 
%However, like the $m=1$ case, the two peaks of maximum growth rate are indistinguishable in the $m=2$ global 
%solution, which are connected by a band of interacting modes. The $k_z>0$ modes are completely stabilized, 
%however the $k_z<0$ modes start slightly further away from the axis than the local prediction.

\item  
%recall that the cyan line indicates the local solution of $\omega_R$ corresponding
 %to the most unstable modes of the left panels, while the magenta lines correspond 
 %to the stable modes. 
From the right panels of Figure \ref{fig_NFW_m012}, we find that
  the most unstable modes are destabilized slow modes that are 
  purely growing, for all $m=0,1$ and $2$ cases. In fact, the local predictions 
  match quite well with the global solutions across all $k_z$ and $m$, with 
  no new branches of slow-Alfv\'{e}n interaction appearing in the $\omega_R-k_z$ plane, 
  unlike those observed in \S \ref{sec_Bphi_PLcons_Bz_cons}.

\item In Figure \ref{fig_NFW_efuncs}, we illustrate the normalized, $v_{1r}$ eigenfunctions corresponding to the most 
unstable $m=1$ modes. The $k_z>0$ modes 
shown in the first panel (from left) are Global II stabilities, which tend to get more localized to the 
inner boundary as $k_z$ increases. The unstable modes in the range 
$-20 \leq k_z \leq -5$ shown in the second panel 
also represent Global II instabilities, however with a larger $\Delta r$ than the first panel. 
The third panel showing $-70 \leq k_z \leq -30$ represent Global III instabilities, i.e. 
the interacting slow modes. Finally, the last panel showing 
$-125<k_z<-85$ represents Local instabilities.

Note that the $m=1$ modes satisfying the magnetic resonance condition for the present field profile 
are given by $k_z = -(v_{A\phi0}/v_{Az0})(1+a_1)/(1+a_1r^2)$. Thus, for $r \in [1,5]$, 
the modes in $-45 \lesssim k_z \lesssim -14$ are supposed to be resonant. 
Again, the most unstable modes in $-45 \lesssim k_z \lesssim -30$ are interacting and non-resonant (third panel),
while the eigenfunctions of the modes in $-20 \lesssim k_z \lesssim -14$ peak far away from the resonant radius 
(second panel).
For example, the eigenfunction for $k_z=-20$ peaks at $r \sim 1.35$, 
which is inside the resonant surface for this mode at 
$r_{\rm res}= \sqrt{(1/a_1)(-1 - v_{A\phi0}(1+a_1)/(k_z v_{Az0}))} \sim 3.82$ (vertical red dashed line in 
second panel of Figure \ref{fig_NFW_efuncs}).
This is similar to the conclusion drawn regarding the unstable modes discussed in \S \ref{sec_PL_m1}.
%As $k_z$ decreases, the corresponding resonant surface shifts toward the inner boundary.

As to the eigenfunctions of the most unstable $m=0$ modes, they represent 
Global I instabilities for $|k_z| <15$, Global II instabilities for 
$15 <|k_z| <30$ and Local instabilities for $|k_z| >30$ ---  
which corroborates the deviation of the global solution 
from the local prediction at small $|k_z|$. The eigenfunctions of the most unstable $m=2$ modes represent 
Global II instabilities for $-40<k_z<-10$, Local instabilities for $-70<k_z<-40$, 
Global III instabilities for $-115<k_z<-70$ (interacting modes), 
and again Local instabilities for $-170<k_z<-115$.

\item Finally, in Figure \ref{fig_NFW_diffa1}, we compare the global 
solutions for the $m=1$ modes corresponding to multiple generic profiles having different $a_1$. 
The toroidal magnetic fields (colored solid lines) 
are shown as a function of radius in the top left panel. 
They are constructed such that they all have the same $v_{A\phi0}=0.447$ and 
$v_{Az} = constant =0.01$ as the fiducial model (see Table \ref{tab_results_NFW_cons} for details). 
We note that as $a_1$ increases, the maximum magnetic field strength decreases as well as 
shifts towards the inner boundary.
The dotted lines are $\propto r^{-0.5}$, which become tangential to the different 
magnetic field curves at different radii. 
The aim here is to establish a correlation 
with the solutions of the power-law profiles discussed in \S \ref{sec_Bphi_PLcons_Bz_cons}.

The bottom panel shows the growth rate as a function of $k_z$ corresponding to the 
cases in the top left panel (global eigenvalue solutions only). 
We observe that as $a_1$ increases, the growth rates of the unstable modes
decrease and the range of instability shrinks. Furthermore, the two growth rate peaks (for $k_z<0$) become 
more distinguishable and the modes start resembling the 
left panel of Figure \ref{fig_PLm1_wir}. 
%, exhibiting similar behavior about the magnetic resonance at $k_z|_{\rm res} \approx -45$ for $r=1$ (see point (iii) above).
Interestingly, 
the $k_z>0$ modes stabilize for $a_1>0.4$, which we now try to explain using the top two panels 
(note that some of the incomplete $k_z>0$ curves are due to inadequate resolution). 

In \S \ref{sec_PL_m1} and Figure \ref{fig_PLm1_alphaeff},
we saw that the $k_z>0$ modes are completely stabilized when $\alpha=-0.5$ as per B98 prediction. 
Keeping that in mind, in the top left panel of 
Figure \ref{fig_NFW_diffa1} we observe that the radius at which the $r^{-0.5}$ curve is tangential to 
the respective field profiles shifts radially inwards as $a_1$ increases. Now, the most unstable $k_z>0$
mode for each $a_1$ roughly corresponds to $k_z=9$ (see bottom panel). The top right panel 
shows the $v_{1r}$ eigenfunctions for $k_z=9$ when $a_1=0.1,0.2,0.3$ and $0.4$. 
All four eigenfunctions represent Global II instabilities, which implies that they 
are influenced by the background toroidal field profile. More 
interestingly, the eigenfunctions also shift radially inwards as $a_1$ increases. Thus, we can conclude that 
if the toroidal field transitions to 
a $r^{-0.5}$  at sufficiently small radii to influence the eigenfunctions of the $k_z>0$ 
unstable modes, they get stabilized. For the present scenario, this happens when $a_1=0.5$. 
Similarly, we can conclude that there will be {\it complete} stabilization (i.e. of all $k_z$ modes) 
with a further increase in $a_1$, as the background toroidal field transitions to a $r^{-1}$ profile. 
Thus, we recover the B98 stability criterion even for a more complex magnetic background, 
precisely due to the local 
nature of the fastest growing internal (pressure-driven) instabilities (provided $v_{Az} \ll v_{A\phi}$).

\end{enumerate}

%\begin{figure*}
%\centering
%\includegraphics[width=\columnwidth]{NFW_compPL_posalp.png}
 %      \caption{}
  %       \label{1}
%\end{figure*}   

%We shall now study the effect of varying $\epsilon$ on the stability of the $m=0$ and $m=1$ modes for 
%$\alpha=-0.3$. The cases are listed in Table \ref{tab_resultsPL_Bzvar}. 
%We use $r \in [1,5]$ for all the cases.

%\subsection{Comparison with 3D jet simulations}

\section{Discussion}
\label{sec_disc}

In this section, we discuss some of our results in the context of recent numerical simulations. 

\begin{figure*}
\centering
\includegraphics[width=\columnwidth]{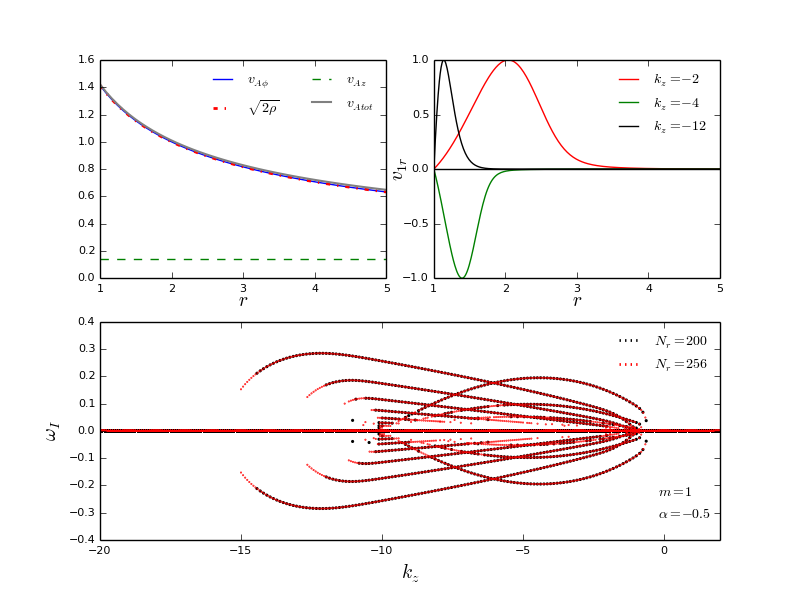}
       \caption{Testing the role of equipartition between thermal and magnetic pressures 
       on the stability of $m=1$ modes. Top left panel: Background radial profiles of a 
       toroidal Alfv\'{e}n velocity $v_{A\phi}=1.41r^{-0.5}$ (blue solid line), constant vertical 
  Alfv\'{e}n velocity $v_{Az}=0.14$ (green dashed line), total Alfv\'{e}n velocity 
  $v_{Atot} = \sqrt{v_{A\phi}^2 + v_{Az}^2}$ (solid gray line) and $\sqrt{2\rho}$, which is a 
  quantity proportional to the thermal pressure (red dot-dashed line), so defined to
  compare with the Alfv\'{e}n velocities. Bottom panel: Growth rate as a function of $k_z$ 
       from the global eigenvalue analysis of the case shown in the top left panel 
       for $m=1$, and radial grid resolutions $N_r=200$ (black) and $N_r=256$ (red).
       Top right panel: Normalized radial eigenfunction $v_{1r}$ as a function of $r$,
 corresponding to the most unstable modes at $k_z = -2,-4,-12$ (see colored legends for $k_z$ values; 
 note that $|k_z|$ increases from right to left).}
         \label{fig_PLm1_equip}
\end{figure*}

%We now compare some of our results with that of recent jet simulations. 
\citet{2016MNRAS.456.1739B} carried out global, 3D relativistic MHD simulations of relativistic, 
Poynting-flux-dominated jets. They simulated 2 kinds of jets --- (1) headless jets that propagate in 
a vacuum funnel and experience no direct influence from the ambient medium; 
and (2) headed jets that propagate through 
an ambient medium. They  found that the ambient medium is responsible for collimating
the headed jets, and that this compression effect triggers the internal kink 
instability (and, also, the external kink instability that we do not discuss here). 
Since these instabilities lead to the dissipation of magnetic energy, 
the simulations showed a build-up of background thermal pressure, which can no longer be ignored 
under the force-free approximation. 
%It is at this regime within the jet that we can compare our findings.
One of the main findings of \citet{2016MNRAS.456.1739B} was that the internal kink instabilities 
saturated once 
equipartition was reached between the thermal and magnetic energies, about which we commented briefly 
in \S \ref{sec_Bphi_PLcons_Bz_var} above. In Figure \ref{fig_PLm1_equip}, we 
reconstruct a situation similar (but not identical) to that in Fig. 14 of \citet{2016MNRAS.456.1739B}, 
to analyze the conditions under which the internal kink 
stabilizes.

In their Fig. 14, \citet{2016MNRAS.456.1739B} showed the 
azimuthally averaged, magnetic field and thermal pressure profiles inside their 3D simulated jet, 
at a height where there was no more dissipation due to the internal kink instability.
The top left panel of our Figure \ref{fig_PLm1_equip} shows a power-law profile for 
the toroidal field (solid blue line) 
such that $\alpha=-0.5$ and $v_{A\phi0}=1.41$ (see equation \ref{eq_rhobg_PL}), in order to 
match the scaling found by \citet{2016MNRAS.456.1739B}. The 
ratio of the vertical to toroidal field strength is roughly similar to that in Fig. 14 of 
\citet{2016MNRAS.456.1739B}, such that we choose $v_{Az}=0.14$ (dashed green line). 
Note that our vertical field is constant, unlike \citet{2016MNRAS.456.1739B} (we shall comment 
on this later). 
%and a (sub-dominant) constant vertical field 
In order to compare the thermal pressure with 
the magnetic fields, we plot the equivalent quantity given by equation 
(\ref{eq_rhoequip}), i.e. $\sqrt{2\rho}$ (red dot-dashed line). 
We see that for this case, the thermal 
pressure is in exact equipartition with the toroidal field, and very nearly in equipartition 
with the total magnetic field, 
$v_{Atot} = \sqrt{v_{A\phi}^2 + v_{Az}^2}$ (gray solid line). Since our computational domain 
is located away from the jet boundaries and core, this plot can be compared with the 
toroidally dominated regime in Fig. 14 of \citet{2016MNRAS.456.1739B}, i.e. the outer 
jet region from $\sim 20-60 R_L$, where $R_L$ is the light cylinder radius in their work 
(we have kept the same color and line-style scheme in the top left panel of our figure as Fig. 14 of 
 \cite{2016MNRAS.456.1739B} for easier comparison). 
%but this does not affect the point we aim to convey.  
The main point of this exercise is to demonstrate 
that despite equipartition, as well as the $r^{-0.5}$ scaling 
for $v_{A\phi}$, we find unstable $m=1$ modes in our analysis 
(with a non-negligible maximum growth rate $\sim 0.3$), 
unlike \cite{2016MNRAS.456.1739B}. 
Note that the bottom panel of Figure \ref{fig_PLm1_equip} presents solutions for two resolutions 
$N_r=200$ and $N_r=256$. This is to show that the apparent gaps in the 
solutions towards higher $|k_z|$ are not real, and they close as the resolution increases.
The eigenfunctions 
of the most unstable modes at $k_z=-2,-4,-12$ are plotted in the top right panel. We shall now discuss 
a few possible explanations for this apparent discrepancy.

First, we check if the the above eigenfunctions can be resolved in the set-up employed by 
\citet{2016MNRAS.456.1739B}, who use a logarithmic radial grid with a uniform (dimensionless) 
spacing $r_{grid}= (r_{br}/r_{in})^{1/N}-1$, where $N$ is their radial grid resolution, and
$r_{in}=0.8R_L$ and $r_{br}=800R_L$ are
the inner and outer radial boundaries (see their Table 1). They use two resolutions, 
namely, $N=128$ and $N=256$, which correspond to $r_{grid} \sim 0.06$ and $0.03$ respectively. 
The dimensionless quantity describing our eigenfunctions,  
equivalent to $r_{grid}$, is therefore $\Delta r/r_1$ (see top left panel of Figure  \ref{fig_PLm0_efuncs}).
%as demonstrated for $k_z=-2$ in the top right panel of Figure \ref{fig_PLm1_equip}. 
We see that as $|k_z|$ increases, the eigenfunctions in the top right panel of Figure \ref{fig_PLm1_equip} 
become more localized towards the inner 
boundary having $\Delta r/r_1 \sim 0.7, 0.4, 0.24$ for $k_z=-2,-4,-12$ respectively. We next 
define a quality factor ${\cal Q} = (\Delta r/r_1)/r_{grid}$, which 
determines how well resolved an eigenfunction is. Although Fig. 14 of \citet{2016MNRAS.456.1739B} 
is for the $N=128$ case, we compare our eigenfunctions with their higher resolution case $N=256$. 
We find that for $r_{grid}\sim0.03$ the eigenfunctions 
appear to be well-resolved, having ${\cal Q} =23,13,8$ for $k_z=-2,-4,-12$ respectively. Thus, in 
principle, the unstable modes we find in our Figure 15 should have led to the internal kink instability in the 3D jets 
simulated by \citet{2016MNRAS.456.1739B}.
%, unless the current resolution is not adequate. 
%However, could there be more fundamental reason for this difference?

 Next, we take a closer look at Fig. 14 of \citet{2016MNRAS.456.1739B} and point out its differences with our 
Figure \ref{fig_PLm1_equip}. 
We observe from Fig. 14 of \citet{2016MNRAS.456.1739B} 
that although the very inner core of the jet (which we exclude from our computation) 
is indeed in equipartition, the outer jet region, $\sim 20-60 R_L$, is not. In fact, the 
thermal pressure dominates over the magnetic pressure in the outer region. Interestingly, the 
thermal pressure profile is nearly flat in this region, which in turn implies a very weak thermal 
gradient.
This is probably a consequence of the rapid thermal pressure build up due to magnetic energy dissipation.
 Moreover, the vertical field has a non-negligible gradient, and its strength, although 
less than the toroidal field, cannot be considered as sub-dominant (hence the B98 stability 
criterion cannot be applied in this case).
%This is clearly unlike the case depicted in our Figure \ref{fig_PLm1_equip}. 
 %Together with the presence of  that can be easily counteracted by the combined vertical and toroidal field gradients. 
Thus, we believe that the main cause for stability seen in  \citet{2016MNRAS.456.1739B} 
is the negligible thermal pressure gradient in the jet region above the dissipation zone, which is easily stabilized by 
the combined vertical and toroidal field gradients present therein (as discussed in \S \ref{sec_Bphi_PLcons_Bz_var}).

%absence of a destabilizing thermal pressure gradient in the outer toroidally dominated region of the jet. 

%We also mention here that 
Our stability analysis is also consistent with the 
``core-envelope'' jet model of \citet{1999MNRAS.308.1069K}, 
which was used to carry out periodic box, 3D simulations by 
\citet{2015MNRAS.452.1089P}. This model consists of a Z-pinched\footnote{A Z-pinch is a 
cylindrical geometry in plasma physics, where an electric current flows in the $z$-direction 
that generates a magnetic field in the $\phi$-direction.} inner jet core, where the toroidal field balances the 
plasma pressure, and a force-free, purely toroidal outer envelope such that $B_\phi \propto 1/r$.
Note that the main objective of \citet{2015MNRAS.452.1089P} was to investigate the stability of 
their jet to external, large-scale instabilities by exploring different pressure distributions for the ambient medium. 
They found that for a critical ambient pressure ($P_{\rm amb} \propto z^{-2}$) the overall jet becomes 
stable to the external instabilities, the outer envelope is featureless, and the internal instabilities are confined to a narrow 
jet core. 
We point out that irrespective of the nature of the ambient medium and the jet core (both of which 
are excluded from our analysis), the outer envelope would be stable to 
all internal instabilities according to the B98 stability criterion (see \S \ref{sec_b98_stabcrit}), 
which has been extensively verified in this work (see \S \ref{sec_Bphi_PLcons_Bz_cons}). 
Thus, if an unstable jet core additionally develops, the magnetic energy would be dissipated to heat 
by the internal instabilities 
therein. This would lead to a bright 
core or spine, which would in turn be surrounded by a stable and darker envelope or sheath, 
giving rise to the so-called ``spine-sheath'' structure, commonly 
invoked to explain observations of jets in 
AGN \citep[e.g.][]{2015MNRAS.450.2824M} and blazars \citep[e.g.][]{2016MNRAS.457.1352S}.

In an earlier study, \citet{2012ApJ...757...16M} also carried out 3D relativistic, 
periodic box, MHD simulations of rotating jets, 
having a poloidal field dominated core and $B_\phi \rightarrow 1/r$ in the envelope. They considered 
a low gas density and pressure background decreasing with radius such that their jets were still 
force-free. They found that the jets were helically distorted due to the 
development of the internal kink instability, whose growth rate depended strongly on the gradient of the poloidal field. 
Their findings were in accord with the linear stability analyses mentioned in 
our Introduction \citep{1996MNRAS.281....1I,1999MNRAS.308.1006L}, which in turn emphasizes the relevance 
of linear studies. \citet{2012ApJ...757...16M} further found that multiple unstable 
wavelengths were excited during the non-linear evolution, 
which coupled to give rise to an external kink instability that disrupted 
the entire jet. 
%Interestingly, their study also demonstrates an important result in support of our theory. 
We consider Fig. 8 of \citet{2012ApJ...757...16M}, which shows the time evolution of 2D density 
slices in the 
$xz$-plane of the jet ($z$ being the jet-axis), for different angular velocities. If we interpret the 
loop-like, high density structures in this figure
(denoted by orange and red) as locations 
where the internal kink instability is triggered, we observe that they are localized in the interior of the jet (for 
all the panels). However, there is no instability seen towards the edge of the jet where $B_\phi \propto 1/r$, 
which is again compatible with the B98 stability criterion.

Finally, we discuss the 3D resistive MHD simulations by \citet{2016MNRAS.462.2970S} 
to highlight the significance of thermal pressure gradients. 
These were performed 
to study reconnection in magnetically confined cylindrical plasma columns. 
\citet{2016MNRAS.462.2970S} modeled both a force-free equilibrium, as well as one with 
a thermal pressure gradient. For the latter case, they found small-scale, finger-like structures 
in the current density triggered by the pressure-driven, internal kink instability 
(see their Fig. 7). These 
features were shown to cause efficient magnetic dissipation via reconnection 
of multiple fragmented current sheets, which in turn reinforces 
the premise of our work.
%and with further growth, to complete disruption of the plasma column in the non-linear regime. 
For the force-free case, \citet{2016MNRAS.462.2970S} observed a bending 
of the plasma column due to the current-driven kink instability as well as the
formation of plume-like structures (see their Fig. 9). 
More interestingly, they found that the currents through several plumes were perpendicular 
to the magnetic field, indicative of a secondary pressure-driven instability (as the
 Lorentz force is required to be balanced by a non-zero thermal pressure gradient in these regions; 
 see their Fig. 10), which led to formation of mixed modes.
Fragmentation of current sheets and an enhanced magnetic dissipation rate were also noted 
in this case. This finding leads to the important inference that initially 
Poynting flux dominated jets
can also develop thermal pressure gradients, which in turn are essential to facilitate 
magnetic reconnection and dissipation.

\section{Summary and conclusions}
\label{sec_conc}

Below we summarize the main findings of our analysis:

\begin{itemize}

\item We have carried out a global eigenvalue analysis of the linearized, ideal 
MHD equations, in order to study the stability of cylindrical jets to both axisymmetric
and non-axisymmetric perturbations. 
Our calculations have been performed in the jet fluid frame, 
and we have ignored velocity shear (both at the surface as well as in the interior), 
rotation and relativistic effects for simplicity. We have focused on 
characterizing the small-scale, internal instabilities that are confined to the interiors of jets, away from 
the effects of both the jet core and boundaries. 
%Unlike external instabilities, which may disrupt the entire jet, the internal instabilities 
%are important as they can lead to small-scale dissipation of magnetic to thermal energy, and can explain why 
%astrophysical jets radiate so efficiently. 
We have further analyzed the importance of a thermal pressure gradient 
for triggering instabilities in a region of the jet dominated by a toroidal 
magnetic field (with a sub-dominant vertical field), as is likely to occur 
at some distance from the jet source. This gives rise to 
predominantly pressure-driven instabilities, which are often overlooked in the literature 
compared to their purely current-driven counterparts arising in force-free jets.  
Since the exact transverse structure of magnetic fields inside jets is not known, we have adopted 
three different background configurations for our stability analysis, 
yielding the results summarized below.

%lead to the formation of current sheets. These are probable sites of magnetic reconnection, 
%which leads to small-scale dissipation of magnetic to thermal energy, causing jets to radiate.

Before we proceed with the summary, 
note that we have further 
classified the {\it internal} instabilities as Local and Global, based 
on the properties of their radial eigenfunctions (see Table \ref{tab_instabilities} for details).
%emphasizing that they are still {\it internal} instabilities. 
The Local instabilities have narrow eigenfunctions and peak 
very close to the inner radial boundary. Hence, these  agree very well with local 
(i.e. radially localized) analytic calculations. The Global instabilities fall into 
several different types, 
%(including the ones arising from the interaction between different $k_z$ modes when $m \neq 0$), 
any of which can be captured accurately only in a global (radially extended) framework as 
they are influenced by the background radial gradients and curvature (again, not 
to be confused with large-scale external instabilities). Note that all the instabilities found 
in this work are purely growing slow magnetosonic modes.

\item First, we have considered a 
power-law profile for the toroidal magnetic field, such that 
$v_{A\phi} \propto r^\alpha$, and a constant vertical magnetic field $v_{Az}$ 
(\S \ref{sec_Bphi_PLcons_Bz_cons}; case $v_{A\phi}{\rm-PL-}v_{Az}{\rm-Cons}$). 
We do a parameter scan for different values of $\alpha$, and find internal instabilities for a 
range of vertical wavenumbers $k_z$ (both $k_z>0$ and $k_z<0$; the sign indicates 
the direction of the helical twist with respect to the magnetic field if $m>0$), for $m=0,1,2$ 
(see Figures \ref{fig_PLm0_wir}, \ref{fig_PLm1_wir} and \ref{fig_PLm2_wir}). 
The maximum growth rate (which is nearly independent of $m$, as is characteristic of 
pressure-driven instabilities) for a given $\alpha$, matches very well with the predictions of
B98, who performed a {\it local}, linear stability analysis of a similar jet model. 
Interestingly, 
we find that in a region where $v_{Az} \ll v_{A\phi}$, {\it all} ($m,k_z$) modes are stabilized for $\alpha \leq -1$
(see e.g. Figures \ref{fig_PLm0_most} and \ref{fig_PLm1_alphaeff}). 
This {\it global} stability criterion is in excellent agreement with the local criterion of 
B98. 
This criterion is better understood by looking at the radial equilibrium equation (\ref{mcb_radeqbm}), 
which shows that the thermal pressure gradient (or equivalently, density gradient in our case) 
has a negative sign in general (for $\alpha> -1$), and is oppositely directed to the restoring magnetic tension force. 
However, when $\alpha=-1$, the thermal pressure gradient vanishes, the toroidal 
magnetic pressure gradient is balanced by its own tension force and the system becomes stable 
irrespective of $m$ and $k_z$. 
Note that for $m=1$, the $k_z>0$ modes get stabilized at a shallower toroidal field profile, i.e.
$\alpha \lesssim -0.5$, which is also in direct corroboration with B98. In a region where $v_{Az}$ 
is comparable to $v_{A\phi}$, however, the modes get stabilized even for $\alpha>-1$ (Figure \ref{fig_PLm1_Bzeff}).

The $m=0$ unstable modes can be classified as Local instabilities across almost 
their entire $k_z$-range (Figure \ref{fig_PLm0_efuncs}). 
The global solutions for the $m \neq 0$ cases, however,  
exhibit a mixture of Local and 
Global instabilities. The Global instabilities tend to occur at 
smaller $|k_z|$, and also include those arising due to 
the interaction between distinct slow modes occupying the same range of $k_z$ 
(Figures \ref{fig_PLm1_efuncs} and \ref{fig_PLm2_efuncs}).
Nevertheless, the fastest growing  modes for all $m$ are
Local instabilities, which  occur predominantly at large negative values of $k_z$.
This is the reason for the excellent agreement between B98's
local analysis and our global solutions. 
The important consequence of this finding is as follows:
in a toroidally dominated region of a jet with 
a negative thermal pressure gradient strong enough to trigger instability (and very weak vertical field), 
{\it internal instabilities can occupy  
any local patch of the jet, irrespective of the exact configuration of the magnetic field in the jet interior}.
According to B98, the growth timescales of these instabilities are inversely proportional to the 
toroidal field strength; thus, they grow fast enough for a substantial region of the jet to become 
internally unstable. This in turn can serve as a likely site for dissipation of magnetic energy to heat 
  (and non-thermally accelerated particles) via current sheet formation.
%These neighboring patches put together can lead to a substantial region of the jet being 
%unstable, which  
The Global instabilities, which are subject to the details of the radial 
profiles, would further enhance this effect.
%B98 showed that these instabilities have growth times of the order 
%of Alfv\'{e}n crossing times, which are 
%short enough in a toroidally dominated jet region to develop non-linearly.
Although internal instabilities can develop close to the base of
Poynting-flux-dominated jets \citep[e.g.][]{2016MNRAS.456.1739B}, we 
propose that a thermal pressure gradient is crucial 
%to trigger these instabilities farther from the source, and 
to sustain them farther from the source, where much of the
particle acceleration leading to radiation takes place.

%We have also found that a stronger vertical field stabilizes these instabilities.

%as $|k_z|$ increases, there is increasing agreement between the two solutions. 

%global solutions is that the $m=0$ modes are essentially Local instabilities 
%across most of the unstable $k_z$ range.
%{\it This verifies the B98 stability criterion for the $m=0$ modes in the global scenario}.
%
%There are, however, some differences between the local and global calculations, especially for $m \neq 0$. 
%For example, the growth rates and the extent of instability for the $k_z>0$ modes (when $m=1$), as well as for the 
%modes with small negative values of $k_z$ (both $m=1$ and $2$), differ significantly from the corresponding local solutions. 

%However, some minor differences with the local prediction arise 
%when one looks closely at the radial eigenfunctions of the most unstable modes. 
%A closer look at the radial eigenfunctions of the most unstable modes reveals some 
%more differences between the local and global solutions.

%This finding has a very important consequence, namely, that 
%any local patch of the jet that satisfies the aforementioned instability criteria will house 
%instabilities irrespective of the exact nature of the magnetic field therein.

%Our results are overall in excellent agreement with that of  with some minor differences. 

\item Next, we have considered the same toroidal field profile as above but with a 
radially varying vertical magnetic field (\S \ref{sec_Bphi_PLcons_Bz_var}; case $v_{A\phi}{\rm-PL-}v_{Az}{\rm-Var}$). 
In order to self-consistently determine the equilibrium, we 
have assumed the vertical magnetic pressure to be proportional to the thermal pressure, 
characterized by the parameter $\epsilon$ 
(see equation \ref{eq_dpdreps}) --- the smaller the value of $\epsilon$, 
the stronger the vertical magnetic field. As the vertical field gradient 
increases (for a given $\alpha$), there is a consequent increase in the vertical field strength relative to the toroidal 
magnetic field, although the thermal pressure gradient remains unchanged 
(as it is independent of $\epsilon$ in this formalism). A decrease in $\epsilon$ induces stabilization of both 
the $m=0$ and $m=1$ modes for a given $\alpha$ (Figure \ref{fig_PL_varBz}). {\it Complete} stabilization 
for the $m=0$ and $m=1$ modes are, however, attained at different values of $\epsilon$, indicating 
that the pinch and kink modes react differently to the background gradients.
Moreover, our results indicate that stability in such a situation is 
{\it not} determined solely by whether equipartition is achieved 
between the thermal and magnetic pressures (see the last point below).

%ensues from the complex interplay between the thermal and magnetic pressure {\it gradients}, and 
%{\it not} solely due to the equipartition between the thermal and magnetic pressures.

\item Finally, we have considered a more complex toroidal field profile (see equation \ref{eq_bphigen}) 
with a constant (weak) vertical magnetic field 
(\S \ref{sec_Bphigen_Bzcons}; case $v_{A\phi}{\rm-Gen-}v_{Az}{\rm-Cons}$). Interestingly, we find that 
for all the $m=0,1$ and $2$ cases, the growth rate and the $k_z$-range of instability of the most unstable 
global solutions show excellent agreement 
with the corresponding local solutions obtained for a power-law toroidal field with $\alpha=0.8$ 
 and the same $v_{A\phi0}$ 
(see Figures \ref{fig_NFW_fiduprof} and \ref{fig_NFW_m012}). 
We have also established a direct correlation between the stability criterion for the 
complex toroidal field configuration and 
the simple power-law configuration. We find that as the background toroidal field becomes 
proportional to $r^{-1/2}$, i.e. $\alpha=-0.5$, the $k_z>0$ modes get completely 
stabilized for $m=1$, as in the power-law case (see Figure \ref{fig_NFW_diffa1}). 
A similar line of reasoning allows us to conclude that the B98 stability criterion for 
all $k_z$-modes (i.e. $\alpha \leq -1$)  would also be valid be in this case.
%, $\alpha \leq -1$, 
%would determine stability of {\it all} $k_z$ modes.
%The instabilities for this case can again be classified as Local 
%and Global, as mentioned above, and the conclusions are very similar to those drawn for the power-law toroidal 
%field case. 
The $m=0$ modes again represent mostly Local instabilities, whereas the $m \neq 0$ modes 
exhibit both Global and Local instabilities, occupying different $k_z$. As $k_z$ assumes increasingly 
negative values, however,  the instabilities become more and more localized close to the 
inner radial boundary for the non-axisymmetric case (see Figure \ref{fig_NFW_efuncs}). 
This further validates our claim that highly localized internal instabilities 
are triggered as long as the B98 instability criterion is satisfied locally inside the jet, irrespective 
of the background magnetic field profile. 
We have also established 
that magnetic resonance does not play any role in triggering the $m \neq 0$ instabilities 
(irrespective of the toroidal field profiles), as 
their eigenfunctions peak far away from the corresponding resonant radii (see \S \ref{sec_resonance}, 
and second panels of Figures \ref{fig_PLm1_efuncs}, \ref{fig_PLm2_efuncs} and \ref{fig_NFW_efuncs}).

\item %We have also compared our results with those of recent 3D numerical simulations 
%by \citet[fully global;][]{2016MNRAS.456.1739B} and 
%\citet[periodic box;][]{2012ApJ...757...16M,2015MNRAS.452.1089P,2016MNRAS.462.2970S},

We have also compared our results with those of recent 3D numerical simulations 
by \citet{2012ApJ...757...16M,2015MNRAS.452.1089P,2016MNRAS.456.1739B} and \citet{2016MNRAS.462.2970S}.
%with the following central question in mind ---
%once triggered, what causes the internal instabilities to saturate? 
We conclude that thermal pressure gradients are indeed key to incite internal instabilities, 
which are fundamental for causing magnetic energy dissipation via reconnection of current sheets.
With regards to the question of what causes internal instabilities to saturate, 
our results indicate two possible answers depending 
upon the region of the jet in question: (1) in a toroidally 
dominated region with a very weak vertical field, complete stability for 
all $m$-modes is attained when the toroidal field behaves as $v_{A\phi} \propto 1/r$, 
as proposed by B98. This supports the ``spine-sheath'' model of jets, with a 
Z-pinched jet core and an envelope stable to internal instabilities, observed in several 
AGN jets;
%This quenches the destabilizing thermal pressure gradient.
(2) in a region with appreciable toroidal field but a non-negligible vertical field strength and 
radial gradient, 
stability ultimately ensues from the complex interplay between the destabilizing 
thermal pressure gradient and the stabilizing magnetic pressure gradients, 
or equivalently, magnetic shear. 
Contrary to the claim by \cite{2016MNRAS.456.1739B}, we conclude that equipartition 
alone may not  guarantee stability (see Figure \ref{fig_PLm1_equip}). Both of these situations are possible in a real jet 
(at different distances from the source), 
depending upon how fast the vertical field falls off relative to the toroidal field due 
to either the lateral expansion of the jet \citep{1984RvMP...56..255B} or 
the divergence of outer flux surfaces \citep{1994ApJ...426..269B}.

% we have attempted to address 
% in this work is, what is the global stability criterion for internal instabilities? 

\end{itemize}

%===========================================================================
%===========================================================================
%\vspace{-7mm}
% ACKNOWLEDGMENTS
\section*{Acknowledgments}
%\vspace{-2mm}

This work was supported in part by NASA Astrophysics
Theory Program grant NNX14AB37G and NSF grant AST-1411879.
UD is immensely grateful to Ben Brown for allocating time on the LCD machines,
where all the calculations
using Dedalus were performed. UD
specially thanks Phil Armitage, Bhupendra Mishra and Vladimir Zhdankin for general
discussions. The authors thank the anonymous referee for their constructive comments.

%UD thanks Greg Salvesen and Ryan O'Leary for useful discussions.
%===========================================================================
%===========================================================================
%\vspace{-7mm}
% REFERENCES
\bibliographystyle{mn2e}
\bibliography{ref_pp}

\appendix

\section{Linearized MHD equations}
\label{sec_linmhd}

In order to carry out a linear stability analysis, we 
perturb the MHD equations (\ref{mhd1}), (\ref{mhd2}), (\ref{mhd4}) and 
(\ref{mhd5}), and 
retain only the terms of linear order such that
\begin{gather}
\rho = \rho_0 + \rho_1 ~, \\
P = P_0 + P_1 ~,\\
{\bf v} = {\bf v_1} ~,\\
{\bf B} = {\bf B_0} + {\bf B_1} ~,\\
{\bf A} = {\bf A_0} + {\bf A_1} ~.
\end{gather}
We can now write the complete set of the linearized MHD equations for a generic case involving non-axisymmetric perturbations,
background radial gradients, compressibility and radial curvature:
\begin{equation}
\partial_t  \rho_1 + \rho_0 \biggl[ \frac{1}{r} \partial_r (r  v_{1r}) + \frac{1}{r} \partial_\phi  v_{1\phi} +
\partial_z  v_{1z} \biggr] +  v_{1r} \partial_r \rho_0 = 0 ~,
\label{eq_cont}
\end{equation}
\begin{align}
\rho_0 \partial_t v_{1r} + \partial_r \rho_1
+ \frac{1}{4 \pi} \biggl[ B_{0\phi} \partial_r B_{1 \phi} &+ B_{1\phi}\partial_r B_{0\phi}
-  \frac{B_{0\phi}}{r} \partial_\phi B_{1r}
- B_{0z} \partial_z B_{1r} + B_{0z} \partial_r B_{1z}   +   
\frac{2 B_{0\phi} B_{1\phi}}{r}  + B_{1z} \partial_r B_{0z} \biggr]
= 0 ~,
\label{eq_radmom}
\end{align}
\begin{equation}
\rho_0 \partial_t v_{1\phi} 
+ \frac{1}{r} \partial_\phi \rho_1 - \frac{1}{4\pi} \biggl[ B_{1r}\partial_r B_{0\phi} + B_{0z} \partial_z B_{1\phi}
- \frac{B_{0z}}{r} \partial_\phi B_{1z} + \frac{B_{0\phi} B_{1r}}{r}    \biggr]  = 0 ~,
\label{eq_phimom}
\end{equation}
\begin{equation}
\rho_0  \partial_t v_{1z}   + \partial_z \rho_1 +
\frac{1}{4\pi} \biggl[ B_{0\phi} \partial_z  B_{1\phi} - \frac{B_{0\phi}}{r}\partial_\phi B_{1z} - 
B_{1r} \partial_r B_{0z}  \biggr]  = 0~,
\label{eq_zmom}
\end{equation}
\begin{equation}
B_{1\phi} - \partial_z A_{1r} + \partial_r A_{1z} = 0~,
\label{eq_bphi}
\end{equation}
\begin{equation}
B_{1z}  +  \partial_r A_{1\phi}  - \frac{1}{r} \partial_r (r  A_{1r}) = 0  ~,
\label{eq_bz}
\end{equation}
\begin{equation}
\partial_t A_{1r}  -  B_{0z} v_{1\phi} + B_{0\phi} v_{1z} = 0  ~,
\label{eq_ar}
\end{equation}
\begin{equation}
\partial_t A_{1\phi}  + B_{0z} v_{1r} = 0  ~,
\label{eq_aphi}
\end{equation}
\begin{equation}
\partial_t A_{1z}  - B_{0\phi} v_{1r} = 0  ~,
\label{eq_az}
\end{equation}
where the relation $P_1 =   c_s^2 \rho_1$ has been used to eliminate $P_1$. Note that $B_{1r}$ is not 
an independent quantity as it can be defined as $B_{1r} = \partial_\phi A_{1z}/r - \partial_z A_{1\phi}$.
This set of 9 equations 
for the 9 perturbed variables 
%$\rho_1, v_{1r}, v_{1\phi}, v_{1z}, B_{1\phi}, B_{1z}, 
%A_{1r}, A_{1\phi}, A_{1z}$ 
is obtained by imposing the {\it Weyl gauge}, $\psi=0$. However, 
we have also solved these equations for a non-zero $\psi$ 
by imposing the {\it Coulomb gauge}, $\nabla \cdot {\bf A} =0$ (which serves as the 10th equation 
for the 10th variable $\psi$). Both gauges lead to identical results as expected. 
%We will use the above system of equations to solve the global eigenvalue problem in \S \ref{sec_numerical} 
%and \S \ref{sec_global_solns}.

\label{lastpage}
\end{document}